\documentclass[preprint]{aastex63}
\usepackage{natbib}
\usepackage{bm}
\usepackage{graphicx}
\usepackage{comment}
\usepackage[utf8]{inputenc}
\usepackage[english]{babel}
\usepackage{booktabs}
\usepackage{lineno}
\usepackage{changepage}

\newcommand{\tit}[1]{\textit{#1}}
\newcommand{\tbf}[1]{\textbf{#1}}
\newcommand\myfontsize{\fontsize{7pt}{8pt}\selectfont}
\newcommand{\revision}[1]{{#1}}
\newcommand{\secrevise}[1]{{#1}}
\newcommand{\joel}[1]{\textcolor{red}{[Joel comments: #1]}}

\begin{document}

\title{Probing the Molecular Hearts of Extreme Bipolar Planetary Nebulae with ALMA}
\author{Paula Moraga Baez}
\affiliation{School of Physics and Astronomy and Laboratory for Multiwavelength Astrophysics, Rochester Institute of Technology, Rochester, NY 14623, USA}
\author{Joel H. Kastner}
\affiliation{Chester F. Carlson Center for Imaging Science and Laboratory for Multiwavelength Astrophysics, Rochester Institute of Technology, 54 Lomb Memorial Drive, Rochester, NY 14623, USA}
\author{Jesse Bublitz}
\affiliation{Green Bank Observatory, Green Bank, WV, USA}
\author{Javier Alcolea}
\affiliation{Observatorio Astronómico Nacional, C/ Alfonso XII 3\&5, E-28014 Madrid, Spain}
\author{Miguel Santander-Garcia}
\affiliation{Observatorio Astronómico Nacional, C/ Alfonso XII 3\&5, E-28014 Madrid, Spain}
\author{Thierry Forveille}
\affiliation{Institut de Planetologie et d'Astrophysique de Grenoble, France}
\author{Pierre Hily-Blant}
\affiliation{Institut de Planetologie et d'Astrophysique de Grenoble, France}
\author{Bruce Balick}
\affiliation{Department of Astronomy, University of Washington, Seattle, WA 98195, USA}
\author{Rodolfo Montez, Jr.}
\affiliation{Harvard-Smithsonian Center for Astrophysics, 60 Garden Street, Cambridge, MA 02138, USA}
%\linenumbers


\begin{abstract}
%\joel{we should think up a more exciting title...I've made some tweaks to abstract, see what you think}

%    The processes that shape planetary nebulae (PNe) and determine their chemistries can be unveiled through multiwavelength observational studies of molecule-rich PNe ranging from young and rapidly evolving nebulae, such as Hb 5, and NGC 6302, to more "mature" bi-lobed nebulae, such as NGC 2899 and NGC 2818. 
%\joel{let's revisit abstract one last time after the Conclusions are finished}

We present results from a program of Atacama Large Millimeter Array (ALMA) 1.3 mm (Band 6) molecular line mapping of a sample of nearby, bipolar/pinched-waist, molecule-rich PNe (NGC 6302, Hubble 5, NGC 2440, NGC 6445, NGC 2899, and NGC 2818). 
%The resulting velocity-resolved, high-resolution 
%radio interferometric 
%ALMA images of these bipolar PNe yield unique insight into their structures, kinematics, and molecular content. 
Maps of $^{12}$CO(2--1) and $^{13}$CO(2--1) emission as well as emission lines of
%various molecular line tracers of high-energy irradiation including 
HCN, HNC, HCO$^+$, CN, and CS -- many of these detected in these PNe for the first time -- reveal the molecular mass distributions, compositions, 
%content line emission distributions
%key details concerning the compositions 
and velocity fields of the equatorial and, in some cases, polar regions of the sample PNe. In each case, the bulk of the molecular gas traces an expanding equatorial torus, with torus expansion velocities ranging from $\sim$15 to $\sim$50 km s$^{-1}$ and molecular masses from $\sim$0.002 to $\sim$0.1 $M_\odot$. The inferred molecular torus dynamical ages, which span the range $\sim$500 yr (Hb 5) to $\sim$11000 yr (NGC 2818), provide support for a model wherein molecular torus ejection precedes bipolar lobe formation.
% We present diagnostic diagrams for $^{12}$C/$^{13}$C and CN hyperfine ratios to probe progenitor mass ranges and emission-line optical depths, and we further explore the use of 
%and expands on previously studied trends in 
%HCN/HNC ratio and HCO$^+$ emission as irradiation tracers in PNe. 
%The low $^{12}$CO/$^{13}$CO ratios of the sample objects are indicative of relatively massive progenitor stars and/or of the effects of interacting binary systems on AGB evolution.
Collectively, these ALMA survey results provide insight into the rapid structural evolution as well as the zones of irradiated molecular gas within 
%dusty, molecule-rich, 
bipolar PNe that are descended from relatively massive progenitors, likely residing in interacting binary systems, over $\sim$10 kyr of the post-AGB evolution of such systems.
%Our survey hence traces the structural evolution of bipolar PNe that are descended from relatively massive progenitors, likely residing in interacting binary systems, over $\sim$20 kyr of the post-AGB evolution of such systems.  
\end{abstract}

\section{Introduction} \label{sec:intro}

The standard model of the formation of planetary nebulae (PNe) describes their origin in terms of asymptotic giant branch (AGB) stars that eject their dusty envelopes, which are then shaped and ionized by fast winds and UV radiation from their remnant hot cores \citep[and references therein]{Kwok1978,Hofner2018}. This process generates strong shocks, nominally producing dense quasi-spherical or ellipsoidal swept-up shells with rarified interior hot bubbles that are diffuse X-ray emission sources \citep{Kastner2012}. A subset of PNe display bipolar or asymmetric geometries that are theorized to be formed by fast, collimated outflows or jets during post-AGB evolution \citep{Balick2002}. The physical mechanisms that are responsible for these collimated PN outflows, and how collimated outflows interact with and shape the ejected AGB envelope, remain open problems. Models proposed to explain the origin of collimated outflows include outflow launching by a binary companion's accretion disk \citep[e.g.,][and references therein]{DeMarco2017} as well as common envelope (CE) binary evolution \citep{Garcia2018}. Hybrid models, such as repeated close binary encounters leading to an Intermediate-Luminosity Optical Transient (ILOT) \citep{Soker2012}, have also been invoked for the formation of PNe that are too complex in shape to be explained by CE events or continuous collimated flows. A prime example of such an object is the extreme bipolar PN NGC 6302, whose array of outflows, covering a range of dynamical ages, 
%structures of varying ages have been found along with other nebular properties that 
cannot be accounted for by isotropic AGB mass loss followed by collimated but steady post-AGB mass ejection \citep{Balick2023}.

Many \revision{if not most} ``classical,'' pinched-waist, butterfly or hourglass-shaped bipolar PNe, like NGC 6302, also harbor large masses of molecular gas. \revision{Indeed, the presence of bright near-IR H$_2$ and mm-wave CO emission from these molecular gas reservoirs is among the defining features of bipolar PNe --- which, furthermore, are likely the descendants of relatively massive ($\gtrsim$2 $M_\odot$) progenitor stars \citep{CorradiSchwarz1995,Kastner1996,Huggins2005}.} These molecule-rich regions can be studied to probe the kinematics and chemistry of highly structured PNe. For decades, millimeter-wave CO emission lines have been used as probes of the molecular gas components of planetary nebulae \citep[e.g.,][]{Huggins1989}. Measurements of $^{12}$CO and $^{13}$CO line intensities allow mass and optical depth measurements of the molecular gas \citep{Cox1992,Huggins1996}. Subsequent to the early single-dish CO surveys of PNe, single-dish mm-wave molecular line surveys detected HCO$^{+}$, HNC, HCN, CN, CS, and other molecules that serve as probes of molecular chemistry and irradiation \citep[e.g.,][]{Huggins2000,Edwards2013,Edwards2014,Bublitz2019,Schmidt2022}. The molecular ions HCO$^+$, N$_2$H$^+$ and CO$^+$ provide tracers of X-ray ionization and UV irradiation of the nebular material \citep{Deguchi1990,Zhang2008,Bublitz2023}, while the HNC/HCN ratio represents a potent yet underutilized diagnostic of incident UV flux \citep{Bublitz2022}. %Other molecules of particular interest include CN, C$_2$H, and HC$_3$N, which trace the production routes of more complex N- and C-bearing molecules \citep{McGuire2018}. 

Interferometric mm-wave molecular-line mapping observations of PNe, which have been fewer and farther between, facilitate studies of the kinematics and masses of their dense equatorial regions \citep[e.g.,][]{Zweigle1997,Peretto2007,Dinh-V-Trung2008,Santander-Garcia2017,Kastner2024,Kastner2025a}, in addition to spatially resolving their molecular irradiation zones \citep{Bublitz2023}. %\joel{we should probably look for more citations to add here...for ex., should cite SMA NGC 6302 mapping paper from a decade ago...have any other Fig 1 objects been mapped interferometrically? I don't think so, but not sure!}
In this work, we present new interferometric Atacama Large Millimeter Array (ALMA) Band 6 (1.3 mm) observations of a sample of a half-dozen objects that are representative of the class of molecule-rich, high-excitation, bipolar PNe: PN Hb 5 (hereafter Hubble~5%\joel{should make this consistent}
), NGC 2440, NGC 6302, NGC 6445, NGC 2818, and NGC 2899. The aim of our survey is to better constrain and understand the progenitor star systems, shaping histories, and irradiation geometries of these PNe. 
%This program constitutes among few molecular line surveys of PNe to exploit ALMA's unique high spatial and spectral resolution mapping capabilities. 
The velocity-resolved line images (data cubes) allow the analysis of molecular chemistry and nebular kinematics, while mm-wave continuum imaging reveals high-density ionized gas within the PNe. \revision{ The main goals of this ALMA molecular-line mapping survey of pinched-waist, bipolar PNe are twofold: \\
(1) Use velocity-resolved mm-wave molecular line mapping to pinpoint the locations and study the structures and kinematics of the regions of cold, dense molecular gas within high-excitation bipolar PNe, so as to investigate the structural evolution of bipolar PNe that are descended from relatively massive progenitor stars. \\
(2) Ascertain the effects of UV vs. X-ray irradiation from PN central stars on the composition and heating of PN molecular gas. }

This paper presents the survey data and initial survey results, focusing on the molecular species and transitions detected, molecular morphology and its spatial relation to the nebula's ionized gas, and the molecular gas kinematics and masses of the sample PNe. The paper is structured as follows. In \S 2, we describe the selection and characteristics of the target PN sample. In \S 3, we outline the ALMA Band 6 program observations. The resulting molecular line detections for our target PN sample are described in \S 4. In \S 5 and \S 6 we present dynamical age measurements and molecular mass estimates, respectively. We summarize our results and conclusions in \S 7.
%Finally, an analysis and discussion of molecular abundances and age trends is presented in \S 7 and \S 8.

\section{Sample: High-excitation, Molecule-rich, Bipolar PNe}

\begin{figure}[h!]
    \centering
    \includegraphics[width=0.9\textwidth]{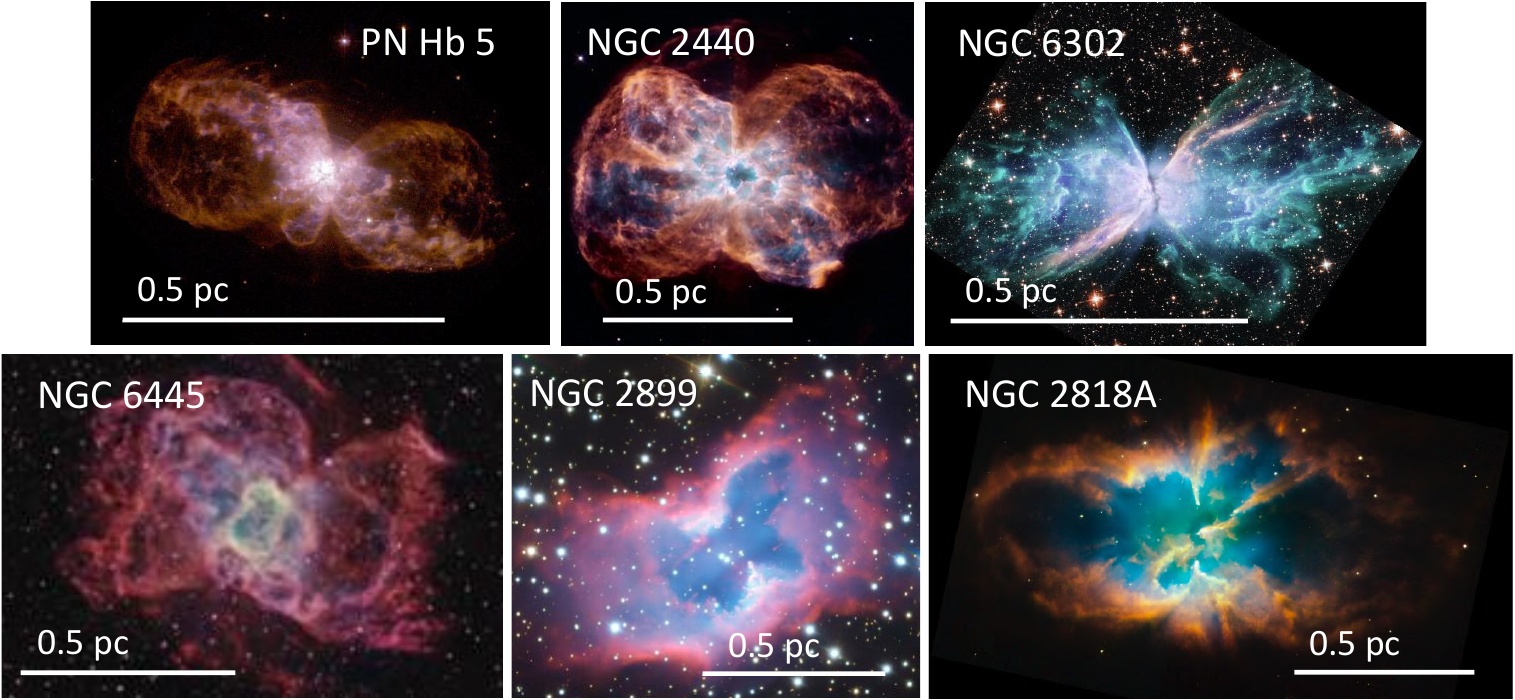}
    \caption{PNe observed in our ALMA Band 6 program. The fiducial 0.5 pc marker is based on the image field of view given the PN distances adopted here (see \S \ref{sec:Dists}). The image of PN Hb 5 was obtained from a Dec. 17, 1997 NASA/ESA/Hubble press release; the image of NGC 6302 was obtained from June 18, 2020 NASA/ESA press release; the image of NGC 6445 was obtained from \citet{Fang2018}; the image of NGC 2440 was obtained from a Feb. 13, 2007 NASA/ESA/STScI press release; the image of NGC 2818 was obtained from a Jan. 28, 2009 NASA/ESA and Hubble Heritage Team (STScI/AURA) press release; the image for NGC 2899 was obtained from a Jan. 15, 2021 ESO/VLT press release through Scientific American.}
    \label{fig:Band6all}
\end{figure}

Figure \ref{fig:Band6all} presents optical images of the six PNe included in our ALMA Band 6 molecular line mapping study. These objects were selected from catalogs of solar neighborhood PNe \citep[e.g.,][]{Frew2016} on the basis of {\it (a)} particularly pronounced bi-lobed and pinch-waisted morphologies, {\it (b)} large masses of molecular gas, as evidenced by detections of bright near-IR H$_2$ \citep{Kastner1996} and mm-wave CO \citep{Huggins1996,Huggins2005}, {\it (c)} exceedingly hot central stars ($T_\mathrm{eff}\gtrsim150$ kK), {\it (d)} large N abundances, and {\it (e)} literature distances of $\lesssim2$ kpc. \revision{Criteria {\it (a)--(d)} are indicative of progenitor stars of mass $\gtrsim$2 $M_\odot$ \citep{CorradiSchwarz1995,Kastner1996,Karakas2016,Henry2018}, providing a sample with which to investigate the structural evolution of bipolar PNe descended from \secrevise{relatively massive progenitors}}. Among these PNe, only NGC 6302 had previously been observed interferometrically in the mm-wave regime \citep{Peretto2007,Dinh-V-Trung2008,Santander-Garcia2017} prior to the ALMA observations presented here.

%While the Figure \ref{fig:Band6all} PNe were selected as representative of the ``extreme'' bipolar, molecule-rich PN class epitomized by NGC 6302, various other PNe would fulfill many or perhaps all of the foregoing criteria. Examples are the very young PN NGC 7027 \citep{Bublitz2023,Moraga2023} and NGC 6537 \citep{Edwards2013}. Indeed, the latter PN was the subject of an ALMA Band 3 molecular line survey (Moraga Baez et al.\ 2025, in preparation). %\joel{let's find a better place for this statement (response to a comment from a coauthor?): As the focus of this paper is PNe belonging to this ``extreme'' bipolar, molecule-rich class, the results should not be extrapolated to the general PN population.}

In the following, we summarize some important properties and characteristics of each of the sample PNe and their central stars, in approximate order of PN lobe dynamical ages. 
%Note that, aside from NGC 6302 \citep{Peretto2007,Santander-Garcia2017}), the sample PNe have never been mapped interferometrically in mm-wavelengths.

\subsection{Hubble 5}\label{sec:Hb5}

Hubble 5 (SIMBAD designation PN Hb 5) is characterized by two main, large lobes that stretch almost 60$''$ across the sky, with projected lobe expansion velocities of $\sim$250 km s$^{-1}$ \citep{Corradi1993,Stanghellini2008,Ortiz2011,Lopez2012}. Additional, smaller lobes are also apparent in the central region of the nebula. %The detection of [S {\sc ii}] was used to constrain the electron densities in Hb 5 to 5$\times$10$^2$ cm$^{-3}$ in its main lobes and 5$\times$10$^3$ cm$^{-3}$ in its central region. 
%While there are some traces of near-IR H$_2$ emission in the lobes of Hb 5, most of the emission is distributed along the central region of the nebula in clumps \citep[and references therein]{Davis2003}. %The NIR spectrum of Hb 5 was modeled to constrain the rotational temperature of the H$_2$ emission in the central and lobed regions, yielding temperatures of 2 kK and 1.1$-$1.5 kK, respectively \citep{Davis2003}. 
In an Infrared Space Observation (ISO) far-IR spectral modeling study, the central star of Hubble 5 was found to have an intrinsic luminosity of 6800 L$_{\odot}$ and an effective temperature of $\sim$170 kK, suggesting a progenitor mass of $>$ 4 M$_{\odot}$  \citep{Pottasch2007}. Diffuse X-ray emission with a luminosity of 0.04 L$_\odot$ was detected in Hubble 5, indicative of the presence of strong shocks \citep{Montez2009,Freeman2014}. 
%\citet{Lopez2012} constructed a morpho-kinematic model of Hb 5 using SHAPE. %The model takes into account the non-homologous lobes by assuming a dense, gaseous core that has areas of low density, thus allowing the lobes to emerge in a point-symmetric manner \citep{Lopez2012}. 
On the basis of morpho-kinematic modeling of Hubble 5, \citet{Lopez2012}  determined the PN to be $\sim$1500 yrs old and to contain 0.0015 M$_\odot$ of ionized gas. \citet{Steffen2013} used a hydrodynamical model to confirm that the model formulated by \citet{Lopez2012} is indeed a plausible formation scenario for Hubble 5, and estimated the mass of its progenitor star as $>$ 5 M$_{\odot}$. 

Single-dish radio observations have detected $^{12}$CO, HCN, HNC, HCO$^+$, and CCH \citep{Schmidt2016,Schmidt2017a,Schmidt2017b}, indicating that Hubble 5 is C-rich. These molecular line studies yield an estimate of the H$_2$ density within the central region of Hubble 5 of a few $\times$10$^5$ cm$^{-3}$.
%$\sim$1.2$-$7.4$\times$10$^5$ cm$^{-3}$ \citep{Schmidt2016,Schmidt2017b}. 

\subsection{NGC 2440}\label{sec:N2440}
%\joel{I streamlined this subsection, and rearranged to be optical then radio work (molecules), like the preceding subsections}

The morphology of NGC 2440 has been described as multi-polar, with its main, bubble-like ionized lobes oriented approximately E--W and a secondary pair of lobe structures oriented more nearly NE--SW \citep[and references therein]{Lopez1998}. 
%The spectroscopic study by \citet{Aller1968} yielded detections of more than two dozen permitted and forbidden lines, and H and He recombination lines along with excited species of N, O, C, Si, S, Fe, F, Ne, Mg, and Ar. 
%\citet{Lopez1998} obtained narrowband ([N {\sc ii}] and [O {\sc iii}]) images of this PN that distinguished three lobe-like structures within the two main outer lobes, and concluded that multiple mass ejections must have occurred. %With a more complete set of panchromatic images, \citet{Cuesta2000} produced total flux, electron density, electron temperature, and extinction maps. 
%\citet{Cuesta2000} concluded that the electron temperature is nearly uniform throughout NGC 2440, peaking in the central region of the PN, with the extinction peaking slightly north of the central region. A long slit 310--690 nm spectroscopy study found that NGC 2440 is abundant in He and N, and that overall O is more abundant than S \citep{Krabbe2006}.
%Near-IR imaging of H$_2$ emission shows that this emission lies primarily near the core of NGC 2440 \citep{Kastner1996}. A \textit{Spitzer} study of archival IRS spectra determined the excitation temperature for the northern region of H$_2$ emission to be $\sim$830 K \citep{Mata2016}. 
%\joel{Miller+2019 obtained spectra of the nebula not the CSPN, so their results (like everyone else's) for CSPN effective temp and luminosity are heavily model-dependent, as they lay out in the paper; but their Table 9 has a nice/handy summary of the various results, so best just to refer to that summary, like so:} 
Estimates of the effective temperature and luminosity of the central star of NGC 2440 obtained from modeling of nebular spectra \citep[][and references therein]{Miller2019} range from $\sim$170 to $\sim$210 kK and $\sim$500 to $\sim$3000 $L_\odot$, respectively. 
%\citet{Miller2019} studied NGC 2440 using HST's Space Telescope Imaging Spectrograph (STIS) and, via modeling, determined central star parameters of $T_{eff} \sim170$ kK and , and 
\citet{Miller2019} estimated progenitor and present-day central star masses of $\sim$2.8 M$_{\odot}$ and $\sim$0.7 M$_{\odot}$, respectively; the latter is in good agreement with the estimate by \citet{Wolff2000}.

\citet{Wang2008} obtained a single-dish map of $^{12}$CO(3$-$2) emission from the nebula that revealed CO within its core region as well as two knot-like structures located at the tips of the secondary lobes; these investigators estimated that the nebula contains a total molecular mass of $\sim$0.01 $M_\odot$.
% \joel{they didn't derive $T_{ex}$, they just assumed $T_{ex} = 25$ K}
%constrained the excitation temperature to $\sim$25 K. 
Other molecules that have been detected near the core nebular region include $^{13}$CO, H$^{12}$CN, H$^{13}$CN, and HCO$^+$ \citep{Schmidt2016,Ziurys2020}. 
%Through analysis of these emission lines, 
On the basis of HCN emission flux being much higher than that of HCO$^+$ emission,
\citet{Schmidt2016} concluded that NGC 2440 is abundant N.

\subsection{NGC 6302}\label{sec:N6302}
Characterized by its bright and highly-excited bipolar lobes, NGC 6302 is known as the classic example of the class of bilobed, pinched-waist PNe \citep{Aller1981}. %The vast range of ionization states present in its bipolar lobes have been the subject of studies that range from far-UV through optical and mid-IR \citep{Cassassus2000,Feibelman2001,Molster2001,Groves2002,Kastner2022}. 
The bipolar lobes display velocities of $\sim$600 km s$^{-1}$, which indicate lobe dynamical ages of $\sim$2000 yrs \citep{Meaburn2005,Meaburn2008,Szyszka2011}. The nebular core is known to harbor one of the hottest central stars of any PN, with $T_{eff}$ estimates ranging from 220$-$400 kK \citep{Cassassus2000,Wright2011,Ashley1988}.
%NGC 6302 has been the subject of HST imaging in order to understand its complex morphology \citep{Szyszka2009,Kastner2022,Balick2023}. The recent \citet{Kastner2022} HST/WFC3 emission-line imaging survey of NGC 6302, which spanned wavelengths from near-UV to near-IR, yielded the discovery of bright 1.64 $\mu$m [Fe {\sc ii}] emission aligned along the two main lobes from SE to NW, tracing shocked regions caused by collimated winds with speeds $\geq$100 km s$^{-1}$. 
Comparison of [N {\sc ii}] images obtained with WFC3 roughly a decade apart demonstrates that various structures and clumps within the lobes span a large range of dynamical ages, from $\sim$300$-$2300 yr, indicating non-homologous expansion or perhaps multiple mass ejection events \citep{Balick2023}.

Within the dusty central torus of NGC 6302, C and O bearing molecules as well as PAHs, crystalline silicates, and H$_2$O ice have been detected \citep{Molster2001,Kemper2002,Santander-Garcia2017}. Emission from CO has been observed in NGC 6302 beginning in the late 1980s \citep{Huggins1989}, and it is the only Fig.~\ref{fig:Band6all} object that has been the subject of interferometric mapping of CO \citep{Peretto2007}. Only recently, however, did ALMA CO mapping reveal the emission as tracing the nebula's dense, dusty molecular torus and the inner lobe regions near the waist of the nebula \citep{Santander-Garcia2017}. The equatorial torus shows slow expansion at $\sim$8 km s$^{-1}$, indicating a dynamical age of $\sim5000-7500$ yrs \citep{Peretto2007,Santander-Garcia2017}, much larger than the dynamical age range of the bipolar lobes.

\subsection{NGC 6445}\label{sec:N6445}
NGC 6445 was first characterized by its bright elliptical center \citep{Aller1973}. Its faint lobes were later captured in deeper H$\alpha$ images, and H$_2$ emission was detected near its core, cementing its classification as a bipolar PN \citep{Balick1987,Schwarz1992,Kastner1996,Cuesta1999}.
% In a study conducted by \citet{Cuesta1999}, the nebular extinction was found to reach its peak along NGC 6445's central elliptical structure, and the electron density followed the same pattern. The strong emission in [O {\sc iii}] and [S {\sc ii}] that NGC 6445 exhibits in this region led \citet{Cuesta1999} to conclude that the elliptical structure may be a highly shocked nebular shell. 
%Traces of Si grains were attributed to the NGC 6445's prominent elliptical shell, which agreed with PAHs detected along its shell with IRAC \citep{vanHoof2000,Phillips2010}. In later studies, it became apparent that NGC 6445 had a more complex H$_2$ morphology, dominating the central ellipse and wrapping around the nebula in an almost S-shaped manner \citep{Fang2018}. \citealt{Fang2018} asserted that the H$_2$ filament structures that wrap around the large [N {\sc ii}]-dominated lobes could be the outer edges of a torus viewed edge-on. %Through the analysis of archival \textit{Spitzer} data, the rotational temperature of the H$_2$ emission was constrained to $\sim$790 K, although previous results indicated a temperature of $\sim$1520 K \citep[and references therein]{Mata2016}.
\citet{vanHoof2000} conducted a UV and far-IR study and determined the effective temperature of the central star of the planetary nebula (CSPN) to be $\sim$188 kK, with a final mass of 0.625$-$0.696 M$_{\odot}$, and a nebular dynamical age of approximately 3300 yrs if a distance of 1.5 kpc is assumed \citep{Phillips1988}. NGC 6445 displays point-source X-ray emission at the position of the central star that may be due either to circumstellar shocks or coronal activity of a main-sequence companion \citep{Montez2015}. 

Molecules were first detected in NGC 6445 by \citet{Huggins1989} in their early survey of $^{12}$CO(2$-$1) emission from PNe. %\joel{the angular diameter listed in their Table 3 is the optical nebula, not CO; HH89's survey used single pointings of a single dish}
%that spanned $\sim$17$''$ across the nebula's center. 
Subsequently, $^{13}$CO, OH, OH$^+$, HCN, HNC, and HCO$^+$ have been detected in various single-dish line surveys \citep{Aleman2014,Bublitz2019,Schmidt2022}.

\subsection{NGC 2899}\label{sec:N2899}
NGC 2899 rivals NGC 2818 as the most evolved PN in our ALMA survey, given its large angular extent ($\sim$145$''$). Although this nebula has been included in multiple surveys dating back five decades \citep{Greig1972,Bohurski1974,Pottasch1984}, there are few dedicated studies of the PN. NGC 2899 is characterized by its large, open bipolar lobes and filamentary structures that appear along its pinched-waist minor axis. Via long-slit echelle spectroscopy, the nebular ionized regions were found to have expansion velocities of $\sim$50 km s$^{-1}$, with high-velocity structures reaching $\sim110-130$ km s$^{-1}$ \citep{Lopez1991}. The effective temperature of the CSPN is evidently extremely high, with literature estimates ranging between $\sim$215--255 kK \citep{Kaler1989,Lopez1991,Drew2014}. Thus, while the central star's initial mass has been estimated as $\sim$1.2 M$_{\odot}$ \citep{Kaler1989}, NGC 2899 is likely the product of a much more massive progenitor. 

The new ALMA data presented here for NGC 2899 constitute the first mm-wave molecular line observations of the PN since its detection in CO by \citet{Huggins1996}.

\subsection{NGC 2818A}\label{sec:N2818}
NGC 2818A (hereafter NGC 2818) is highly unusual among PNe for its apparent association with an open cluster, which actually carries the official designation NGC 2818 \citep{Tifft1972,Kohoutek1986,Bonatto2008,Rani2023,Fragkou2025}. In H$\alpha$, the morphology of the PN NGC 2818 consists of two distinct lobes with a diameter of nearly 110$''$, and what appears to be the remnants of a torus at its center. NGC 2818 is one of the more evolved PN in our target group, with kinematical age measurements ranging from 6500$-$12300 yr \citep{Vazquez2012,Derlopa2024}. The initial and present-day mass of the nebula's central star have been estimated as $\sim$2 M$_{\odot}$ and $\sim$0.6 M$_{\odot}$, respectively \citep{Dufour1984,Derlopa2024,Fragkou2025}. Estimates of the central star temperature range from $\sim$130 kK \citep{Fragkou2025} to $\sim$160 kK \citep{Vazquez2012,Mata2016}.
%Early ground-based spectroscopy revealed a deficit of C and large abundances of N, O, Ne, Ar, Si, and Cl \citep{Dufour1984}. HST/WFPC2 images confirmed the presence of strong [N {\sc ii}], [O {\sc iii}], and [S {\sc ii}] emission lines \citep{Vazquez2012}. There was some correlation found between near-IR H$_2$ emission and H$\alpha$ with the strongest molecular hydrogen emission lying outside the hydrogen recombination emission \citep{Huggins1996,Schild1995,Kastner1996}. \citealt{Mata2016} failed to detect PAHs in NGC 2818, and found the T$_{ex}$ of H$_2$ to be $\sim$850 kK. 
A recent study of the 3D kinematics of the ionized lobes of NGC 2818 by \citet{Derlopa2024} indicates the lobes contain filamentary arms that are expanding at 180 km s$^{-1}$. Its central torus was modeled as an equatorial structure expanding at 70 km s$^{-1}$ with an inclination angle of 80$-$70 deg with respect to the plane of the sky \citep{Derlopa2024}. 

Like NGC 2899, the PN NGC 2818 was detected in the \citet{Huggins1996} single-dish CO survey but, prior to the ALMA mapping presented here, had not been the subject of subsequent mm-wave molecular line observations.

\subsection{Distances to ALMA targets}\label{sec:Dists}

\begin{table}
%\begin{center}
\caption{\sc ALMA Target Distances}
\vspace{.1in}
\label{tbl:AllDist}
\footnotesize
%\begin{adjustwidth}{-2.1cm}{}
%\begin{adjustbox}{width=1.1\textwidth}
\begin{tabular}{lcccccc}
\toprule
PN G & Name & BS2023 D$^a$ & Lit. D$_{lower}$ & Lit. D$_{upper}$ & Adopted D$^b$ & References$^c$\\
& & (kpc) & (kpc) & (kpc) & (kpc)\\
\midrule
\midrule
359.3-00.9 & PN Hb 5 & 0.9$^{+0.1}_{-0.1}$ & 1.7%$^{+0.3}_{-0.3}$
& 2.0 & 0.9$^{+0.1}_{-0.1}$ & SH10, S86\\
234.8+02.4 & NGC 2440 & 1.0$^{+0.1}_{-0.1}$ & 0.8 & 2.2 & 1.0$^{+0.1}_{-0.1}$ & G97, W08\\
349.5+01.0 & NGC 6302 & 0.5$^{+0.1}_{-0.1}$ & 0.8%$^{+0.1}_{-0.1}$
& 1.17%$^{+0.01}_{-0.01}$ 
& 1.0$^{+0.1}_{-0.1}$ & L19, M08\\
008.0+3.9 & NGC 6445 & 1.1$^{+0.1}_{-0.1}$ & 1.4 & 1.5 & 1.1$^{+0.1}_{-0.1}$ & S08, SH10\\
277.1-03.8 & NGC 2899 & 0.8$^{+0.2}_{-0.2}$ & 0.9 & 1.4 & 0.8$^{+0.2}_{-0.2}$ & D82, R99\\
261.9+08.5 & NGC 2818 & 1.5$^{+0.1}_{-0.1}$ & 3.1 & 4.2%$^{+1.7}_{-1.3}$ 
& 3.3$^{+0.1}_{-0.1}$ & P21, BJ21\\
% old D to NGC 2818: 2.908$^{+0.454}_{-0.343}$
\midrule
\end{tabular}
%\end{adjustbox}
%\end{adjustwidth}
%\end{center}

{\sc Notes:} (a) Gaia DR3 statistically-calibrated distances from \citet{Bucciarelli2023}. (b) Distances adopted for this work. (c) References for literature distances (columns Lit. D$_{lower}$ and Lit. D$_{upper}$): S08 \citep{Stanghellini2008}, SH10 \citep{Stanghellini2010}, G97 \citep{Gorny1997}, W08 \citep{Wang2008}, P21 \citep{Poggio2021}, BJ21 \citep{BailerJones2021}, D82 \citep{Daub1982}, R99 \citep{Rauch1999}, L19 \citep{Lago2019}, M08 \citep{Meaburn2008}, S86 \citep{Sabbadin1986}. 
\end{table}

\revision{The basis for the adopted distance to each ALMA Band 6 target PN is summarized in Table \ref{tbl:AllDist}, with individual cases discussed below. As is apparent from the Table --- and as is typical for PNe whose central stars lack Gaia parallaxes --- previously derived distances vary widely for most of the sample objects.
%The distance to NGC 2440 has been estimated at 0.8 kpc \citep{Gorny1997}, with much larger estimates of 1.9 kpc \citep{Frew2013} and 2.19 kpc \citep{Wang2008}. In the case of NGC 2818, distance estimates range from 1.79 kpc \citep{Phillips2004} %, 2.1 kpc \citep{Kastner1996} to as large as 4.2 kpc \citep{Derlopa2024}. Under the assumption that the PN is in fact associated with the NGC 2818 open cluster, \revision{its distance} would be 3.25 kpc \citep[and references therein]{Rani2023,Fragkou2025}. Distance estimates for NGC 2899 range from 0.86 kpc to 1.4 kpc \citep{Cahn1971,Daub1982,Rauch1999,Stanghellini2008}.
%has been estimated to be 0.86--1.1 kpc \citep{Cahn1971,Daub1982,Stanghellini2008} with an upper limit of 1.4 kpc \citep{Rauch1999}. Distance estimates for NGC 6302 range from 0.805 kpc to 1.170 kpc \citep{Meaburn2008,Lago2019,Gomez2020}. Hubble 5 has been estimated to be as distant as 2.0 kpc \citep{Sabbadin1986}, with more recent distance estimates of 1.685 kpc \citep{Stanghellini2008} and 1.706 kpc \citep{Stanghellini2010}, while estimates for NGC 6445's distance range between 1.38 kpc and 1.5 kpc \citep{Stanghellini2008,Stanghellini2010}. 
Recent} analysis of Gaia Data Release 3 (DR3) data has yielded a distance scale based on a far larger set of calibrators \citep{Bucciarelli2023}, and provides revised distance estimates for all of our sample PNe. 
%\joel{are any of the ``Gaia D's'' in Table 1 actually Gaia parallax distances, as opposed to Gaia-calibrated D's? If so, need to call those Gaia parallax D's out explicitly, via footnote as well as here in text.}
%, with reliable error bars produced by error propagation. 
\revision{These Gaia DR3 statistically-calibrated distances --- which are not to be confused with Gaia DR3 parallax distances to the PN central stars --- are listed in Table \ref{tbl:AllDist} in column ``BS2023 $D$''.}
%\footnote{We stress that the ``Gaia $D$'' distances listed in Table \ref{tbl:AllDist} are not directly derived from Gaia parallaxes; see \citet{Bucciarelli2023} for details.}. 
We adopted these distances and distance uncertainties for all PNe except NGC 2818 and NGC 6302. 
%In the case of NGC 2818, the parallax measured by Gaia is smaller than its measured error, which also points to an unreliable distance. \paula{Mention in this par that the parallax was statistically corrected which is why the distance errors don't reflect the large parallax errors. The initial parallax measurement is not trusted, which is why we are choosing to ignore this distance.} 
%\joel{I think the point for NGC 2818 is that the association w/ the cluster has become more secure of late, so the D has to be more like 3 kpc:} 
Given that recent analyses have placed the association of NGC 2818 with the open cluster NGC 2818 on more solid footing \citep{Rani2023,Fragkou2025}, we follow \citet{Fragkou2025} and adopt the Gaia-DR3-based distance to the cluster, 3.3 kpc, as the distance to the PN. 
In the case of NGC 6302, the \citet{Bucciarelli2023} DR3-based distance calibration places this nebula at $\sim0.545$ kpc, \revision{a factor $\sim$2 closer than estimates based on detailed morpho-kinematic modeling \citep[e.g.,][]{Meaburn2008,Lago2019,Gomez2020}. Hence, instead of adopting the \citet{Bucciarelli2023} distance, we take the mean of these reliable literature distances. Although the \citet{Bucciarelli2023} distances to Hb 5 and NGC 6445 are also significantly smaller than the minimum literature distances, most of these previous literature estimates were obtained via pre-Gaia parallax-based statistical methods; we hence adopt the \citet{Bucciarelli2023} Gaia DR3 statistically-calibrated distances for these two PNe.}

\revision{The distances adopted for purposes of the analysis in this paper are listed in column ``Adopted $D$'' of Table \ref{tbl:AllDist}.}

\section{Observations} \label{sec:data}
Our ALMA Band 6 survey, conducted under ALMA programs 2021.1.00456.S, 2021.2.00004.S, and 2022.1.00401.S, used both the 12-meter telescope array and the Atacama Compact (7-meter telescope) Array (hereafter referred to as 12-m and ACA, respectively). %\joel{I think it's best to use those two 'abbreviations' throughout, esp ACA as opposed to 7-m...check for consistency, through the rest of the paper} 
We configured the 12-m and ACA spectrometers in a half-dozen spectral setups in the 220--270 GHz frequency range to target emission lines of CO, CN, HCN, HNC, HCO$^{+}$, CO$^{+}$, CS, SO, and isotopologues of CO, HCN, and HCO$^{+}$ in the sample PNe. Details of the science goal (spectrometer) setups, including targeted emission lines, velocity resolution, and bandwidths, are listed in Table \ref{tbl:allConfig}. \revision{The combination of 12-m array and ACA observations facilitates mapping of molecular gas structures over a wide range of nebular angular size scales and, in some cases, confirms the detection of weak emission lines that appear only in finer structures (resolved with the 12-m array) or extended, diffuse structures (mapped with ACA).} Array configurations used in this work are Band 6 ACA (angular resolution of 5.45$''$), Band 6 12-m C43-1 (1.47$''$), Band 6 12-m C43-3 (0.62$''$), and Band 6 12-m C43-4 (0.40$''$) (see ALMA Cycle 11 Technical Handbook; \citealt{ALMA11}). The velocity resolution for each spectral window ranges from 0.163 to 1.40 km s$^{-1}$ for the targeted emission lines.

A total of 43 observations were made using these ACA, C43-1, C43-3, and C43-4 configurations. Integration times varied between $\sim$5--46 minutes depending on the configuration that was being used, with ACA and 12-m configurations of typical duration $\geq$19 minutes and $\leq$12 minutes, respectively. Table \ref{tbl:allCal} lists details of the ALMA arrays and data processing, including the acquisition dates, number of antennas, time elapsed during observations, baselines for each configuration, angular resolution, maximum recoverable scale (MRS), and phase and flux calibrators.

Data presented and analyzed here were obtained directly from the ALMA pipeline, run with the Common Astronomy Software Applications (CASA)\footnote{CASA (https://casa.nrao.edu/) is the primary data processing software for the Atacama Large Millimeter/submillimeter Array (ALMA) and NSF's Karl G. Jansky Very Large Array (VLA), and is also frequently used for other radio telescopes. CASA software can process data from both single-dish and aperture-synthesis telescopes, and one of its core functionalities is to support the data reduction and imaging pipelines for ALMA, VLA and the VLA Sky Survey (VLASS). CASA is being developed by an international team of scientists based at the National Radio Astronomical Observatory (NRAO), the European Southern Observatory (ESO), and the National Astronomical Observatory of Japan (NAOJ), under the guidance of NRAO. For more information see \citealt{CASA2022}} package version 6.2.1.7 for programs 2021.1.00456.S and 2021.2.00004.S, and CASA version 6.4.1.12 for program 2022.1.00401.S. Cleaning was performed with \texttt{tclean} using a Briggs-tapered weighting (\texttt{brigsbwtaper}) for emission lines with a 0.5 robust parameter. Briggs-weighting and a robust parameter of 0.5 were used for continuum images. The resulting molecular line spatial-spectral data cubes, spanning the full spectrometer bandwidths, were then processed via the ALMA Data Mining Toolkit (ADMIT)\footnote{The ADMIT package products and usage guide are located at https://admit.astro.umd.edu.} software package to identify molecular emission lines and extract spatial-spectral data subcubes isolating the line emission, as well as to generate velocity-integrated (``moment 0'') as well as velocity-weighted (``moment 1'') emission-line images. In cases of non-detections of targeted lines (i.e., no detection during ADMIT processing), spectral windows were examined visually to verify that no emission is present above the noise, with the detected $^{12}$CO(2--1) emission defining the velocity range over which to search for signal. 
%In those cases, non-detection is defined as the ratio of mean flux (over the relevant velocity range) to root mean square noise being $\approx$1. 

\begin{table}
%\begin{center}
\caption{\sc ALMA Band 6 Science Goal Spectrometer Setups}
\vspace{.1in}
\label{tbl:allConfig}
\footnotesize
\begin{tabular}{lrccc}
\toprule
Setup & Center Freq. & Line Targets & Resolution & Bandwidth \\
 & (GHz) & & (km s$^{-1}$) & (MHz)\\
\midrule
\midrule
B6-1 & 213.000 & Continuum & 43.979 & 1875.00\\
    & 226.697 & CN $N=$2$-$1, $J=$3/2$-$1/2, $F=$3/2$-$1/2 & 0.747 & 937.50\\
    & 226.874 & CN $N=$2$-$1, $J=$5/2$-$3/2, $F=$7/2$-$5/2 & 0.747 & 937.50\\
    & 230.538 & $^{12}$CO $J=$2$-$1 & 0.734 & 937.50 \\
    & 241.016 & C$^{34}$S $J=$5$-$4 & 1.400 & 1875.00\\
    & 244.936 & CS $J=$5$-$4 & 1.382 & 1875.00\\
\midrule
B6-2 & 250.500 & Continuum & 37.395 & 1875.00\\
    & 251.857 & SO 3$\Sigma$ $N_J=5_6-4_5$ & 0.672 & 468.75\\
    & 265.886 & HCN $J=$3$-$2 & 0.550 & 937.50 \\
    & 267.557 & HCO$^+$ $J=$3$-$2 & 0.632 & 937.50 \\
\midrule
B6-3  & 256.302 & H(29)$\alpha$ & 0.165 & 117.19 \\
    & 259.011 & H$^{13}$CN $J=$3$-$2 & 0.163 & 117.19\\
    & 260.255 & H$^{13}$CO$^+$ $J=$3$-$2 & 0.163 & 117.19\\
    & 271.981 & HNC $J=$3$-$2 & 0.622 & 937.50\\
\midrule
B6-4 & 219.560 & C$^{18}$O $J=$2$-$1 & 0.771 & 468.75 \\
    & 219.949 & SO 3$\Sigma$ $N_J=6_5-5_4$ & 0.769 & 468.75\\
    & 220.398 & $^{13}$CO $J=$2$-$1 & 0.768 & 937.50 \\
    & 231.901 & H(30)$\alpha$ & 0.730 & 937.50 \\
    & 233.500 & Continuum & 40.118 & 1875.00\\
    & 235.789 & CO$^+$ $J=$2$-$1 $F=$3/2$-$1/2 & 0.718 & 468.75 \\
\midrule
\end{tabular}
%\end{center}
\end{table}

\begin{table}
\caption{\sc Details of Band 6 Observations}
\label{tbl:allCal}
\begin{center}
\scriptsize
%\begin{adjustwidth}{-0.7cm}{}
\begin{tabular}{lrccccccccc}
\toprule
Setup & Target & Date & Configuration & No. Ant. & Int. & Baselines & Res. & MRS & Phase Cal. & Flux Cal. \\
& & & & & (min.) & (meters) & ($''$) & ($''$) & &\\
\midrule
\midrule
B6-1 & PN Hb 5 & Oct 25, 2022 & 12-m C43-1 & 43 & 5 & 14--312 & 1.47 & 12.4 & J1820--2528 & J1924--2914\\
    & & Dec 28, 2022 & ACA & 8 & 37 & 9--47 & 5.45 & 29.0 & J1820--2528 & J1617--5848\\
    & & Mar 19, 2023 & 12-m C43-4 & 36 & 12 & 15--918 & 0.40 & 4.89 & J1820--2528 & J1924--2914\\
    & NGC 2440 & Apr 03, 2022 & ACA & 10 & 19 & 8--44 & 5.45 & 29.0 & J0730--1141 & J0538--4405\\
    & & Jan 15, 2023 & 12-m C43-4 & 46 & 6 & 15--783 & 0.40 & 4.89 & J0748--1639 & J0750$+$1231\\
    & NGC 6302 & Aug 31, 2022 & 12-m C43-4 & 44 & 11 & 15--783 & 0.62 & 7.02 & J1717--3342 & J1617--5848\\
    & & Jan 01, 2023 & ACA & 9 & 19 & 8--48 & 5.45 & 29.0 & J1717--3342 & J1427--4206 \\
    & NGC 6445 & Oct 25, 2022 & 12-m C43-1 & 43 & 5 & 14--312 & 1.47 & 12.4 & J1820--2528 & J1924--2914\\
    & & Dec 28, 2022 & ACA & 8 & 37 & 9--47 & 5.45 & 29.0 & J1820--2528 & J1617--5848\\
    & & Mar 19, 2023 & 12-m C43-4 & 36 & 12 & 15--918 & 0.40 & 4.89 & J1820--2528 & J1924--2914\\
    & NGC 2899 & Apr 24, 2022 & ACA & 10 & 42 & 8--44 & 5.45 & 29.0 & J0940--6107 & J1058$+$0133\\
    & & Jan 10, 2023 & 12-m C43-4 & 46 & 6 & 15--783 & 0.40 & 4.89 & J0852--5755 & J1058$+$0133\\
    & NGC 2818 & Mar 28, 2023 & ACA & 10 & 19 & 8--44 & 5.45 & 29.0 & J0922--3959 & J1058$+$0133\\
    & & Jan 15, 2023 & 12-m C43-4 & 46 & 6 & 15--783 & 0.40 & 4.89 & J0922--3959 & J0854$+$2006\\
\midrule
B6-2 & PN Hb 5 & Jan 01, 2023 & ACA & 9 & 19 & 8--48 & 5.45 & 29.0 & J1717--3342 & J1517--2422\\
    & NGC 2440 & Dec 25, 2022 & 12-m C43-3 & 42 & 9 & 15--500 & 0.62 & 7.02 & J0748-1639 & J0750$+$1231\\
    & & Mar 23, 2023 & ACA & 9 & 19 & 8--44 & 5.45 & 29.0 & J0730--1141 & J0538--4405\\
    & NGC 6302 & Aug 29, 2022 & 12-m C43-3 & 43 & 11 & 15--783 & 0.40 & 4.89 & J1717--3342 & J1617--5848\\
    & & Jan 12, 2023 & ACA & 10 & 28 & 8--48 & 5.45 & 29.0 & J1717--3342 & J2258--2758\\
    & NGC 2899 & Apr 28, 2022 & ACA & 9 & 48 & 8--44 & 5.45 & 29.0 & J0940--6107 & J1058$+$0133\\
    & & Jan 07, 2023 & 12-m C43-3 & 43 & 11 & 15--650 & 0.62 & 7.02 & J0852--5755 & J1107--4449\\
    & NGC 2818 & Apr 22, 2022 & ACA & 10 & 28 & 8--44 & 5.45 & 29.0 & J0922--3959 & J0854$+$2006\\
    & & Dec 24, 2022 & ACA & 10 & 28 & 8--48 & 5.45 & 29.0 & J0922--3959 & J1058$+$0133\\
    & & Jan 10, 2023 & 12-m C43-3 & 46 & 9 & 15--783 & 0.62 & 7.02 & J0922--3959 & J1037--2934\\
\midrule
B6-3 & PN Hb 5 & Dec 29, 2022 & 12-m C43-3 & 36 & 9 & 15--500 & 0.62 & 7.02 & J1744--3116 & J1427--4206\\
    & NGC 2440 & Mar 23, 2022 & ACA & 9 & 19 & 8--44 & 5.45 & 29.0 & J0730--1141 & J0538--4405\\
    & & Dec 24, 2022 & 12-m C43-3 & 41 & 9 & 15--500 & 0.62 & 7.02 & J0748--1639 & J0750$+$1231\\
    & NGC 6302 & Aug 30, 2022 & 12-m C43-4 & 46 & 11 & 15--783 & 0.62 & 7.02 & J1717--3342 & J1517--2422 \\
    & & Jan 01, 2023 & ACA & 9 & 28 & 8--48 & 5.45 & 29.0 & J1717--3342 & J1427--4206\\
    & NGC 6445 & Apr 07, 2023 & ACA & 10 & 28 & 8--47 & 5.45 & 29.0 & J1733--1304 & J1924--2914\\
    & NGC 2899 & Apr 14, 2022 & ACA & 10 & 40 & 8--44 & 5.45 & 29.0 & J0940--6107 & J1058$+$0133\\
    & & Jan 10, 2023 & 12-m C43-3 & 43 & 9 & 15--783 & 0.62 & 7.02 & J0852--5755 & J0519--4546\\
    & NGC 2818 & May 18, 2022 & ACA & 9 & 28 & 8--44 & 5.45 & 29.0 & J0922-3959 & J1058$+$0133\\
    & & Oct 15, 2022 & 12-m C43-3 & 44 & 11 & 14--500 & 0.62 & 7.02 & J0922--3959 & J1058$+$0133\\
    & & Dec 23, 2022 & ACA & 11 & 28 & 8--48 & 5.45 & 29.0 & J0922--3959 & J0854$+$2006\\
\midrule
B6-4 & NGC 2440 & Mar 27, 2022 & ACA & 8 & 37 & 8--44 & 5.45 & 29.0 & J0730--1141 & J0854+2006\\
    & & Jan 18, 2023 & 12-m C43-3 & 43 & 34 & 15--782 & 0.62 & 4.89 & J0730--1141 & J0854$+$2006\\
    & NGC 6302 & Aug 30, 2022 & 12-m C43-3 & 46 & 32 & 15--783 & 0.40 & 4.89 & J1717--3342 & J1617--5848\\
    & & Oct 26, 2022 & 12-m C43-1 & 43 & 11 & 14--312 & 1.47 & 12.4 & J1717-3342 & J1924-2914\\
    & NGC 2899 & May 23, 2022 & ACA & 9 & 42 & 8--44 & 5.45 & 29.0 & J0852--5755 & J1058$+$0133\\
    & & Jan 11, 2023 & 12-m C43-4 & 45 & 26 & 15--783 & 0.40 & 4.89 & J0852--5755 & J1058$+$0133\\
    & NGC 2818 & Mar 29, 2022 & ACA & 10 & 46 & 8--44 & 5.45 & 29.0 & J0922--3959 & J1058$+$0133\\
    & & Jan 16, 2023 & 12-m C43-4 & 44 & 23 & 15--783 & 0.40 & 4.89 & J0922--3959 & J1058$+$0133\\
\midrule
\end{tabular}
%\end{adjustwidth}
\end{center}
\end{table}

\section{Results} \label{sec:Results}

%\begin{flushleft}
\begin{table}
\caption{\sc Band 6 Observations}
%\vspace{0.2in}
\begin{center}
\label{tbl:allObs}
\myfontsize
\begin{tabular}{lrccccccc}
\toprule
& & & \multicolumn{6}{c}{Flux$^a$ (Error$^b$)} \\
\cmidrule(lr){4-9}
ALMA&Mol.&$\nu$&PN Hb 5&NGC 6302&NGC 2440&NGC 6445&NGC 2818&NGC 2899\\
Config.& &(GHz)&(Jy km s$^{-1}$)&(Jy km s$^{-1}$)&(Jy km s$^{-1}$)&(Jy km s$^{-1}$)&(Jy km s$^{-1}$)&(Jy km s$^{-1}$)\\
\midrule
\midrule
ACA & $^{12}$CO & 230.538 & 359.078 (0.968) & 799.576 (0.688) & 92.957(0.872) & 830.698 (0.931) & 17.167 (0.728) & 94.044 (0.925)\\
& $^{13}$CO & 220.398 & -- & -- & 44.230 (0.391) & -- & \tbf{3.466 (0.404)} & \tbf{23.009 (0.284)}\\
& C$^{18}$O & 219.560 & -- & -- & ND (0.037)$^c$ & -- & ND (0.034)$^c$ & ND (0.035)$^c$\\
& CO$^+$ & 235.789 & -- & -- & ND (0.072)$^c$ & -- & ND (0.059)$^c$ & ND (0.051)$^c$ \\
& CN & 226.697 & \tbf{118.239 (0.789)} & \tbf{26.864 (0.976)} & \tbf{22.324 (0.933)} & \tbf{116.502(1.164)} & \tbf{2.375 (0.587)} & \tbf{8.063(0.409)}\\
&    & 226.874 & \tbf{279.184 (0.789)} & \tbf{79.536 (0.738)} & \tbf{57.088(0.739)} & \tbf{277.213 (0.977)} & \tbf{9.340(0.508)} & \tbf{24.294 (0.416)}\\
& CS & 244.936 & ND (0.070)$^c$ & \tbf{13.852 (0.866)} & -- & ND (0.061)$^c$ & -- & -- \\
& C$^{34}$S & 241.016 & -- & \tbf{3.054 (0.667)} & -- & -- & -- & --\\
& HCN & 265.886 & 101.676 (1.030) & -- & 23.210 (0.629) & -- & \tbf{2.883 (0.238)} & \tbf{9.799 (0.345)}\\
& H$^{13}$CN & 259.011 & -- & \tbf{15.319 (0.413)} & 7.690 (0.656) & ND (0.097)$^c$ & ND (0.091)$^c$ & ND (0.088)$^c$ \\
& HNC & 271.981 & -- & \tbf{24.116 (0.763)} & 5.991 (0.606) & 13.285 (0.819) & \tbf{0.949 (0.208)} & \tbf{2.457 (0.354)}\\
& HCO$^+$ & 267.557 & 139.286 (1.419) & -- & 9.208 (0.923) & -- & \tbf{0.492 (0.122)} & \tbf{1.836 (0.362)}\\
& H$^{13}$CO$^+$ & 260.255 & -- & \tbf{8.401 (0.434)} & ND (0.153)$^c$ & ND (0.143)$^c$ & ND (0.108)$^c$ & ND (0.139)\\
& SO & 251.857 & ND (0.067)$^c$ & -- & ND (0.028)$^c$ & -- & ND (0.029)$^c$ & ND (0.067)$^c$ \\
C43-1 & $^{12}$CO & 230.538 & 517.688 (0.436) & -- & -- & 982.035 (1.156) & -- & --\\
& $^{13}$CO & 220.398 & -- & 286.858 (0.363) & -- & -- & -- & --\\
& C$^{18}$O & 219.560 & -- & \tbf{5.812 (0.104)} & -- & -- & -- & --\\
& CO$^+$ & 235.789 & -- & ND (0.007)$^c$ & -- & -- & -- & --\\
& CN & 226.697 & \tbf{55.984 (0.843)} & -- & -- & \tbf{92.246 (0.675)} & -- & --\\
&    & 226.874 & \tbf{114.224 (0.789)} & -- & -- & \tbf{222.366 (0.628)} & -- & --\\
& CS & 244.936 & ND (0.009)$^c$ & -- & -- & \tbf{5.982 (0.524)} & -- & --\\
& SO & 219.949 & -- & 6.344 (0.098) & -- & -- & -- & --\\
C43-3 & $^{13}$CO & 220.398 & -- & -- & 9.762 (0.221) & -- & -- & --\\
& C$^{18}$O & 219.560 & -- & -- & ND (0.003)$^c$ & -- & -- & --\\
& CO$^+$ & 235.789 & -- & -- & ND(0.004)$^c$ & -- & -- & --\\
& HCN & 265.886 & -- & \tbf{97.244 (0.477)} & 25.143 (0.296) & -- & \tbf{1.962 (0.124)} & \tbf{9.658 (0.277)}\\
& H$^{13}$CN & 259.011 & ND (0.029)$^c$ & \tbf{23.432 (0.500)} & 8.268 (0.221) & -- & ND (0.012)$^c$ & --\\
& HNC & 271.981 & ND (0.013)$^c$ & \tbf{19.634 (0.473)} & 4.023 (0.303) & -- & \tbf{2.446 (0.133)} & --\\
& HCO$^+$ & 267.557 & -- & \tbf{140.005 (1.699)} & 9.247 (0.503) & -- & \tbf{0.778 (0.210)} & \tbf{2.049 (0.329)}\\
& H$^{13}$CO$^+$ & 260.255 & \tbf{4.529 (0.050)} & \tbf{11.111 (0.847)} & \tbf{2.298 (0.196)} & -- & ND (0.018)$^c$ & --\\
& SO & 251.857 & -- & 1.702 (0.191) & ND (0.009)$^c$ & -- & -- & ND (0.005)$^c$\\
&    & 219.949 & -- & 4.556 (0.123) & \tbf{1.775 (0.169)} & -- & -- & --\\
C43-4 & $^{12}$CO & 230.538 & -- & 818.523 (0.389) & 104.494 (0.537) & 396.250 (3.156) & 5.291 (0.244) & 65.674 (0.326)\\
& $^{13}$CO & 220.398 & -- & 262.820 (0.257) & -- & -- & \tbf{1.187 (0.101)} & \tbf{19.334 (0.215)}\\
& C$^{18}$O & 219.560 & -- & \tbf{2.545 (0.162)} & -- & -- & ND (0.002)$^c$ & ND (0.003)$^c$ \\
& CO$^+$ & 235.789 & -- & ND (0.004)$^c$ & -- & -- & ND (0.004)$^c$ & ND (0.004)$^c$ \\
& CN & 226.697 & -- & -- & \tbf{20.843 (0.451)} & \tbf{43.218 (2.916)} & \tbf{0.667 (0.086)} & \tbf{5.878 (0.409)}\\
&    & 226.874 & -- & -- & \tbf{50.520 (0.480)} & \tbf{145.479 (2.978)} & \tbf{1.729 (0.113)} & \tbf{20.579 (0.618)}\\
& CS & 244.936 & -- & -- & ND (0.004)$^c$ & ND (0.040)$^c$ & \tbf{2.218 (0.282)} & ND (0.004)$^c$ \\
& SO & 251.857 & -- & -- & -- & -- & ND (0.004)$^c$ & -- \\
&    & 219.949 & -- & -- & -- & -- & -- & ND (0.003)$^c$ \\
& H30$\alpha$ & 231.901 & -- & \tbf{25.828 (0.235)} & -- & -- & -- & --\\
\midrule
\end{tabular}
\end{center}
%\vspace{0.2in}
%\footnotesize
{\sc Notes:} \\
a) Flux measurements are listed for all ALMA configurations in which data were obtained and the emission line was detected. Cases where configurations were not requested or observations in a given array configuration ran out of observation time are marked with `--'; flux measurements for new molecule detections are denoted by bold font.\\
%A nondetection is defined to occur when the emission line flux is less than 3$\sigma$. 
b) Listed errors are obtained from image noise levels and do not include calibration uncertainties.\\
\revision{c) `ND' indicates the line was observed but not detected; in parentheses are the root-mean-square values at the frequency of non-detected lines, reported in units Jy/beam.}
\end{table}
%\end{flushleft}

Details of the detections and nondetections of molecular species and lines for each PN are listed in Table \ref{tbl:allObs}, including molecule names, frequencies, and the correlator setup to which the detection belongs. Tables listing molecular line detections, quantum numbers, Einstein coeffecients, energy upper limits for their quantum states, and integrated line fluxes for individual PNe, \revision{as well as the PN systemic velocities ($V_{LSR}$) and molecular line velocity half-widths at half-power ($\Delta V_{1/2}$) obtained from the ALMA data, are presented in Appendix \ref{sec:app1}. Spatially integrated molecular line profiles are presented in Appendix \ref{sec:app4}. }

\begin{figure}[h!]
    \centering
    \includegraphics[width=1.0\textwidth]{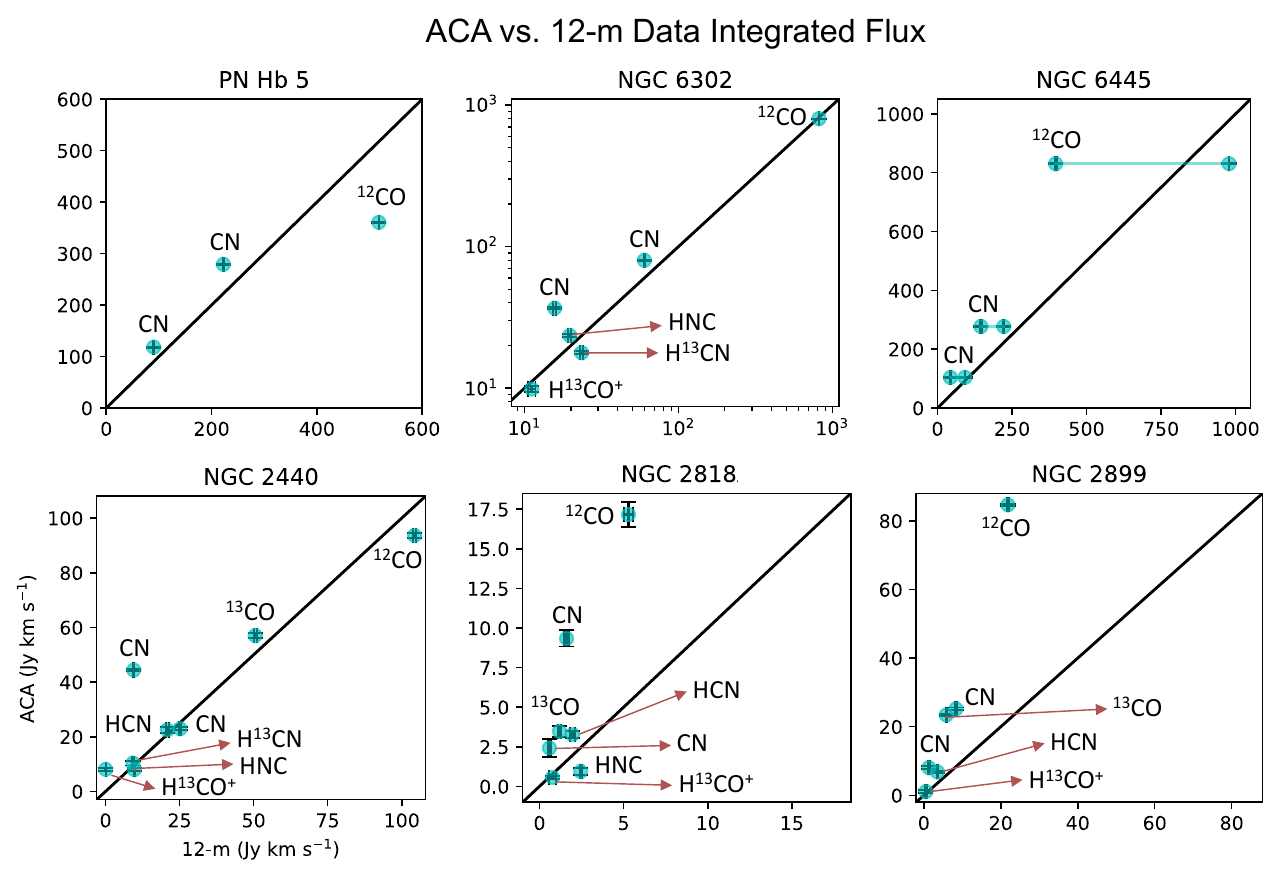}
    \caption{Comparison between the integrated fluxes measured with ACA and 12-m data for each molecular line for which observations were made in both ACA and 12-m configurations. Consistent flux measurements would follow along the black lines. The horizontal line in the plot for NGC 6445 indicates results from the C43-1 and C43-4 12-m configuration observations of $^{12}$CO(2--1) and CN hyperfine structure. Error bars were placed for all integrated flux measurements with calibration uncertainties included.}
    \label{fig:intFluxComp}
\end{figure}

The integrated intensities of molecular lines detected in ALMA Band 6 ACA and 12-m configurations are compared in Fig.~\ref{fig:intFluxComp} to determine whether there is a loss of flux as the imaging resolution increases. Overall, the brighter molecular emission line sources (NGC 6302, NGC 6445, and NGC 2440) display little flux loss. On the other hand, there appears to be systematic flux loss in the 12-m configurations for the weaker, more fragmented  molecular line sources, NGC 2818 and NGC 2899 (see \S \ref{sec:ResN2818} and \S \ref{sec:ResN2899}, respectively). An exception is PN Hubble 5, where $^{12}$CO appears brighter in the 12-m configuration than the ACA configuration. This flux loss in the ACA observation is most likely an artifact of continuum subtraction errors that occurred during the pipeline image cleaning process, as is evident when comparing spectral profiles of $^{12}$CO for 12-m and ACA observations (see App.~\ref{sec:app4}), and is not observed when comparing other molecular lines observed for this nebula.

\begin{figure}[p]
    \centering
    \includegraphics[width=0.9\textwidth]{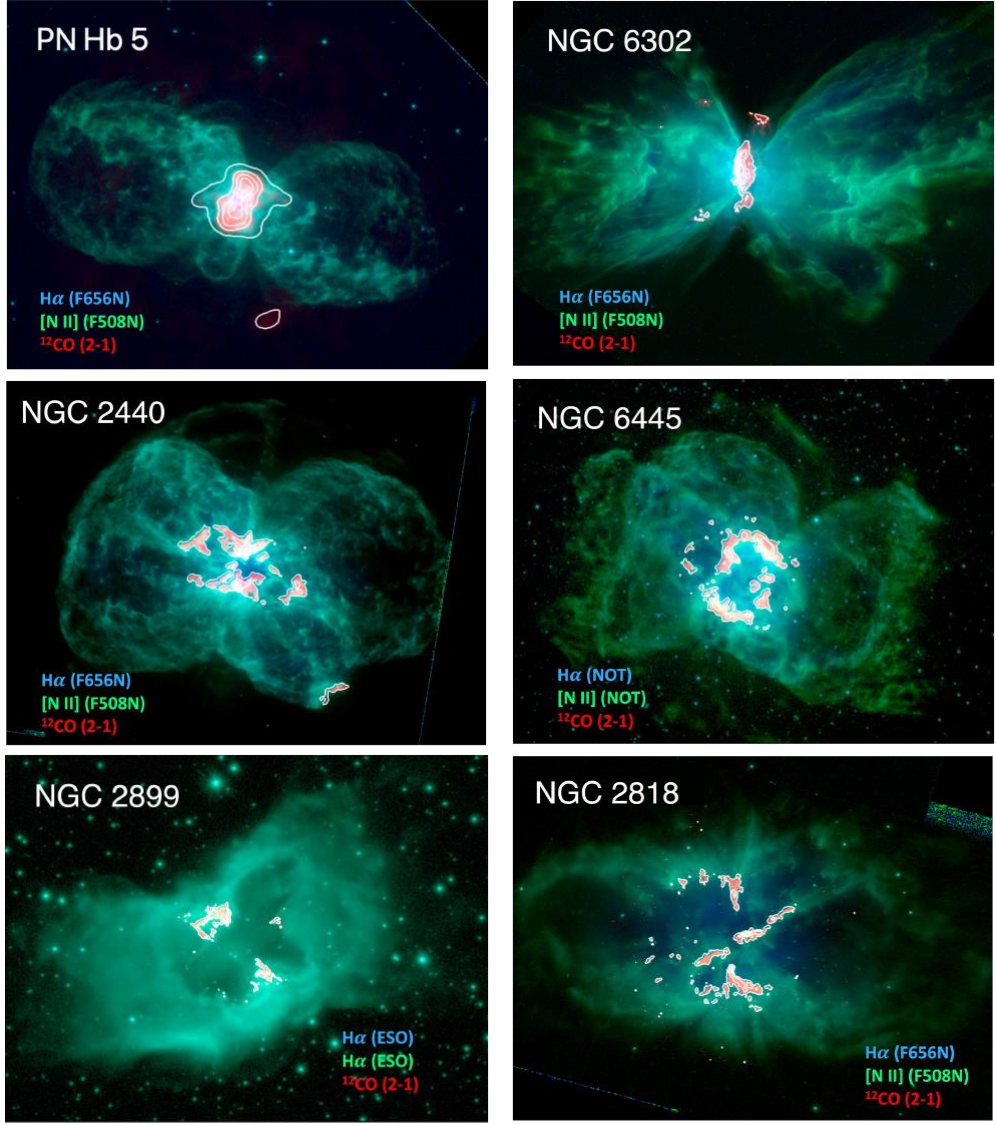}
    \caption{Color overlay images for all ALMA Band 6 targets. H$\alpha$ (blue) and [N {\sc ii}] (green) images are overlaid with ALMA Band 6 12-m $^{12}$CO zeroth moment (red) images. White contours are present to show the full extent of $^{12}$CO emission with respect to optical emission lines. H$\alpha$ and [N {\sc ii}] emission lines were observed with HST/WFPC2 for PN Hb 5, NGC 2440, and NGC 2818. These emission lines were observed with HST/WFC3 for NGC 6302, and NOT for NGC 6445. In the case of NGC 2899, only H$\alpha$ ESO images (blue and green) are displayed.}
    \label{fig:OptVsRadio}
\end{figure}

In Figure \ref{fig:OptVsRadio}, $^{12}$CO zeroth moment images are overlaid on optical (H$\alpha$ and [N {\sc ii}]) emission-line imaging to reveal the location of the molecular gas with respect to the optical lobes for each PN. Each PN displays a distinct CO emission morphology; however, most of the detected emission is observed to arise from or near the central waist regions of each bipolar PNe, indicating that, as in the case of NGC 6302 \citep{Santander-Garcia2017}, the bulk of the molecular gas observed in each case is tracing a central toroidal structure. 

In the following subsections, we summarize the molecular transitions detected in each PN in order of inferred molecular torus dynamical age (\S ~\ref{sec:ageCalc}), and we discuss in more detail how these ALMA Band 6 data reveal the structures of the dense waists of the survey PNe.

\subsection{Hubble 5 (PN Hb 5)}\label{sec:ResHb5}

\begin{figure}[h!]
    \centering
    \includegraphics[width=1.0\textwidth]{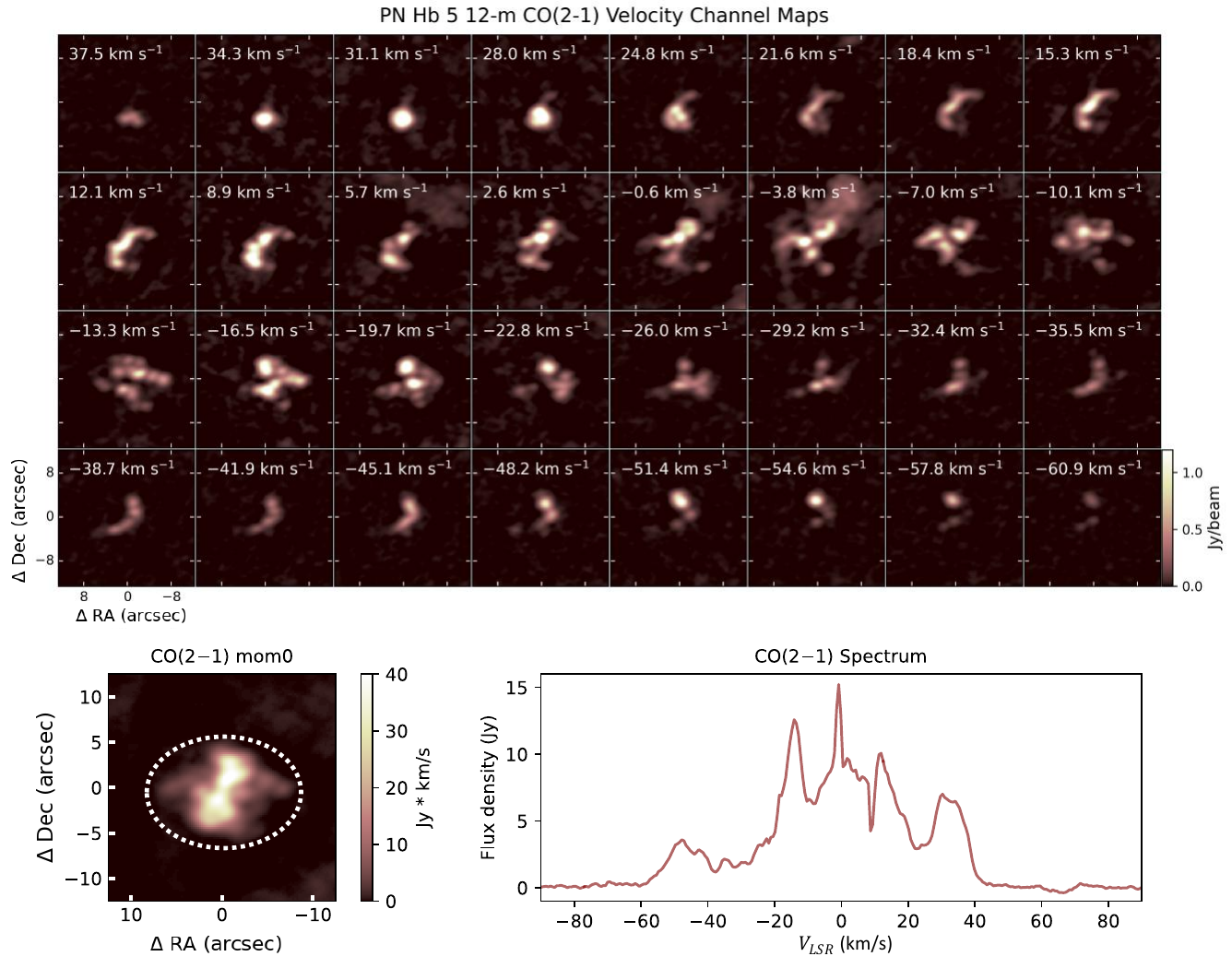}
    \caption{Images and spectrum extracted from 12-m $^{12}$CO(2--1) data cube obtained for Hubble 5. \revision{\textit{Top panel}:} PN Hb 5 $^{12}$CO(2$-$1) emission in units Jy/beam across multiple velocity channels in increments of 3.2 km s$^{-1}$. \revision{\textit{Bottom left panel}:} $^{12}$CO(2$-$1) zeroth moment in units Jy km s$^{-1}$ corresponding to the data cube displayed above. The white dotted oval surrounding the CO(2$-$1) emission represents the region selected for spectral data extraction from the same data cube. \revision{\textit{Bottom right panel}: the resulting spatially integrated spectrum.}}
    %The CO(2$-$1) emission spectrum is displayed in units of flux density (Jy) along velocity space (km s$^{-1}$).
    \label{fig:Hb5DatCube}
\end{figure}

\begin{figure}[h!]
    \centering
    \includegraphics[width=0.99\textwidth]{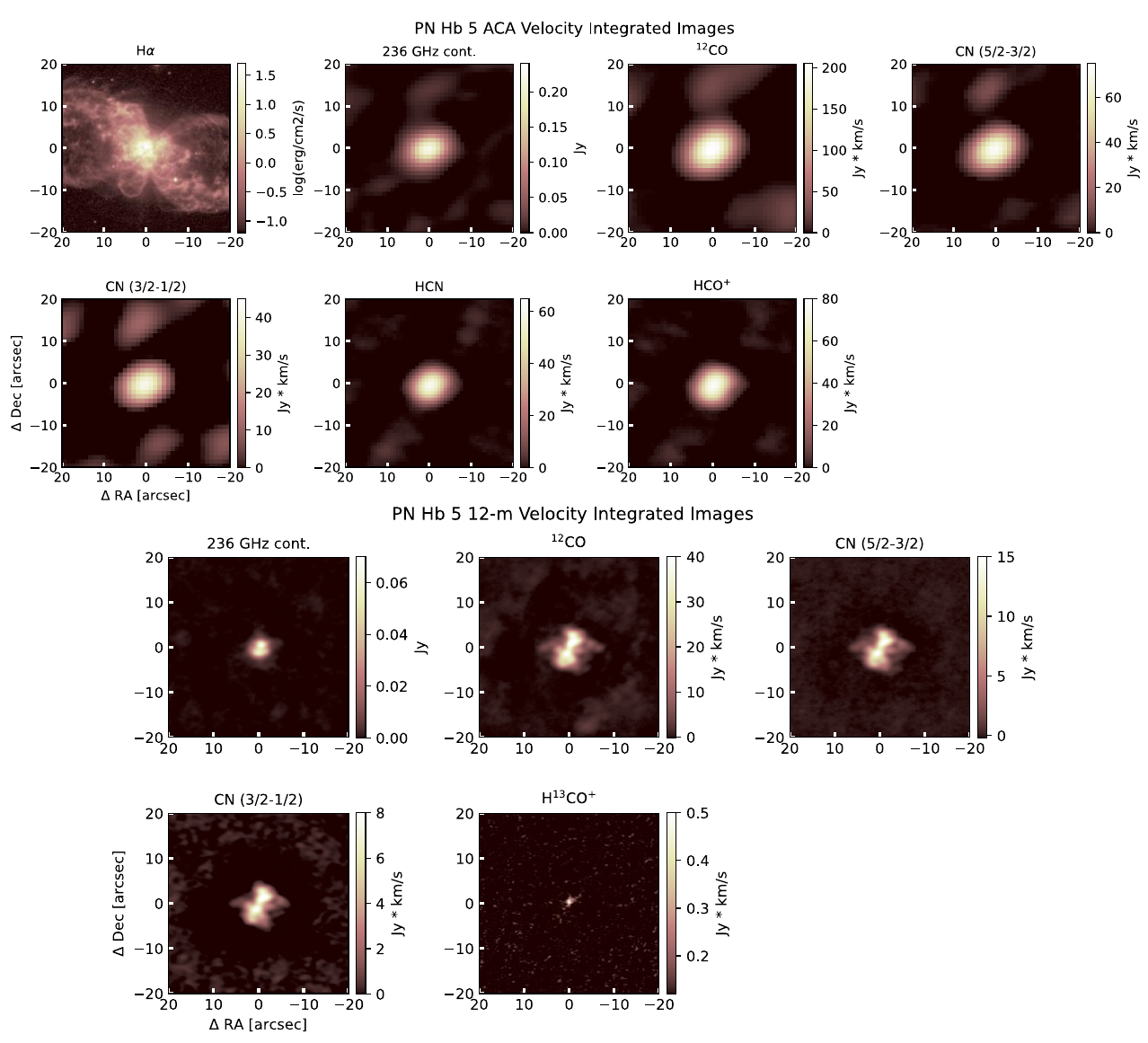}
    \caption{ALMA Band 6 mom0 images of emission lines detected in PN Hb 5 with the ACA array (top 7 panels) and 12-m (bottom 6 panels). \textit{Top left panel}: archival HST/WFPC2 H$\alpha$ image. \textit{Second panel from left}: ACA image of the continuum emission at 236 GHz as received from the ALMA pipeline. \textit{Following 5 panels}: $^{12}$CO, CN (5/2$-$3/2), CN (3/2$-$1/2), HCN, and HCO$^+$ mom0 images, respectively. \textit{Following 5 (12-m) panels}: 12-m array continuum emission at 236 GHz as received from the ALMA pipeline, $^{12}$CO, CN (5/2$-$3/2), CN (3/2$-$1/2), and H$^{13}$CO$^+$, respectively.  All images have a field of view of 40$''$ x 40$''$. }
    \label{fig:Hb5Mont}
\end{figure}

Figure \ref{fig:Hb5DatCube} presents extractions from the 12-m $^{12}$CO(2--1) observation data cube for Hubble 5, including channel maps (top panels), zeroth moment image (bottom left), and spatially-integrated spectral line profile (bottom right). Detections in Hubble 5, presented in the form of moment 0 images (hereafter mom0) in  Figure~\ref{fig:Hb5Mont}, include the $J=2\rightarrow{}1$ transition of $^{12}$CO, $N=2\rightarrow{}1$ hyperfine transitions of CN ($J=5/2\rightarrow{}3/2$ and $J=3/2\rightarrow{}1/2$), and $J=3\rightarrow{}2$ transitions of HCN, HCO$^+$, and H$^{13}$CO$^+$. Emission from $^{12}$CO and CN was detected in both 12-m and ACA data. 
%(displayed in Fig.~\ref{fig:Hb5Mont}). %\joel{I forget why we decided to only include these figs in the main text for Hb 5, and not for the others...} 
However, due to observing time limitations, HCN and H$^{12}$CO$^+$ data were only obtained in the ACA observations, while H$^{13}$CO$^+$ data were only obtained with the 12-m array. 
%This is, unfortunately, due to the allocated time for ACA observations expiring before all science goal observations could be made.
Fig~\ref{fig:Hb5Spect} displays the spectral line profiles extracted for the detections obtained with each ALMA array. %\joel{may as well put these line profile figs into the Appendix, with the rest...as it is, a bit confusing to just have Hb 5's line profiles here}

In Fig.~\ref{fig:Hb5Mont} (lower panels), the 12-m observations of $^{12}$CO, CN (5/2$-$3/2), and CN (3/2$-$1/2) emission reveal a 10$''$ structure oriented roughly north to south, with fainter extended ``wings'' that appear nearly orthogonal to this main N--S structure, running E to W. The brightest emission in the 12-m velocity channel maps (Fig.~\ref{fig:Hb5DatCube}, upper panels) appears to trace an elliptical or toroidal structure that wraps around the nebula's pinched waist. This main torus has a large projected expansion velocity, $\sim$50 km s$^{-1}$ (see also Fig~\ref{fig:Hb5Spect}), while the fainter, extended ``wing'' structures are confined to smaller projected velocities of $\sim$7 km s$^{-1}$. The 12-m observations also reveal more extended $^{12}$CO, \revision{in the form of diffuse emission surrounding the molecular torus (see Fig.~\ref{fig:Hb5Mont}),} that may trace the circular ring system detected in HST imaging of Hb~5.
%In this paper, Band 6 ACA observations of Hb 5 will not be discussed due to flux lost in the $^{12}$CO data cube continuum subtraction. Moreover, 
The ACA observations of Hubble 5, while yielding detections of HCN and H$^{12}$CO$^+$ at the nebular core, offer little additional information regarding the morphology of the molecular emission region, due to the nebula's compact nature (Fig.~\ref{fig:Hb5Mont}, upper panels). 

Based on the color composite of ALMA and HST imaging (Fig.~\ref{fig:OptVsRadio}), the brightest $^{12}$CO emission indeed appears to be a nearly edge-on view of the torus. The southern region of the $^{12}$CO emission overlaps with a small optical emission lobe (relative to the main bipolar lobes) extending in that direction, while the extended ``wings'' detected in the $^{12}$CO and CN maps likely coincide with the front wall of the E optical lobe and the back wall of the W optical lobe, respectively. In the case of Hubble 5, the central torus traced by $^{12}$CO and CN molecular emission appears entirely intact as it wraps around the nebula's axis of symmetry. 

\subsection{NGC 2440}\label{sec:ResN2440}

\begin{figure}[h!]
    \centering
    \includegraphics[width=0.9\textwidth]{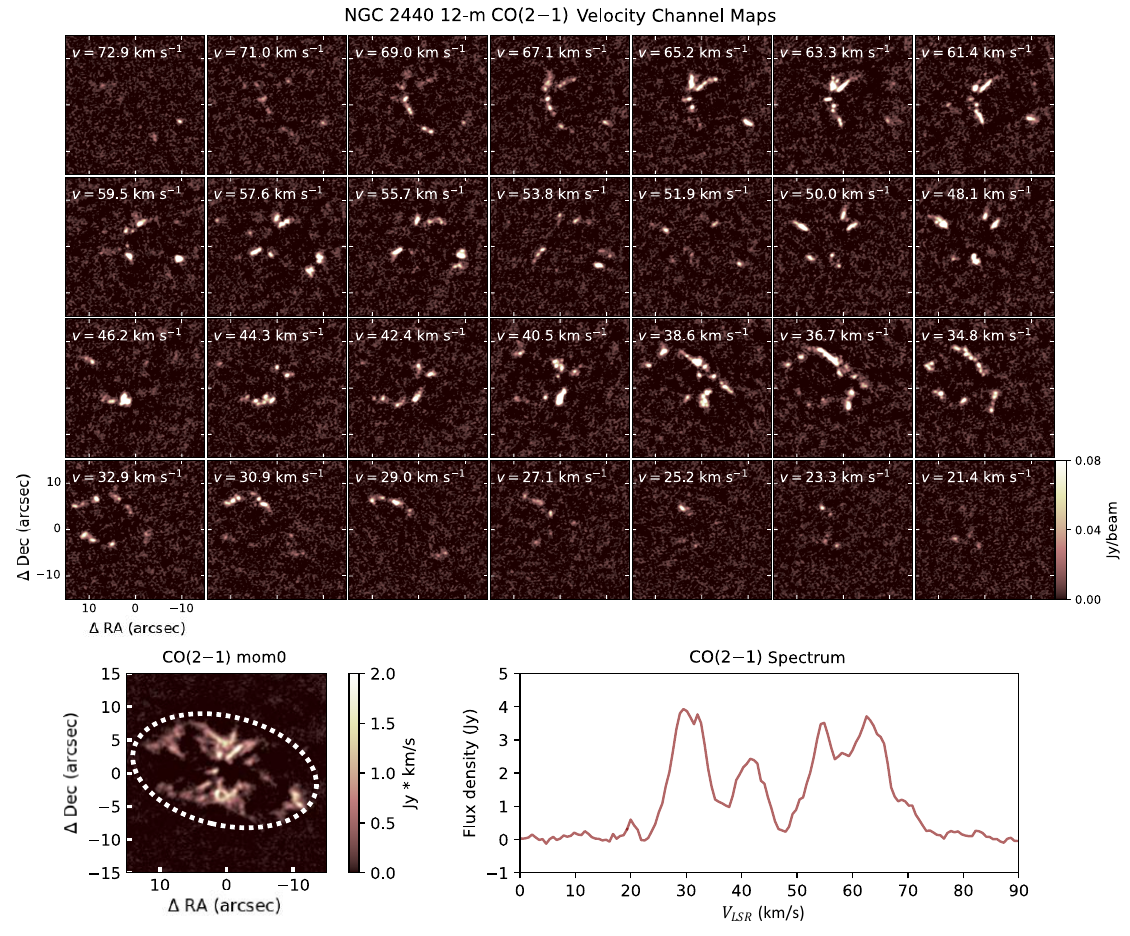}
    \caption{Same as Fig.~\ref{fig:Hb5DatCube}, but displaying 12-m Band 6 data for NGC 2440. Velocity channels are displayed in increments of 1.9 km s$^{-1}$.}
    \label{fig:N2440DatCube}
\end{figure}
\begin{figure}[h!]
    \centering
    \includegraphics[width=1\textwidth]{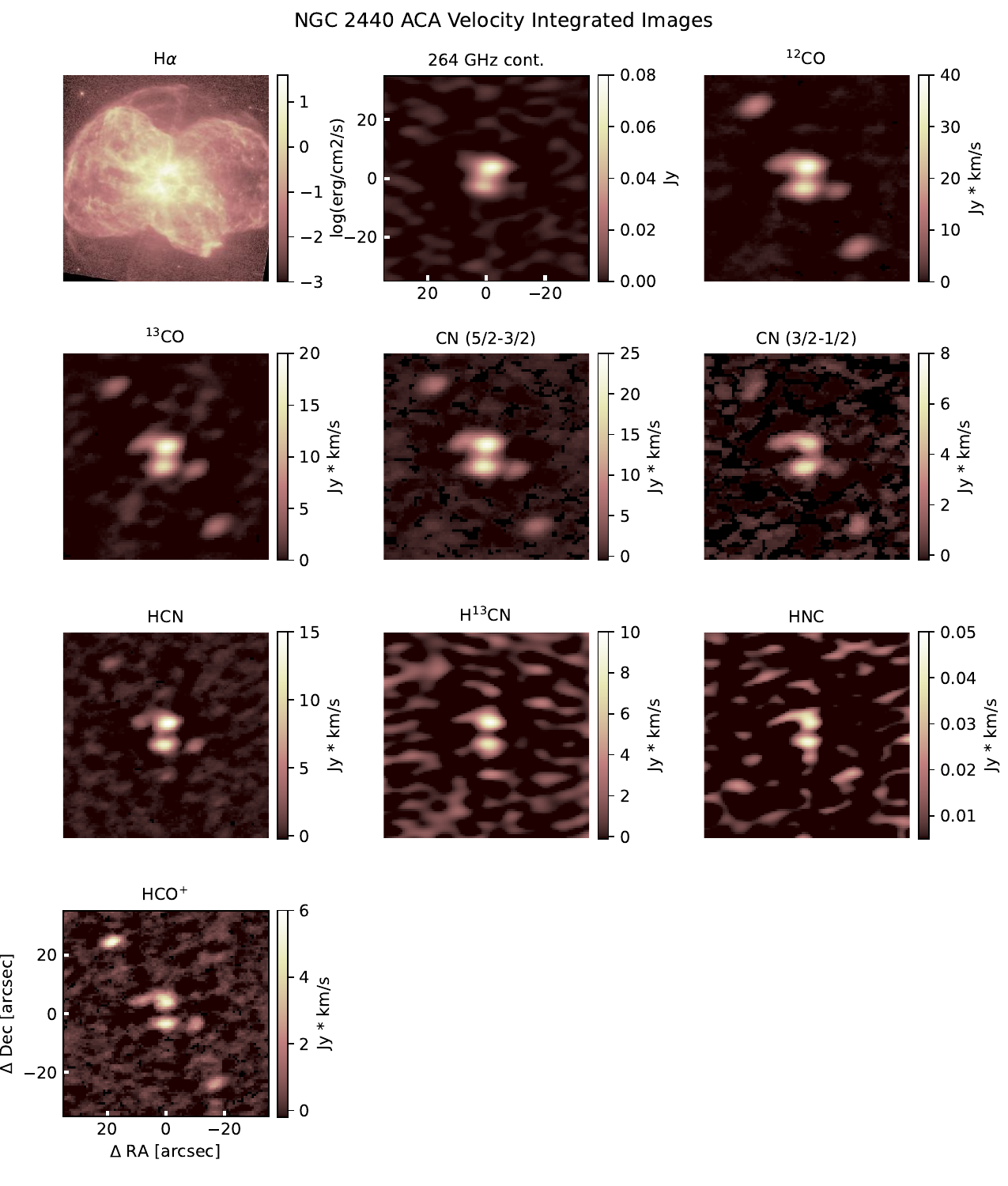}
    \caption{As in Fig.~\ref{fig:Hb5Mont}, but displaying ACA Band 6 mom0 images for NGC 2440. \textit{Top left panel}: an archival HST/WFPC2 H$\alpha$ image of NGC 2440. \textit{Second panel}: an image of the continuum emission at 264 GHz as received from the ALMA pipeline. \textit{Following panels}: $^{12}$CO, $^{13}$CO, CN (5/2$-$3/2), and CN (3/2$-$1/2), HCN, H$^{13}$CN, HNC, and HCO$^+$, respectively. All images have a field of view of 70$''$ x 70$''$.}
    \label{fig:N2440Mont1}
\end{figure}
\begin{figure}[h!]
    \centering
    \includegraphics[width=1\textwidth]{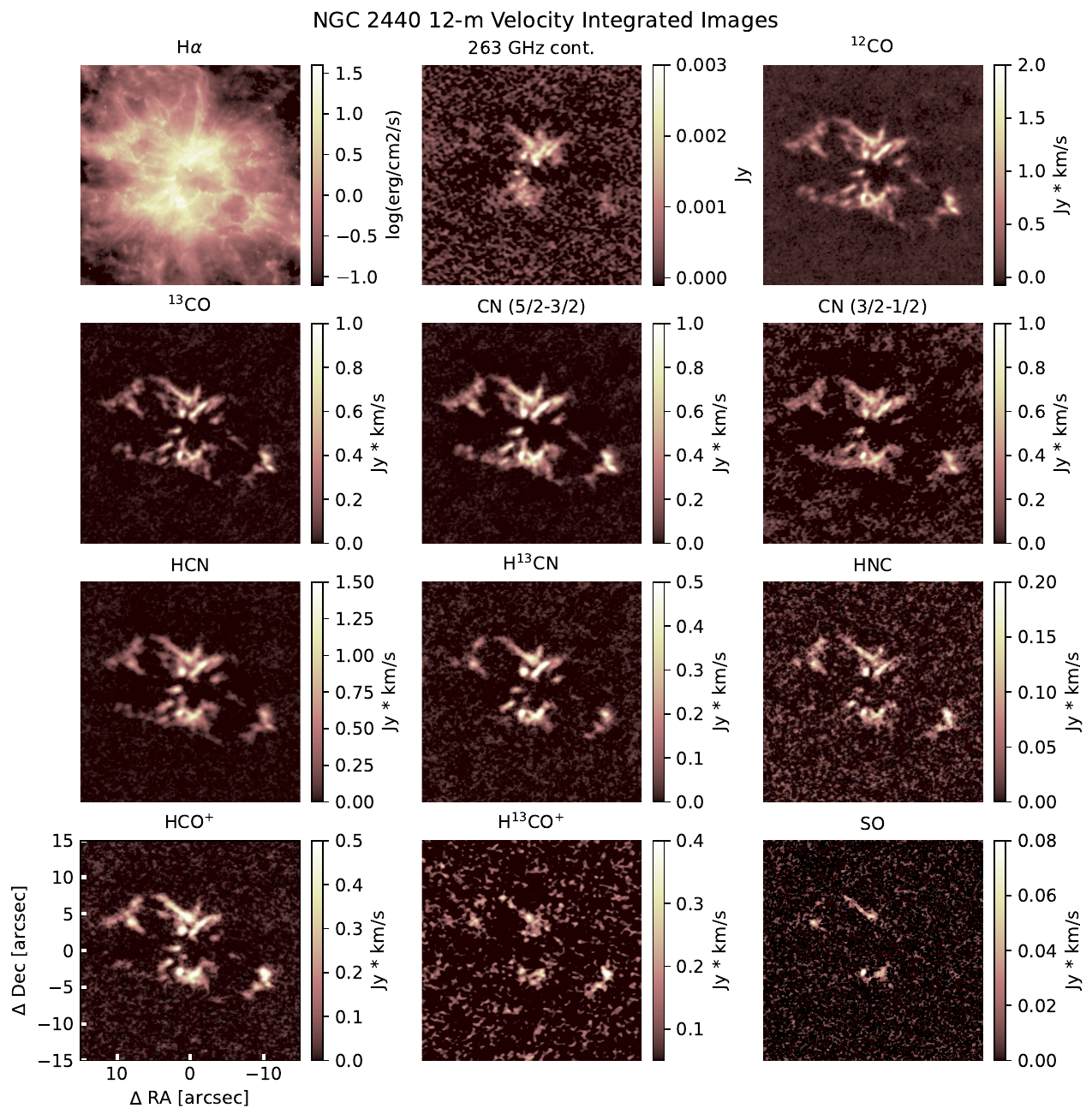}
    \caption{Same as Fig.~\ref{fig:Hb5Mont}, but displaying 12-m Band 6 mom0 images for NGC 2440. \textit{Top left panel}: an archival HST/WFPC2 H$\alpha$ image of NGC 2440. \textit{Second panel}: an image of the continuum emission at 263 GHz as received from the ALMA pipeline. \textit{Following panels}: $^{12}$CO, $^{13}$CO, CN (5/2$-$3/2), and CN (3/2$-$1/2), HCN, H$^{13}$CN, HNC, HCO$^+$, and SO 6(5)--5(4), respectively. All images have a field of view of 30$''$ x 30$''$.}
    \label{fig:N2440Mont2}
\end{figure}
%\joel{double-check that each of these sections includes specific references to the relevant figures (after moving them up into the main text), including spectra; here I've added these refs to your text:}

Channel maps, mom0 image, and spectral line profile extracted from the 12-m Band 6 observations of $^{12}$CO emission from NGC 2440 are presented in Fig.~\ref{fig:N2440DatCube}.
The ACA observations of NGC 2440 yielded detections of $^{12}$CO, $^{13}$CO, HNC, HCN, H$^{13}$CN, HCO$^+$, and hyperfine transitions of CN (Fig.~\ref{fig:N2440Mont1} and Fig.~\ref{fig:N2440Spect}, left). The 12-m array observations yielded detections of these same transitions as well as H$^{13}$CO$^+$ and SO (Figs.~\ref{fig:N2440Mont2} and Fig.~\ref{fig:N2440Spect}, right). 

The ACA observations show bright molecular emission from the core region of NGC 2440, as well as two isolated knots of molecular emission offset by $\sim$35$''$ to the NE and SW from the bright core (Fig.~\ref{fig:N2440Mont1}). 
%that the morphology of mm-wave emission from cold molecular gas appears similar to the morphology of H$_2$ emission around the core of NGC 2440 \citep{Kastner1996}. \joel{refer to ACA mom0 montage figure here} However, in addition to the bright emission from the core region of the PN, the ACA molecular line mapping reveals 
All three structures were previously detected in single-dish CO mapping by \citet{Wang2008}. In our ACA data, the polar lobe knot structures are detected in $^{12}$CO, $^{13}$CO, CN, and HCO$^+$, and appear faintly in HCN emission. 

The 12-m observations were centered on the core region of NGC 2440, leaving the NE and SW knot structures just outside of the field of view (Fig.~\ref{fig:N2440DatCube}). In the 12-m array $^{12}$CO mom0 map, the core emission region of NGC 2440 is resolved into what appears to be a dense torus and molecular gas filaments that form a bi-lobed structure that is similar to, but larger than, that revealed in ALMA mapping of PN Hubble 5 (\S~\ref{sec:ResHb5}). The bi-lobed molecular gas structure, which is 20$''$ in extent, is oriented roughly E to W. The $^{12}$CO velocity channel maps and spectrum indicate that the W lobe is redshifted at velocities up to $\sim$70 km s$^{-1}$ and the E lobe is blueshifted at velocities up to $\sim$20 km s$^{-1}$. Thus, the projected expansion velocity of the central molecular gas structure is $\sim$25 km s$^{-1}$. 

The projected relative expansion speed of the NE and SW knots obtained from the ACA channel maps (not shown), assuming these features constitute a pair, is $\sim$14 km s$^{-1}$. This indicates that the knots are moving nearly perpendicular to our line of sight. We note that these knot structures 
%visible in Fig.~\ref{fig:N2440Mont1} 
lie at the tips of two interior lobes of ionized gas oriented at $\sim$45$^\circ$ with respect to NGC 2440's larger, E--W aligned lobe pair (Fig.~\ref{fig:OptVsRadio}). These structures hence could be clumps of progenitor AGB star ejecta that were shot out as projectiles along the NE--SW direction, before the central star was unveiled and the nebula became ionized; their ejection may have generated NGC 2440's second, misaligned lobe pair. 

\subsection{NGC 6302}\label{sec:ResN6302}
\begin{figure}[h!]
    \centering
    \includegraphics[width=1.0\textwidth]{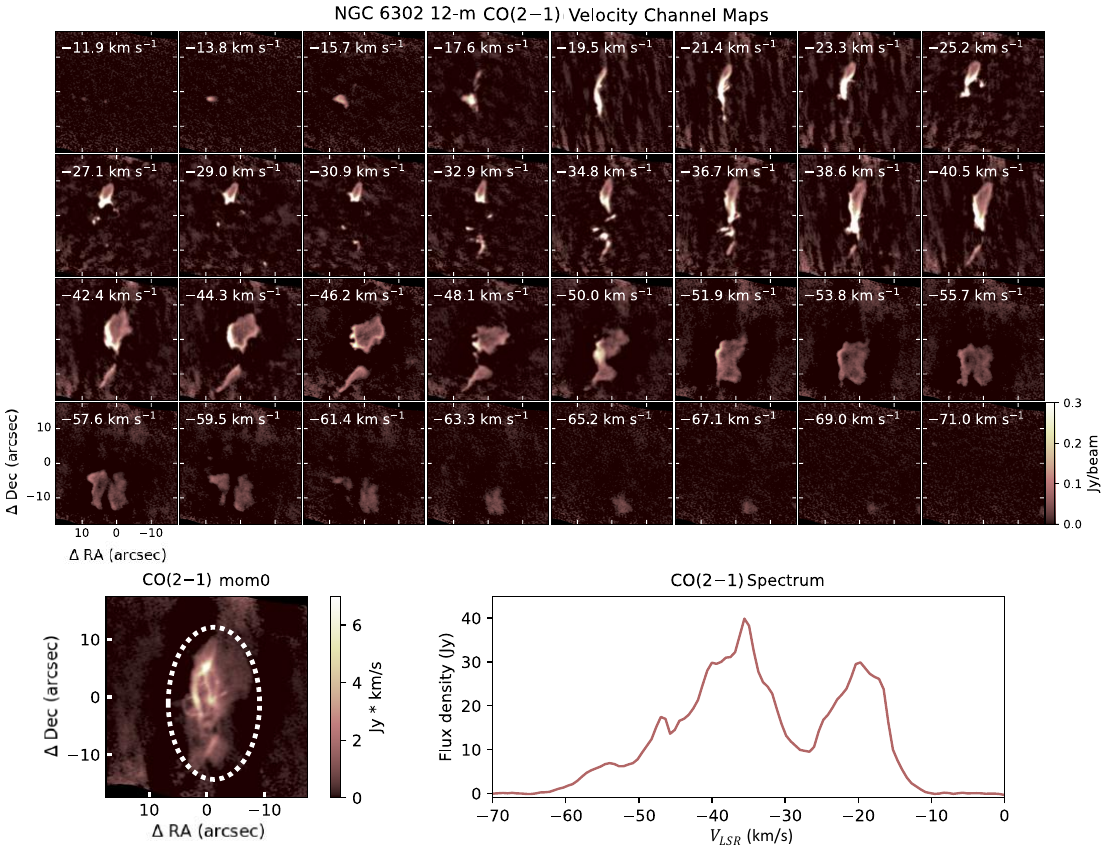}
    \caption{Overview of 12-m Band 6 $^{12}$CO data for NGC 6302, presented as in Fig.~\ref{fig:Hb5DatCube}. Velocity channels are displayed in increments of 1.9 km s$^{-1}$.}
    \label{fig:N6302DatCube}
\end{figure}
\begin{figure}[h!]
    \centering
    \includegraphics[width=1\textwidth]{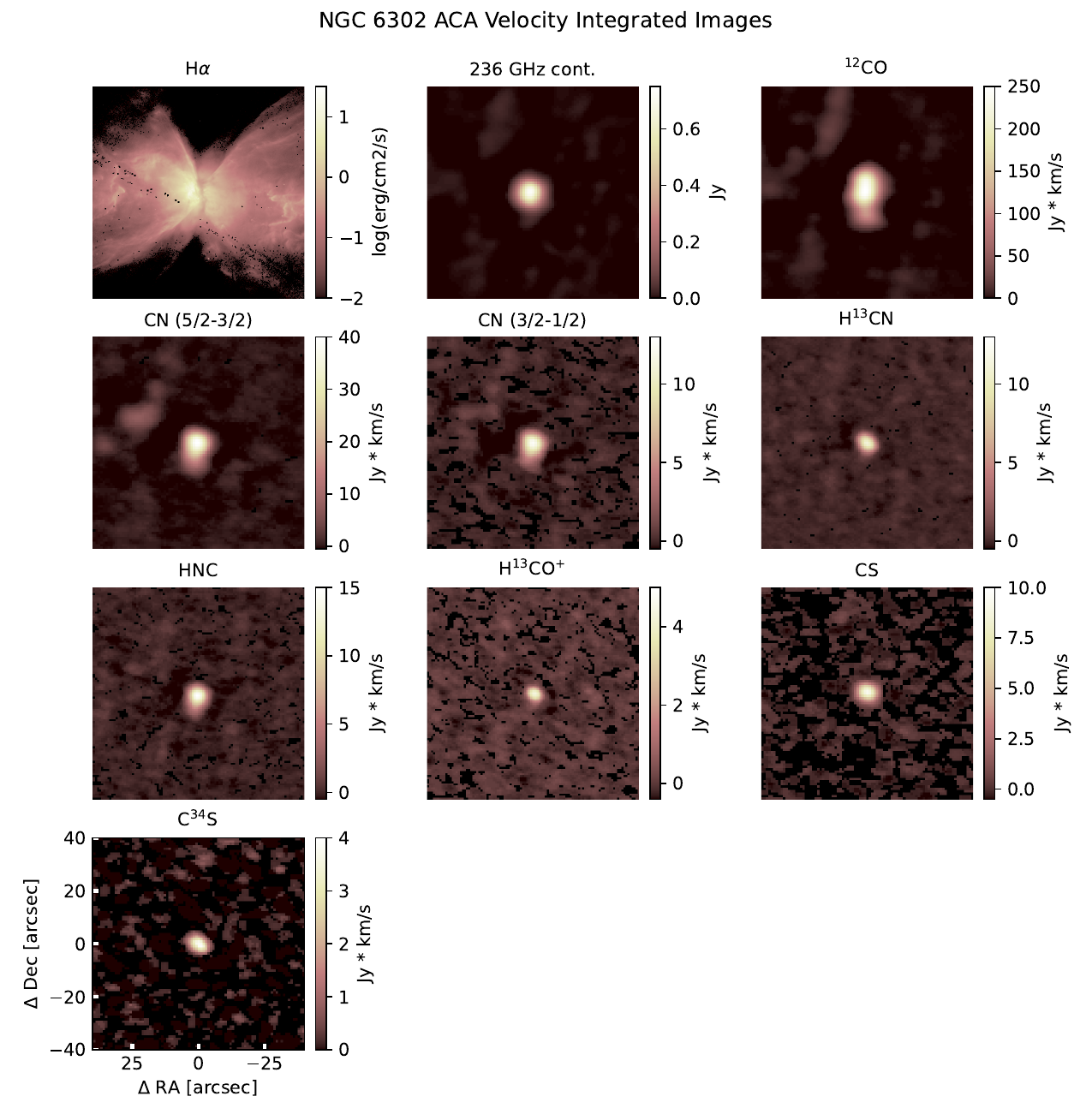}
    \caption{ACA Band 6 mom0 images for NGC 6302. \textit{Top left panel}: an archival HST/WFC3 H$\alpha$. \textit{Second panel}: continuum emission at 236 GHz as received from the ALMA pipeline. \textit{Following panels}: $^{12}$CO, CN (5/2$-$3/2), CN (3/2$-$1/2), H$^{13}$CN, HNC, H$^{13}$CO$^+$, CS, C$^{34}$S respectively. All images have a field of view of 80$''$ x 80$''$.} %\joel{use this sort of terse text for remaining mom0 montage fig captions, for rest of the PNe mom0 montage figs}}
    \label{fig:N6302Mont1}
\end{figure}
\begin{figure}[h!]
    \centering
    \includegraphics[width=1\textwidth]{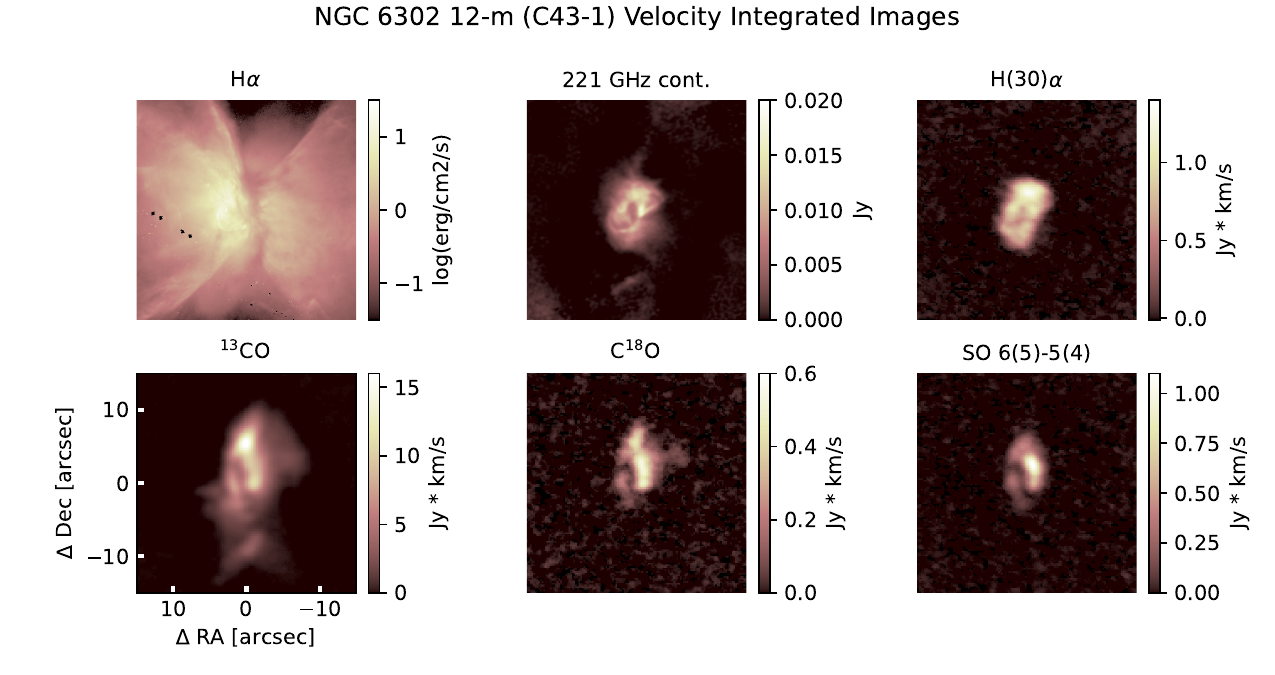}
    \caption{12-m (C43-1) Band 6 mom0 images for NGC 6302. \textit{Top left panel}: an archival HST/WFC3 H$\alpha$ image. \textit{Second panel}: 12-m array C43-1 image of the continuum emission at 221 GHz as received from the ALMA pipeline. \textit{Following panels}: H(30)$\alpha$, $^{13}$CO, C$^{18}$O, and SO 6(5)--5(4). All images have a field of view of 30$''$ x 30$''$.}
    \label{fig:N6302Mont2}
\end{figure}
\begin{figure}[h!]
    \addtocounter{figure}{-1}
    \centering
    \includegraphics[width=0.95\textwidth]{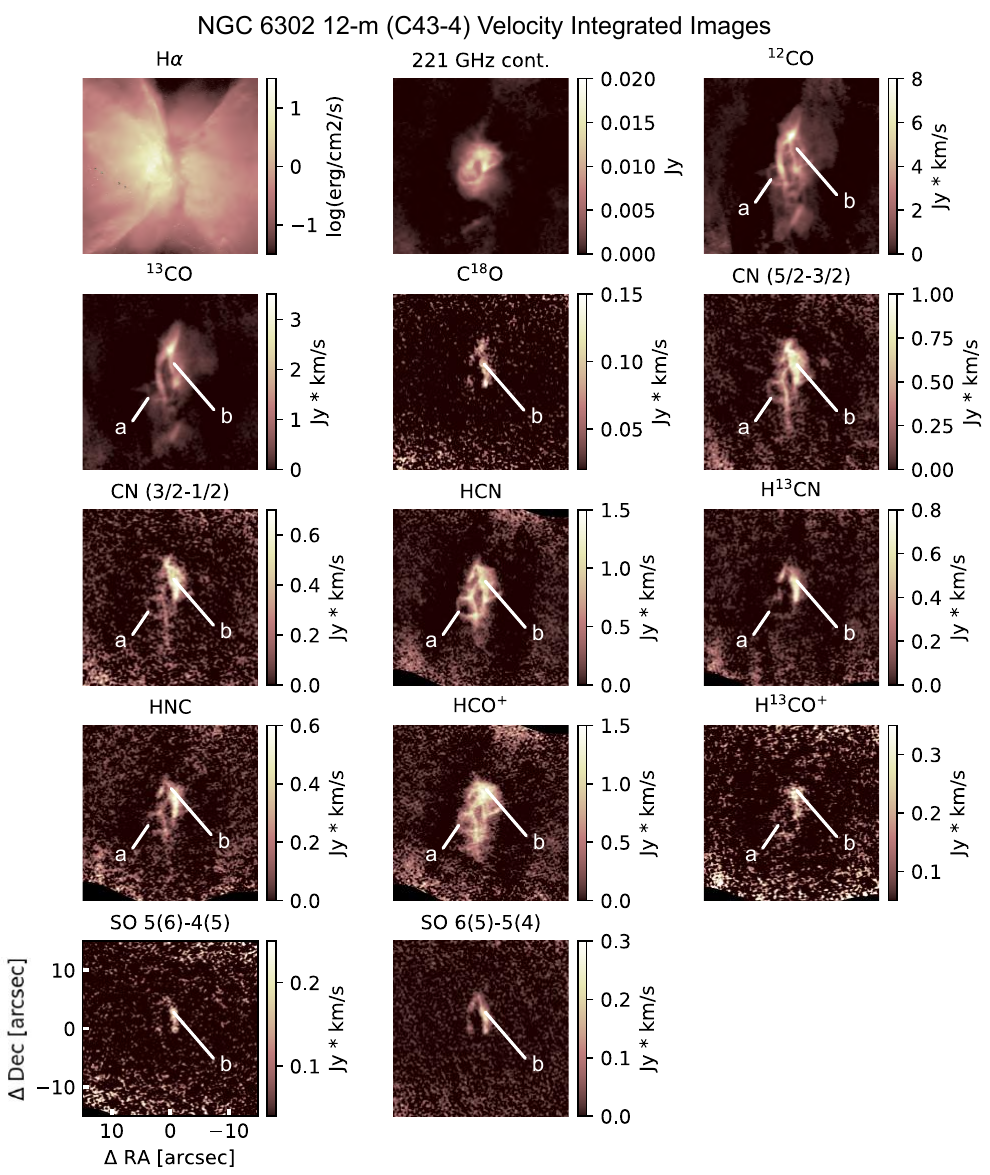}
    \caption{(cont.) 12-m (C43-4) Band 6 mom0 images for NGC 6302. \textit{Top left panel}: an archival HST/WFC3 H$\alpha$ image. \textit{Second panel}: continuum emission at 221 GHz as received from the ALMA pipeline. \textit{Following panels}: $^{12}$CO, $^{13}$CO, C$^{18}$O, CN (5/2$-$3/2), CN (3/2$-$1/2), HCN, H$^{13}$CN, HNC, HCO$^+$, H$^{13}$CO$^+$, SO 5(6)--4(5), and SO 6(5)--5(4). All images have a field of view of 30$''$ x 30$''$. Two important structures in the nebula, denoted by `a' and `b', are discussed in more detail in \S~\ref{sec:ResN6302}.}
    \label{fig:N6302Mont3}
\end{figure}

%\joel{I edited the following to follow the new figure placements in text...also note use of the latex addtocounter command; had to look that one up to remember how to do the multi-page figure continuation thing...it's the right command to use!}

Among the PNe included in our ALMA Band 6 survey, NGC 6302 has yielded the largest set of detections of molecular species, with high signal-to-noise measurements of almost all lines targeted. A notable exception is the nondetection of CO$^+$; NGC 7027 remains the only PN in which this molecule has been detected thus far \citep[][and references therein]{Bublitz2023}. In our ACA observations of NGC 6302, the $J=2\rightarrow{}1$ transition of $^{12}$CO, $N=2\rightarrow{}1$ hyperfine transitions of CN, $J=3\rightarrow{}2$ transitions of HNC, H$^{13}$CN, and H$^{13}$CO$^+$, and $J=5\rightarrow{}4$ transitions of CS and C$^{34}$S were all detected. Additional detections of the $J=2\rightarrow{}1$ transitions of $^{13}$CO and C$^{18}$O, $J=3\rightarrow{}2$ transitions of HCN and HCO$^+$, $N_J=6_5\rightarrow{}5_4$ and $N_J=5_6\rightarrow{}4_5$ transitions of SO, and H30$\alpha$ were obtained with the 12-m array. Channel maps, mom0 image, and spectral line profile extracted from the 12-m Band 6 observations of $^{12}$CO emission from NGC 6302 are presented in Fig.~\ref{fig:N6302DatCube}. The zeroth moment images obtained from the ACA and 12-m array data cubes are displayed in Fig.~\ref{fig:N6302Mont1} and Fig.~\ref{fig:N6302Mont2}, respectively; spectral profiles obtained from these data cubes are presented in Fig.~\ref{fig:N6302Spect}.
%--\ref{fig:N6302Mont4} of Appendix \ref{sec:app2} 
% \joel{reference(s) to spectral line profile figs may have to be updated once we decide where they should go in text}

The 12-m data reveals the detailed morphology of the molecular gas that traces NGC 6302's central region, demonstrating its complexity (Fig.~\ref{fig:N6302DatCube}). Along with a nearly edge-on toroidal structure tracing the dusty central waist of the nebula, the $^{12}$CO (2--1) emission traces a loop that extends to the east \revision{(denoted as structure `a' in Fig.~\ref{fig:N6302Mont3})}. Upon closer inspection of the $^{12}$CO velocity channel maps, however, it is clear that this latter structure is blue-shifted and stretches from the back edge of the central torus, looping around the front. These distinct kinematic structures were previously detected in ALMA $^{12}$CO (3--2) mapping of NGC 6302 obtained, at lower sensitivity, by \citet{Santander-Garcia2017}; they concluded that the CO traces two rings of molecular gas that are inclined approximately 60$^\circ$ to each other. Similar two-loop structures have been observed in velocity-resolved CO maps of the ring-like PNe NGC 7293 \citep{Young1999}, NGC 3132, and NGC 6720 \citep{Kastner2024,Kastner2025a}. 

The same double-looped structure appears in the ALMA 12-m maps of CN, HCN, HNC, HCO$^+$ emission from NGC 6302, and is very also faintly present in $^{13}$CO and H$^{13}$CO$^+$ emission. However, the southern region of the central torus is faint in all other emission lines detected, leaving only the northern region visible. Several of the detected emission line maps exhibit a locally bright, thin structure (denoted as structure \revision{`b'} in Figure \ref{fig:N6302Mont3}) that coincides with a local minimum in the $^{12}$CO emission line map. This local minimum is likely due to a particularly dense, cold clump of molecular gas causing self-absorption in $^{12}$CO; this cold, dense gas likely sits in the foreground of the molecular torus, where it is shielded from direct UV radiation from the planetary nebula central star.

When comparing optical and radio emission-like morphologies of NGC 6302, the molecular gas is almost completely concentrated in the central torus of the nebula, closely tracing the dark lane observed in HST/WFC3 images (Fig.~\ref{fig:OptVsRadio}). However, a few compact knots of CO emission are detected within the east optical lobe of the nebula. These molecular knots are spatially closely in conjunction with dusty clumps seen in HST imaging of the polar lobes of NGC 6302 \citep{Kastner2022}.
The molecular emission morphology seen in the ACA mom0 maps appears marginally extended N--S in $^{12}$CO, CN, and HCN emission, and appears round and possibly unresolved in all other emission lines (Figs.~\ref{fig:N6302Mont1}, ~\ref{fig:N6302Mont2}); the emission detected by the ACA remains compact and close to the central torus, with no additional extended emission apparent from the lobe regions. 

\subsection{NGC 6445}\label{sec:ResN6445}
\begin{figure}[h!]
    \centering
    \includegraphics[width=1.0\textwidth]{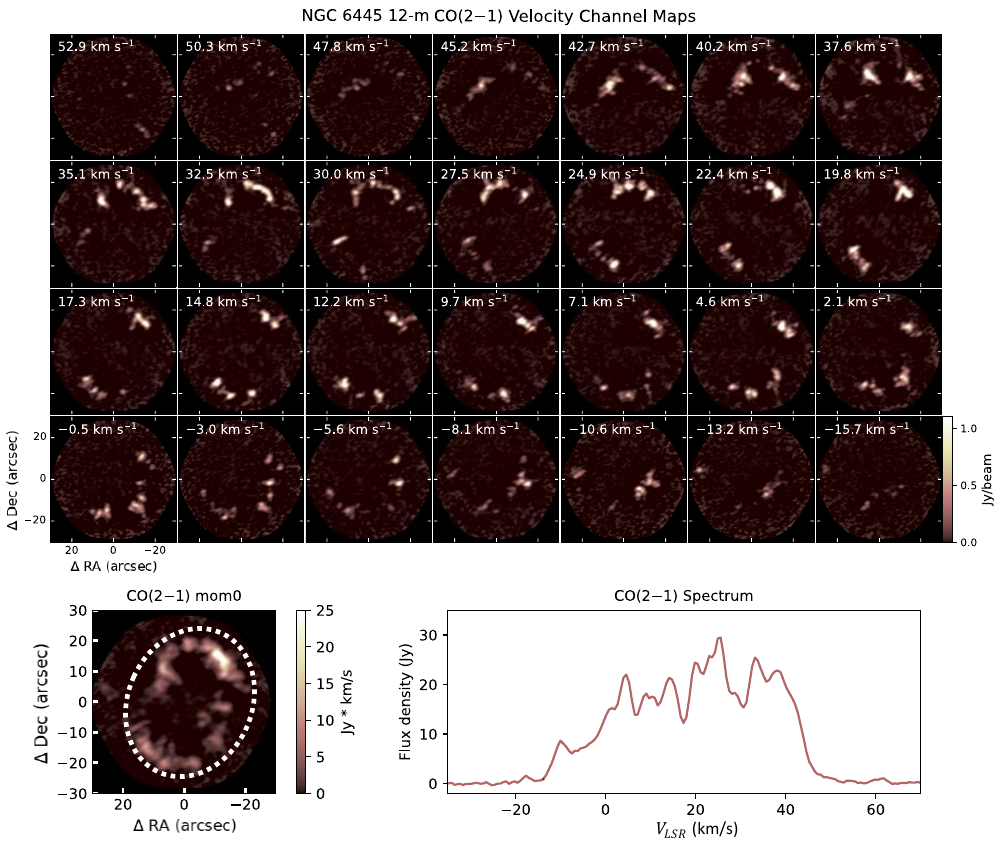}
    \caption{Same as Fig.~\ref{fig:Hb5DatCube}, but displaying ALMA Band 6 data for NGC 6445. Velocity channels are displayed in increments of 2.6 km s$^{-1}$.}
    \label{fig:N6445DatCube}
\end{figure}
\begin{figure}[h!]
    \centering
    \includegraphics[width=1\textwidth]{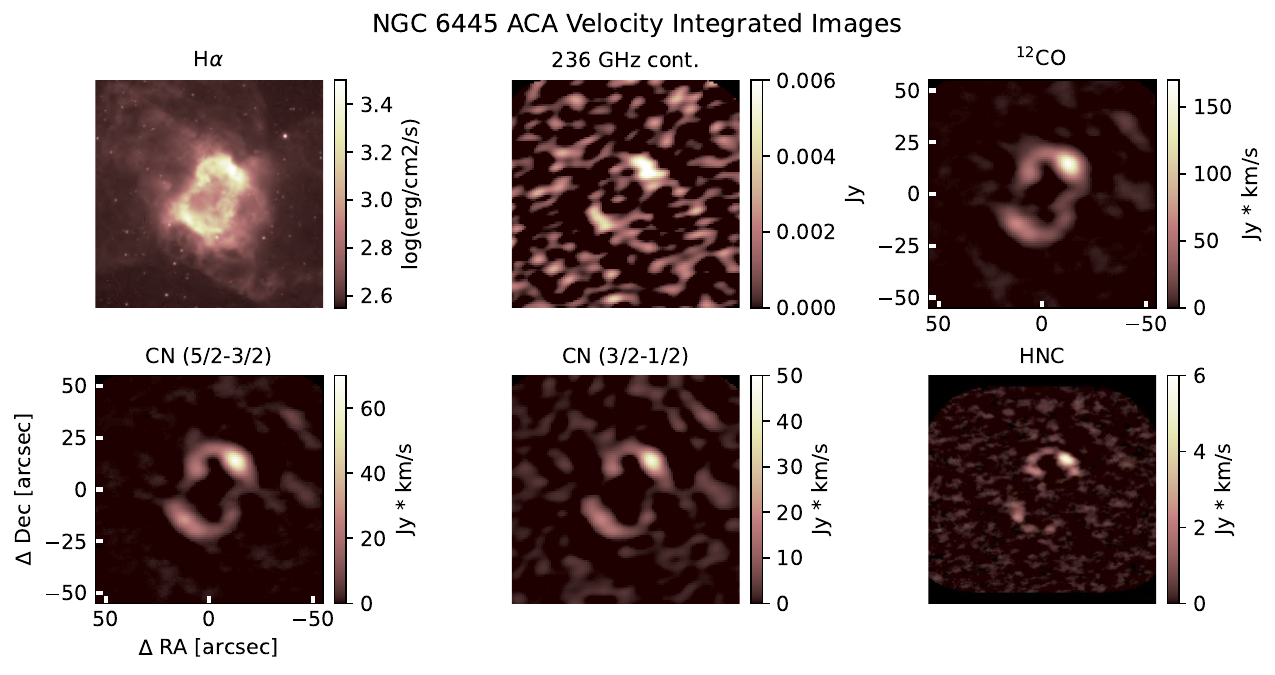}
    \caption{As in Fig.~\ref{fig:Hb5Mont}, but displaying ALMA Band 6 ACA mom0 images for NGC 6445. \textit{Top left panel}: an archival NOT H$\alpha$ image of NGC 6445. \textit{Second panel}: continuum emission at 236 GHz (C43-1) as received from the ALMA pipeline. \textit{Following panels}: $^{12}$CO, CN (5/2$-$3/2), CN (3/2$-$1/2), and HNC. All images have a field of view of 120$''$ x 120$''$.}
    \label{fig:N6445MontACA}
\end{figure}
\begin{figure}[h!]
    \centering
    \includegraphics[width=1\textwidth]{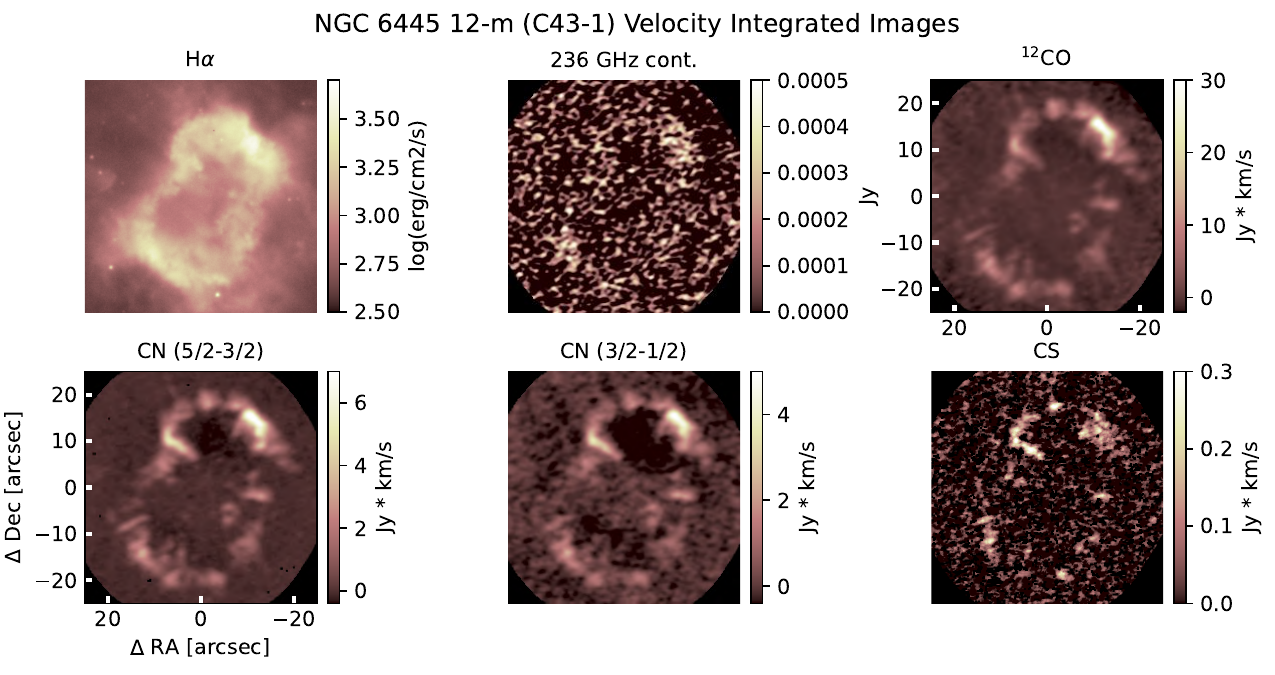}
    \caption{As in Fig.~\ref{fig:Hb5Mont}, but displaying ALMA Band 6 12-m mom0 images for NGC 6445. \textit{Top left panel}: an archival NOT H$\alpha$ image of NGC 6445. \textit{Second panel}: continuum emission at 236 GHz (C43-1) as received from the ALMA pipeline. \textit{Following panels}: $^{12}$CO, CN (5/2$-$3/2), CN (3/2$-$1/2), and CS, respectively. All images have a field of view of 50$''$ x 50$''$.}
    \label{fig:N6445Mont1}
\end{figure}
\begin{figure}[h!]
    \addtocounter{figure}{-1}
    \centering
    \includegraphics[width=1\textwidth]{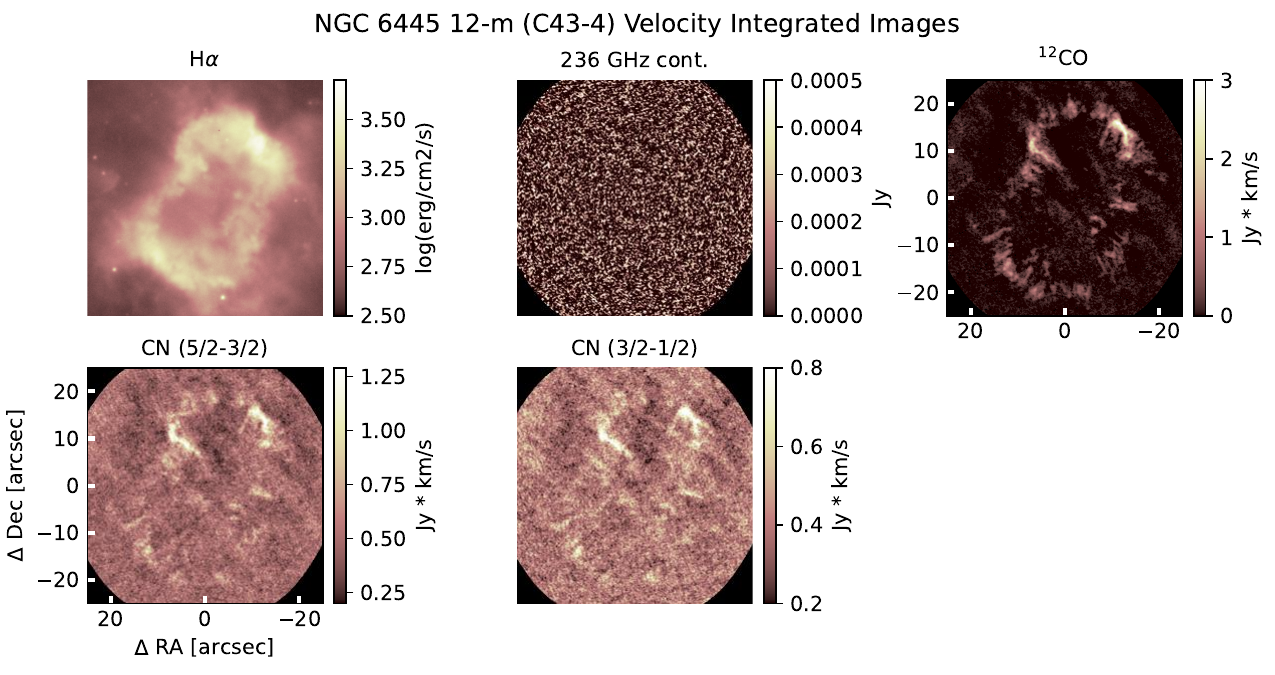}
    \caption{(cont.) \textit{Top left panel}: an archival NOT H$\alpha$ image of NGC 6445. \textit{Second panel}: continuum emission at 236 GHz (C43-4) as received from the ALMA pipeline. \textit{Following panels}: $^{12}$CO, CN (5/2$-$3/2), and CN (3/2$-$1/2), from left to right. All images have a field of view of 50$''$ x 50$''$.}
    \label{fig:N6445Mont2}
\end{figure}

Channel maps, mom0 image, and spectral line profile extracted from the 12-m Band 6 observations of $^{12}$CO emission from NGC 6445 are presented in Fig.~\ref{fig:N6445DatCube}. Other molecular line detections include HNC and hyperfine transitions of CN in the ACA observations (Fig.~\ref{fig:N6445MontACA}) and CS in the 12-m array observations (Fig.~\ref{fig:N6445Mont1}). As in the case of Hubble 5, the small number of molecules observed reflects observing time limitations for both ACA and 12-m arrays. 

It is immediately apparent from Fig.~\ref{fig:N6445DatCube} that the $^{12}$CO emission traces the bright ring of optical line emission surrounding the nebula's inner cavity. All molecular emission lines detected in NGC 6445 follow this same overall morphology. In the $^{12}$CO velocity channel maps, the molecular gas appears in the form of an elliptical ring of clumps tracing the pinched waist of the nebula, with clump velocities extending from roughly $-10$ km s$^{-1}$ to $+45$ km s$^{-1}$. Thus --- with the benefit of ALMA's spatial resolution --- we can conclude that the molecular gas within NGC 6445 is largely confined to an expanding, clumpy molecular torus, as opposed to multiple polar outflow components, as previously suggested on the basis of single-dish molecular line spectroscopy \citep{Schmidt2022}. The $^{12}$CO channel maps and integrated spectrum indicate that the molecular torus is expanding at roughly 30 km s$^{-1}$, with the SW region of the shell moving towards the observer and the NE region moving away. All spatially-integrated spectral profiles for NGC 6445 are shown in Appendix \ref{sec:app4}. %\joel{need to decide where/how the spectra for all PNe will be presented}

\subsection{NGC 2899}\label{sec:ResN2899}
\begin{figure}[h!]
    \centering
    \includegraphics[width=1.0\textwidth]{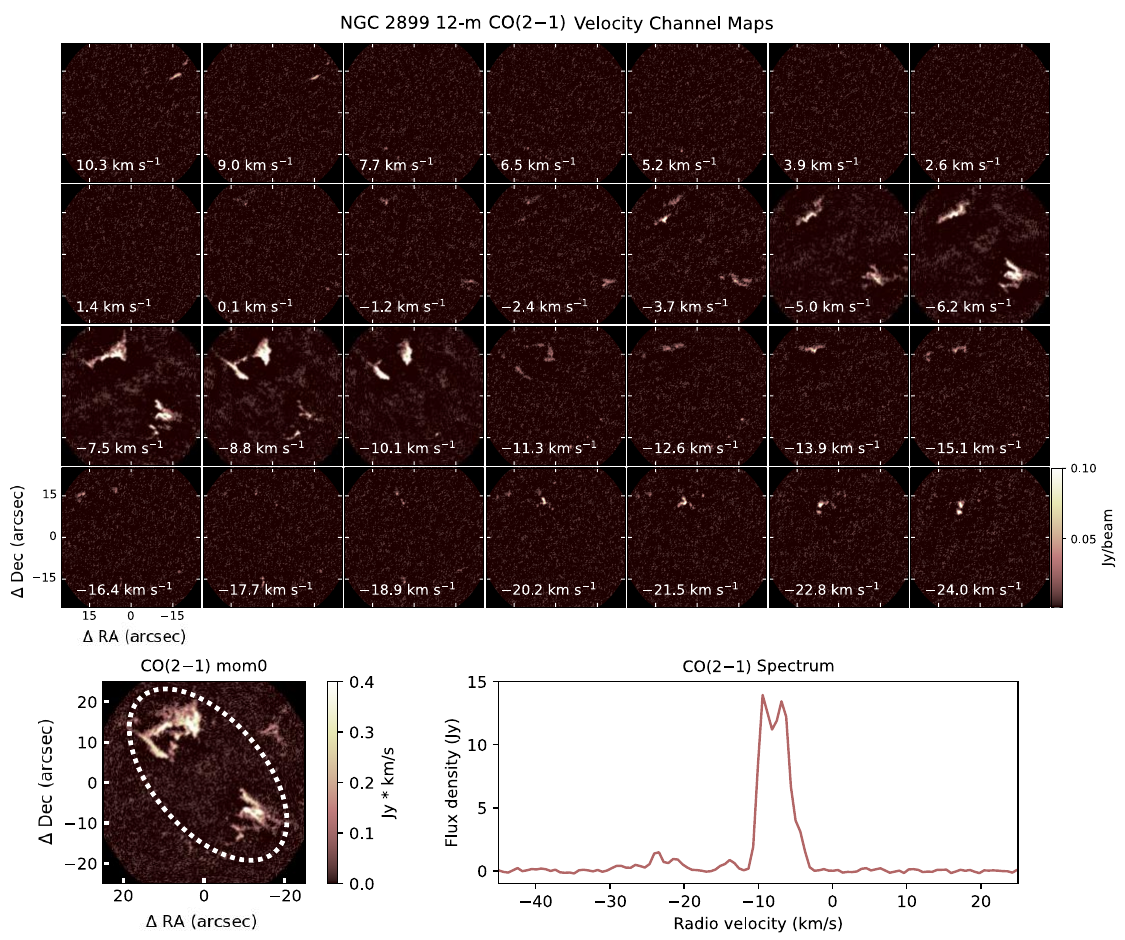}
    \caption{Same as Fig.~\ref{fig:Hb5DatCube}, but displaying 12-m Band 6 data for NGC 2899. Velocity channels are displayed in increments of 1.3 km s$^{-1}$.}
    \label{fig:N2899DatCube}
\end{figure}

\begin{figure}[h!]
    \centering
    \includegraphics[width=1\textwidth]{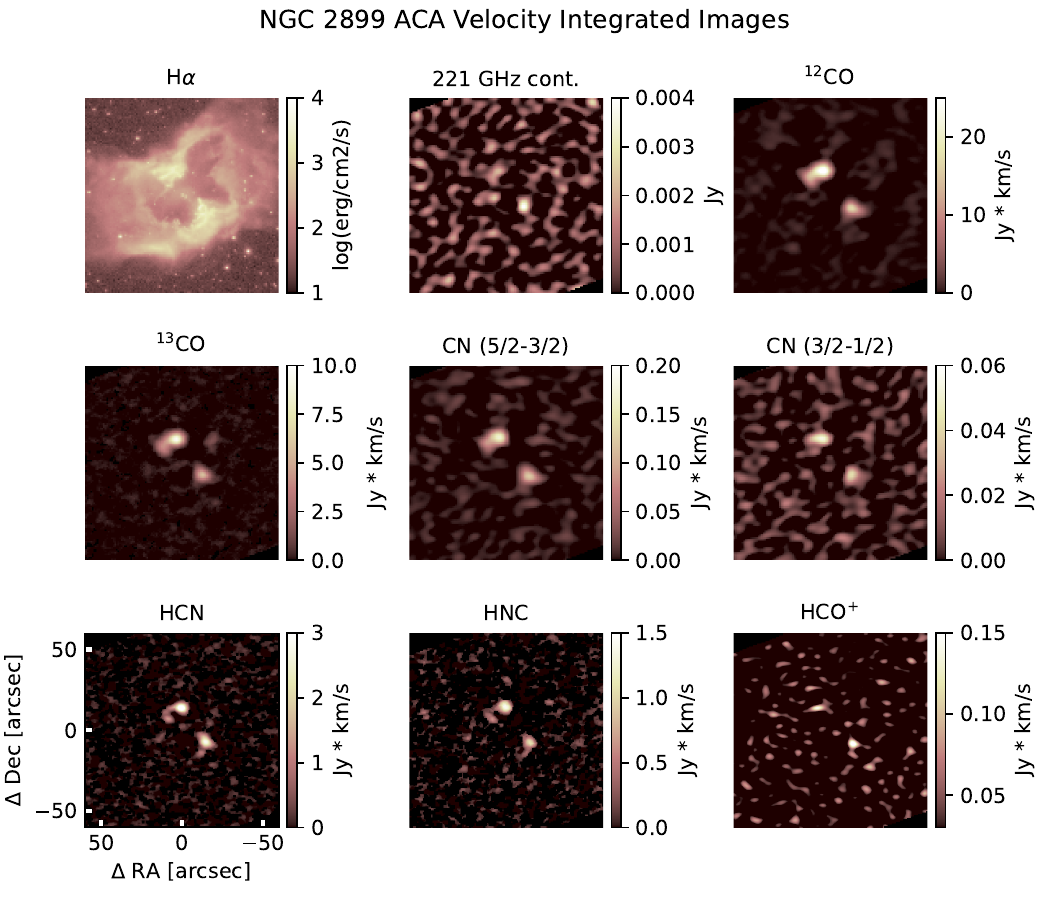}
    \caption{Same as Fig.~\ref{fig:Hb5Mont}, but displaying ACA Band 6 data for NGC 2899. \textit{Top left panel}: an archival ESO H$\alpha$ image. \textit{Second panel}: continuum emission at 221 GHz. \textit{Following panels}: $^{12}$CO, $^{13}$CO, CN (5/2$-$3/2), CN (3/2$-$1/2), HCN, HNC, and HCO$^+$. All images have a field of view of 120$''$ x 120$''$.}
    \label{fig:N2899Mont1}
\end{figure}

\begin{figure}[h!]
    \centering
    \includegraphics[width=0.8\textwidth]{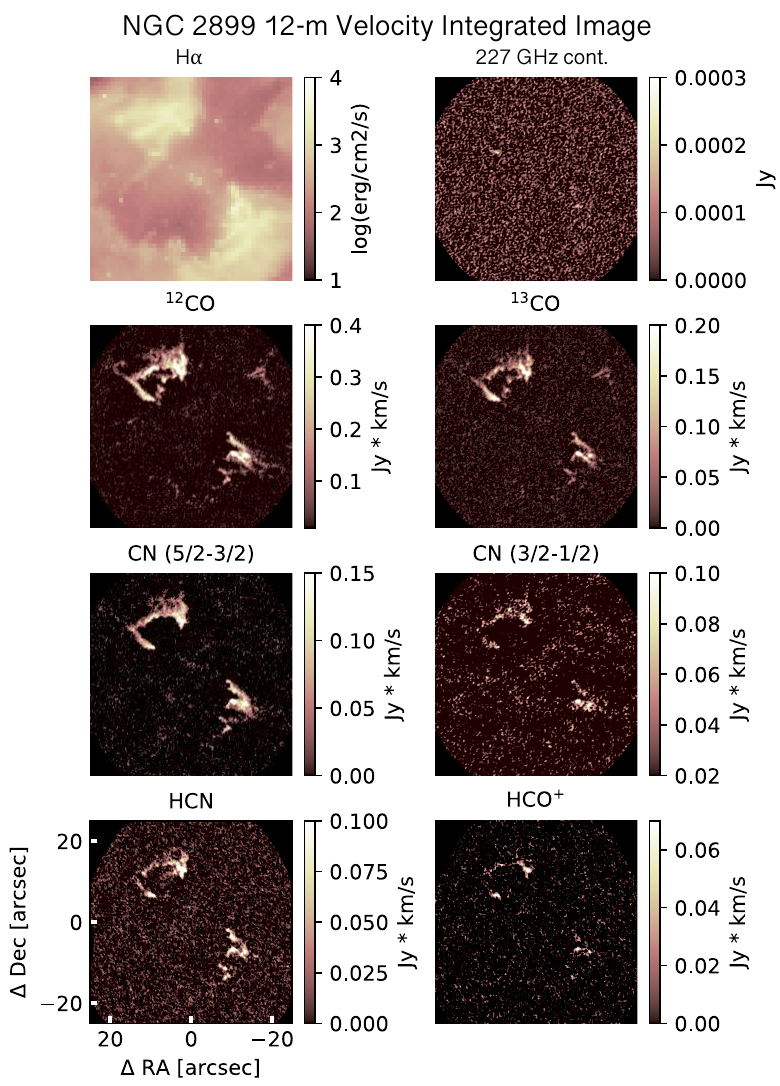}
    \caption{Same as Fig.~\ref{fig:Hb5Mont}, but displaying 12-m Band 6 data for NGC 2899. \textit{Top left panel}: is an archival ESO H$\alpha$ image. \textit{Second panel}: continuum emission at 227 GHz. \textit{Following panels}: $^{12}$CO, $^{13}$CO, CN (5/2$-$3/2), CN (3/2$-$1/2), HCN, and HCO$^+$. All images have a field of view of 50$''$ x 50$''$.}
    \label{fig:N2899Mont2}
\end{figure}
Channel maps, mom0 image, and spectral line profile extracted from the 12-m Band 6 observations of $^{12}$CO emission from NGC 2899 are presented in Fig.~\ref{fig:N2899DatCube}.
Molecular detections in NGC 2899 include $^{12}$CO, $^{13}$CO, CN, HCN, and HCO$^+$ in 12-m and ACA observations with a detection of HNC in ACA observations only (see Figs.~\ref{fig:N2899Mont1}--\ref{fig:N2899Mont2}). All spatially-integrated spectral profiles for NGC 2899 are presented in Appendix \ref{sec:app4}. 

Unlike the PNe that have been discussed thus far in this section, the molecular emission from NGC 2899 appears in the form of clumps along the pinched waist of the nebula rather than a complete toroidal structure. In Fig.~\ref{fig:N2899DatCube}, the $^{12}$CO emission resides mainly in two structures along the central cavity. There is a smaller clump on the W edge of the cavity, which appears in ACA as well as 12-m observations, along with faint, short filaments that trail along the S edge of the cavity, only visible in the 12-m data (see Figs.~\ref{fig:N2899Mont1} and \ref{fig:N2899Mont2}). The two main knot structures are expanding away from each other at $\sim$13 km s$^{-1}$, while the fainter structures have an expansion speed of $\sim$22 km s$^{-1}$. Whether or not all of these structures were once part of the same ring of molecular gas is unclear; the knots may be the remnants of a multiple ring system. Regardless, the fragmented structure of the molecular gas found in NGC 2899 indicates that this nebula is quite evolved, consistent with the conclusions of \citet{Lopez1991}. %\joel{is this accurate? did Lopez+91 find an old dynamical age?}

\subsection{NGC 2818}\label{sec:ResN2818}

\begin{figure}[h!]
    \centering
    \includegraphics[width=1.0\textwidth]{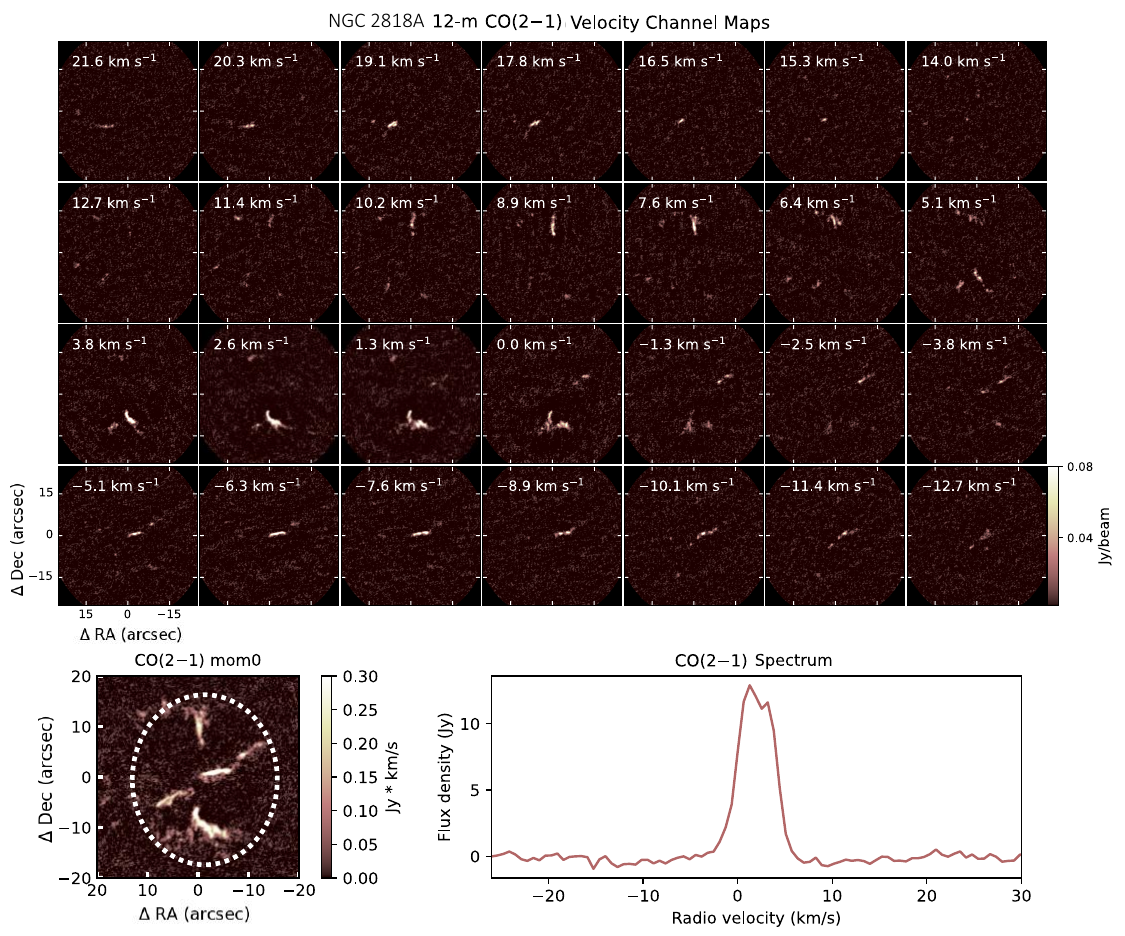}
    \caption{Same as Fig.~\ref{fig:Hb5DatCube}, but displaying 12-m Band 6 data for NGC 2818. Velocity channels are displayed in increments of 1.2 km s$^{-1}$.}
    \label{fig:N2818DatCube}
\end{figure}
\begin{figure}[h!]
    \centering
    \includegraphics[width=0.8\textwidth]{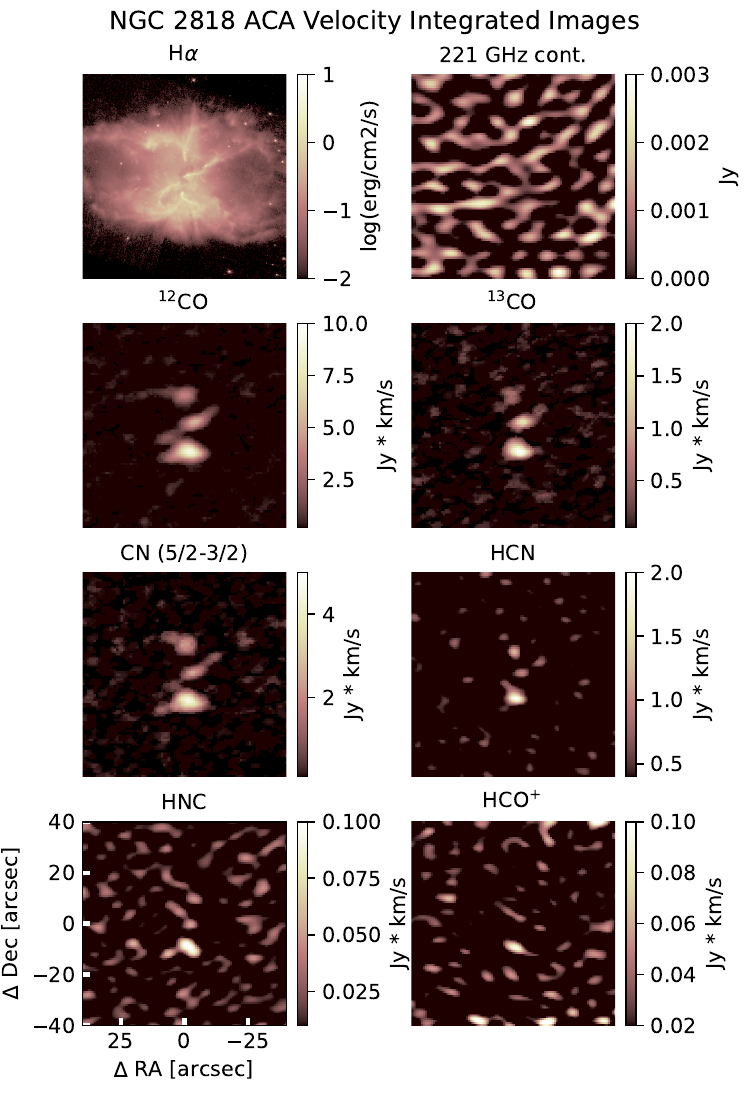}
    \caption{Same as Fig.~\ref{fig:Hb5Mont}, but displaying ACA Band 6 data for NGC 2818. \textit{Top left panel}: an archival HST/WFPC2 H$\alpha$ image of NGC 2818. \textit{Second panel}: continuum emission at 221 GHz (ACA) as received from the ALMA pipeline. \textit{Following panels}: $^{12}$CO, $^{13}$CO, CN (5/2$-$3/2), HCN, HNC, and HCO$^+$. All images have a field of view of 80$''$ x 80$''$.}
    \label{fig:N2818Mont1}
\end{figure}
\begin{figure}[h!]
    \centering
    \includegraphics[width=1\textwidth]{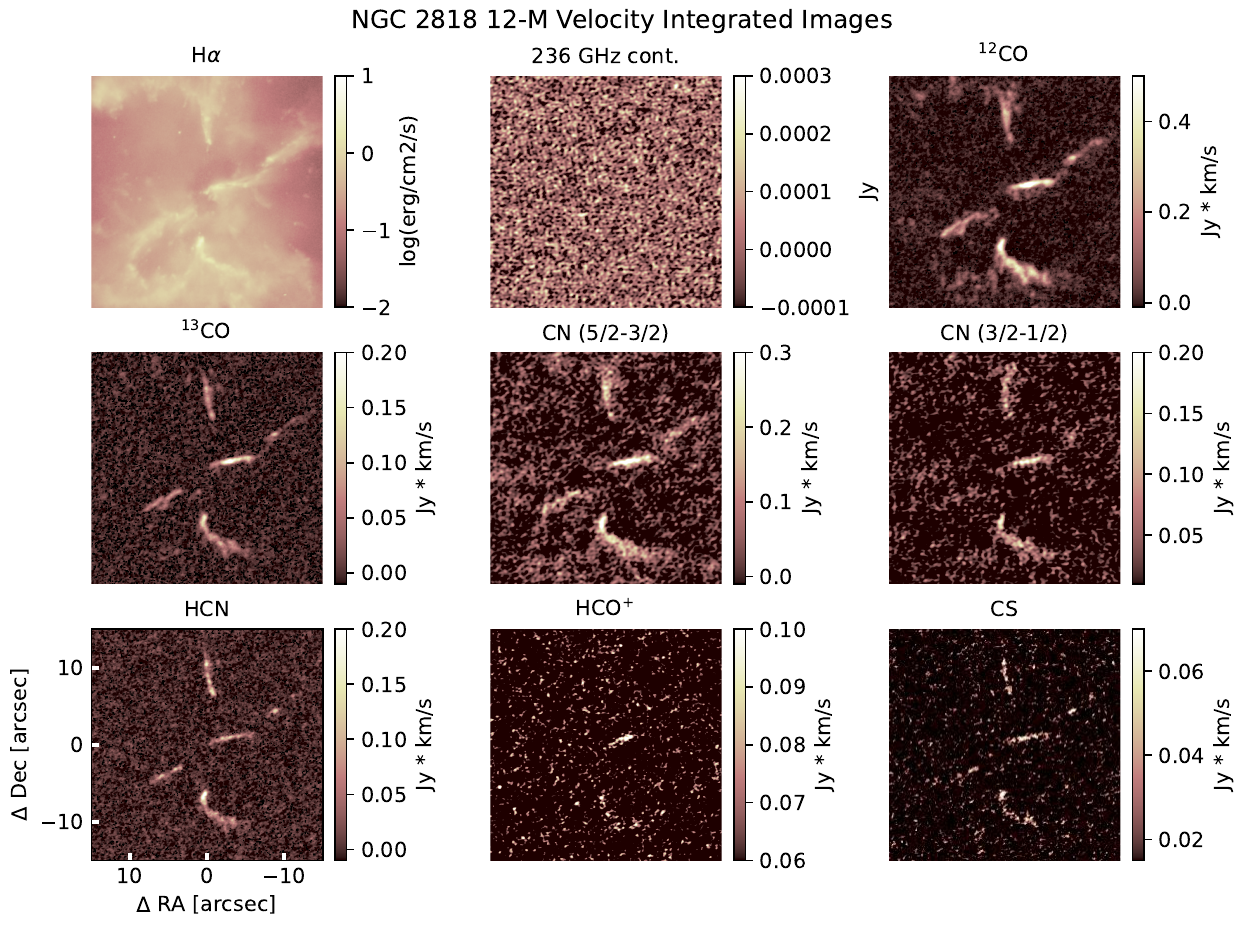}
    \caption{Same as Fig.~\ref{fig:Hb5Mont}, but displaying 12-m Band 6 data for NGC 2818. \textit{Top left panel}: an archival HST/WFPC2 H$\alpha$ image. \textit{Second panel}: continuum emission at 236 GHz (C43-4) as received from the ALMA pipeline. \textit{Following panels}: $^{12}$CO, $^{13}$CO, CN (5/2$-$3/2), CN (3/2$-$1/2), HCN, HCO$^+$, and CS. All images have a field of view of 30$''$ x 30$''$.}
    \label{fig:N2818Mont2}
\end{figure}

Channel maps, mom0 image, and spectral line profile extracted from the 12-m Band 6 observations of $^{12}$CO emission from PN NGC 2818 are presented in Fig.~\ref{fig:N2818DatCube}.
Molecules detected in NGC 2818 include $^{12}$CO, $^{13}$CO, CN, HCN, and HCO$^+$ in both ACA and 12-m arrays along with a detection of CS in 12-m array observations only \revision{and a detection of HNC in ACA observations only} (see Figures \ref{fig:N2818Mont1} and \ref{fig:N2818Mont2} for ACA and 12-m array observations, respectively). As in the case of NGC 2899, just discussed, the molecular torus of NGC 2818 appears to be highly fragmented. In Fig.~\ref{fig:N2818DatCube}, the $^{12}$CO emission traces the filament structures that spread across the center of the nebula in optical emission-line images (Fig.~\ref{fig:OptVsRadio}). The 12-m array observations of $^{12}$CO reveal finer structures that wrap around the central cavity of ionized gas. From these observations, we infer that the clumps of molecular gas were once part of a nearly edge-on,  molecular torus, suggesting that NGC 2818 is indeed an evolved PN. The remnant of this molecular torus is expanding at $\sim$17 km s$^{-1}$, with the eastern-most clumps moving towards the observer and the western-most clump moving away. All spatially-integrated spectral line profiles for NGC 2818 are shown in Appendix \ref{sec:app4}.

\section{Dynamical Ages}\label{sec:ageCalc}

\subsection{Measurements}
\begin{figure}[p!]
    \centering
    \includegraphics[width=0.95\textwidth]{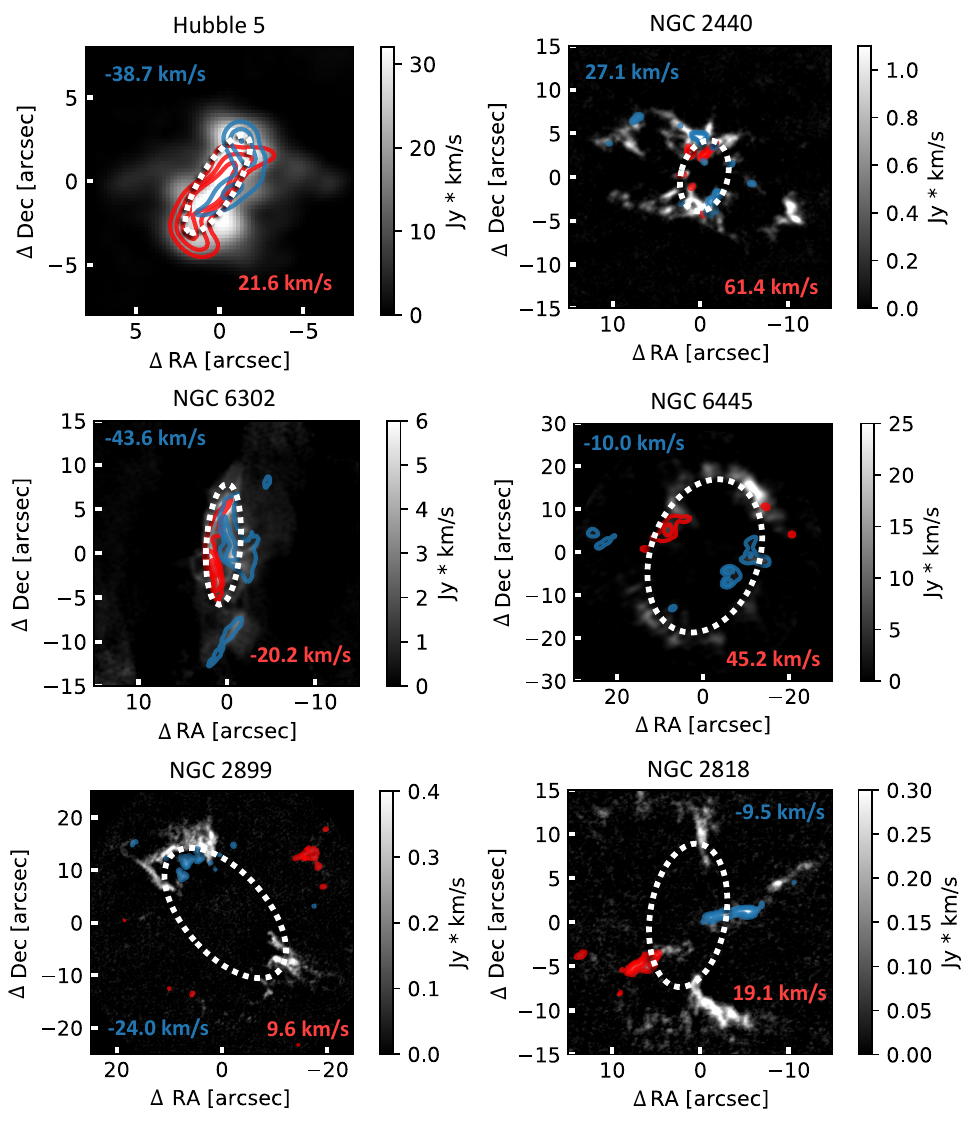}
    \caption{Contours of redshifted (red) and blueshifted (blue) emission components overlaid on greyscale mom0 images for Hubble 5, NGC 2440, NGC 6302, NGC 6445, NGC 2899, and NGC 2818, as extracted from the 12-m array $^{12}$CO(2$-$1) data cubes. The central $V_{\mathrm{LSR}}$ used to highlight the red- and blueshifted components are noted in each panel with red and blue text, respectively. \revision{A set of 3 channels centered around the $V_{\mathrm{LSR}}$ velocities labeled in the figure were used to create the blue and red contour maps with a total velocity range \secrevise{of} 1.9 km s$^{-1}$ for all PNe.} White dotted ellipses overlaid on the emission components denote the (projected) molecular torus geometry adopted for each PN (with ellipse parameters listed in Table~\ref{tbl:allAge}).}
    \label{fig:Band6age}
\end{figure}

In order to calculate the dynamical age of a PN's central torus, we assume that the molecular emission regions have simple toroidal geometries, radially expanding with uniform velocities. By assuming that the intrinsic shape of the torus is perfectly circular, the problem is simplified to one dimension of motion with one inclination angle $i$, defined as the angle of inclination with respect to the plane of the sky (where $i=90^\circ$ is defined as edge-on). Fig.~\ref{fig:Band6age} shows how \revision{the orientation of each molecular torus was defined with respect to the mom0 image and the projected (redshifted and blueshifted) velocity components of the molecular gas for the target PNe, which in turn are used to estimate the projected expansion velocity of each torus}. \revision{The dynamical age $t$ is then
%\revision{Solving for dynamical age $t$ as a function of major axis measurement and expansion velocity,  we obtain $t=x/V\cos{i}$; taking account the conversion of the measured minor axis of the defined ellipse from arcseconds to km units using and the conversion from seconds to years, the formula for} dynamical age in years is then \revision{defined} as 
\begin{equation}
%    t \: (\mathrm{yr})=1.496\times10^{15} \, x('') \, D(\mathrm{pc}) \, \cos{i}/V(\mathrm{km/s)}
%    t \: (\mathrm{yr})=4.8\times10^{15} \, a \, D \, /(V_{\mathrm{exp}} \, \sin{i})
t \: (\mathrm{yr})=4.8 \, a \, \sin{i} \: D \, V_{\mathrm{exp}}^{-1}
\end{equation}
\noindent
where $a$ is the angular size of the projected ellipse's semi-major axis in arcseconds; $i$ is given by $\cos^{-1}{(b/a)}$, where $b/a$ is the projected ellipse semi-minor to semi-major axis ratio; $D$ is the adopted distance to the PN in pc; and $V_{\mathrm{exp}}$ is the estimated projected expansion velocity in km s$^{-1}$.}
%and the coefficient ($4.8\times10^{15}$) accounts for the conversion from angular to linear size scales as well as from seconds to years.}
%\joel{Paula, please double-check the equation...I tweaked the $V \cos{i}$ factor, I think this is correct?}

The $^{12}$CO ALMA Band 6 data in 12-m configuration are used to estimate the ellipse parameters of each nebula because these data resolve the tori more readily than the ACA data. The inferred ellipse parameters used for dynamical age estimates, as well as the resulting ages, are listed in Table \ref{tbl:allAge}. The listed uncertainties in these dynamical ages reflect the uncertainties in our adopted distances, and do not reflect systematic errors in the determination of the dimensions and inclinations of the molecular tori (i.e., the ellipse parameters).
%\joel{add the adopted $V$ to Table, and reduce all quantities to 2 sig figs} 
The resulting dynamical ages of the molecular tori range from $\sim$500 yr (Hubble 5) to $\sim$11000 yr (NGC 2818).

\begin{table}
\begin{center}
\caption{\sc Molecular Torus Dynamical Age Measurements }
\vspace{.2in}
%\begin{adjustwidth}{-3cm}{}
%\begin{adjustbox}{width=1.2\textwidth}
\label{tbl:allAge}
\scriptsize
\begin{tabular}{lcccccccc}
\toprule
PN G & Name & Distance & Ellipse & Incl. & Expansion & Dynamical & Optical Lobe & Optical Lobe \\
 & & & Dimensions & Angle $i^a$ & Velocity$^a$ & Age$^a$ & Incl. Angle & Dyn. Age\\
 & & (kpc) & ($a$, $b$) & ($^{\circ}$) & (km/s) & (yrs) & ($^{\circ}$) & (yrs)\\
\midrule
\midrule
359.3-00.9 & PN Hb 5 & 0.9$^{+0.1}_{-0.1}$ & (4.5$''$, 2.5$''$) & 55 & $30$ & 500$^{+120}_{-120}$ & 62$^b$ & 800$^b$\\
234.8+02.4 & NGC 2440 & 1.0$^{+0.1}_{-0.1}$ & (5.0$''$, 2.5$''$) & 65 & $17$ & 1500$^{+360}_{-360}$ & 65$^c$ & 1000$^d$\\
349.5+01.0 & NGC 6302 & 1.0$^{+0.1}_{-0.1}$ & (11.0$''$, 3.5$''$) & 70 & $14$ & 3300$^{+660}_{-660}$ & 70$^c$ & 900--2300$^e$\\
008.0+03.9 & NGC 6445 & 1.1$^{+0.1}_{-0.1}$ & (18.0$''$, 13.0$''$) & 50 & $22$ & 3700$^{+550}_{-550}$ & 50$^c$ & 3400$^f$\\
277.1-03.8 & NGC 2899 & 0.8$^{+0.2}_{-0.2}$ & (16.0$''$, 6.0$''$) & 70 & $8$ & 7600$^{+2200}_{-2200}$ & 75$^c$ & 5400$^g$\\
261.9+08.5 & NGC 2818 & 3.3$^{+0.1}_{-0.1}$ & (10.0$''$, 5.0$''$) & 60 & $12$ & 11000$^{+2000}_{-2000}$ & 60$^h$ & 7400$-$11000$^{h,i}$\\
\midrule
\end{tabular}
%\end{adjustbox}
%\end{adjustwidth}
\end{center}
{\sc Notes:} (a) This work; age uncertainties are determined by distance errors and uncertainties in ellipse dimensions of 0.5$''$; \secrevise{the expansion velocities listed in this table correspond to emission in the molecular torus only, these are different from the expansion velocities listed in Appendix~\ref{sec:app1} which are derived using all observed emission for each molecular line}. (b) \citealt{Lopez2012}. 
%(c) \citealt{Wang2008}. \joel{remove this value and reference (see below)...this is the PA of the flow with the lobe CO clumps, not inclination angle} 
(c) \citealt{Schwarz2008}. 
%(d) \citealt{Baessgen1995}. 
%(e) \citealt{Gorny1997}. 
(d) See text. 
%(g) \citealt{Santander-Garcia2017}. 
(e) \citealt{Balick2023}. 
%(i) \citealt{Schmidt2022}. 
(f) Dynamical age of central ionized region as estimated by \citealt{vanHoof2000}. 
(g) Based on data presented in \citealt{Lopez1991}; see text. 
(h) \citealt{Vazquez2012}. 
(i) \citealt{Derlopa2024}. All literature ages for optical lobes are corrected using this work's adopted distances; see text. 
\end{table}

\subsection{Comparisons with Previous Estimates}

%\joel{for the PNe where our $D$ differs from that assumed by others who estimated ages, we should correct their ages for our adopted $D$ in the Table, and note in text, when making the comparison (in the Table footnotes for these PNe, you can say ``corrected for adopted $D$'')...see example for NGC 2440, below (though it will need to be further revised...}

%\joel{I added this little preamble...}

Although mm-wave transitions of CO (and, in some cases, other trace molecules) had been previously detected in all six of the PNe included in our ALMA Band 6 mapping survey (see \S\S~\ref{sec:ResHb5}--\ref{sec:ResN2899}), only NGC 6302 has previously been mapped at sufficient spatial and spectral resolution (by ALMA) to resolve its central torus and thereby estimate a torus dynamical age \citep[2200 yr;][]{Santander-Garcia2017}. The estimated molecular torus dynamical ages of the other five nebulae reported in Table~\ref{tbl:allAge} are hence the first such direct determinations for these PNe. In the following, we compare these estimated torus dynamical ages with PN dynamical ages reported in the literature, almost all of which describe the expansion ages of the bipolar lobes (as opposed to the equatorial tori) of the PNe.

%CO was first detected in Hb 5 by \citet{Schmidt2016}, with no conclusive age derived, making this determination of molecular torus age using ALMA Band 6 data the first. 
{\it Hubble 5:} On the basis of long-slit optical spectroscopy data, \citet{Lopez2012} determined an age for the optical lobes of $1500\pm540$ yr assuming a distance of 1.7 kpc. \revision{This age estimate decreases to $800\pm300$ yr for our revised distance of 0.9 kpc. The molecular torus dynamical age derived in this work, $\sim$$500$ yr, hence falls at the lower end of the range of uncertainty in the \citet{Lopez2012} estimate.} The molecular torus inclination angle we derive from the ALMA CO mapping ($55^\circ$) is similar to that estimated for the inclination of the nebula's polar axis \citep[$62^\circ$;][]{Lopez2012}.
%\joel{Corradi \& Schwarz 1992 or 1993 did the same sort of long-slit spectroscopy; I vaguely recall they didn't offer an age because of a large $D$ uncertainty at that point...what $D$ did Lopez use? (see comment above)}\paula{Lopez uses a distance estimated in a 2010 Stanghellini paper, $\sim$1.7 kpc.}
    
%The age of NGC 2440's molecular gas has been calculated, using an ellipse with major axis $a=4.34''$ and minor axis $b=1.99''$ with inclination $i=63^\circ$, to have an age of $\sim2300$ years. 
{\it NGC 2440:} We find a molecular torus dynamical age of $\sim$1500 yr and inclination angle of $i=65^\circ$. Previous dynamical age estimates for the optical lobes of NGC 2440 range between 2100 and 4200 yr  \citep[and references therein]{Baessgen1995,Gorny1997,Schmidt2016}, after correcting to our revised distance estimate of 1.0 kpc. However, there is evidence the lobes may be much younger dynamically.
\citet{Lago2016} used comprehensive long-slit spectroscopy as the basis for detailed 3D spatio-kinematical modeling of the nebula, determining inclinations of $\sim$63$^\circ$ and $\sim$90$^\circ$ for its two main lobe pairs, oriented at position angles of 85$^\circ$ and 35$^\circ$, respectively; the former agrees well with the polar axis inclination of 65$^\circ$ 
%for the lobes that this work focuses on 
listed in \citet{Schwarz2008}, as well as our inferred torus inclination. Although \citet{Lago2016} do not estimate dynamical ages for the optical lobe pairs, their inferred expansion velocities of 184 km s$^{-1}$ and 160 km s$^{-1}$ for the PA 85$^\circ$ and 35$^\circ$ lobe pairs (respectively) would imply very young dynamical ages of $\sim$800--1000 yr for these structures, given our adopted distance of 1.0 kpc to NGC 2440.
Our estimated molecular torus dynamical age, $\sim$1500 yr, hence indicates that this equatorial torus structure was ejected well before the polar lobe pairs.
%\citet{Wang2008} used their single-dish $^{12}$CO ($J=3-2$) mapping observations, an assumed distance of $\sim$2 kpc, and an assumed inclination angle of 35$^\circ$ to estimate a dynamical age of 15000 years for the molecular emission region. 
%as in our ALMA Band 6 12-m data, and a 2D orientation was measured rather than a 3D orientation along the plane of the sky, as is done in this work. 
%This estimate corresponds to a dynamical age of $\sim$8000 yr, for our adopted $D$ of $\sim$1.0 kpc. However, as \citet{Wang2008} noted, it is not possible to ascertain the molecular emission morphology of NGC 2440 from their 14$''$ resolution CO mapping data. Our estimated molecular torus dynamical age is hence the first reliable estimate, for this nebula. 

%\joel{after I edited the above I looked back at  Wang+2008 and I see their estimate of 15000 yr (actually 17300 yr) is for the two NE--SW CO knots, not the torus. (Also, the 35$^\circ$ estimate is the position angle in the sky, not the inclination.) They actually give two different age estimates, assuming 18 km/s and 180 km/s as the possible deprojected velocities of the CO knots; if 180 km/s then the age is only 1730 yr, which becomes about 800 yr for our $D$ (wow -- very recent!).  So the above will need to be revised...and should take into account the results of SHAPE 3D modeling by \citet{Lago2016}, which indeed yielded 160 km/s for the outflow $V$ of the NE--SW lobe pair}

{\it NGC 6302:} 
%is calculated to have a torus with an age of $\sim2800$ years, using an ellipse with major axis $a=5.69''$ and minor axis $b=1.98''$ with inclination $i=70^\circ$. 
Our estimated molecular torus inclination of $i=70^\circ$ and dynamical age of $\sim3300$ years are in good agreement with previous estimates of 75$^\circ$ and 2500--3700 years, respectively, as obtained from far more detailed structural modeling of early-cycle ALMA $^{12}$CO(3--2) mapping data \citep{Santander-Garcia2017}. 
The torus inclination is similar to that inferred for the bipolar lobe symmetry axis \citep[70$^\circ$;][]{Schwarz2008}. Recent analysis of proper motions of knots and features in the optical (ionized) nebula, as imaged by HST/WFC3 over a $\sim$10 yr interval, revealed multiple, point-symmetric zones within its bipolar lobe system; these lobe zones (knot systems) have dynamical ages ranging from $\sim$900 yr to $\sim$2300 yr, strongly suggestive of episodic, misaligned polar outflows \citep{Balick2023}. The estimated dynamical age of the oldest of these episodic polar lobe ejections revealed by multi-epoch HST imaging, which is very similar to that deduced from ground-based proper motion measurements of clumps in the outermost regions of the NW lobe \citep[2200 yr;][]{Meaburn2008}, would make the lobes somewhat younger than the molecular torus, based on the dynamical ages estimated from ALMA CO data both via our simple analysis and the detailed modeling of \cite{Santander-Garcia2017}. 
%In optical wavelengths, the torus is visible as a dark lane and its inclination angle was calculated to be 70$^\circ$ \citep{Schwarz2008}. The age of NGC 6302 has been more difficult to pin down in optical wavelengths because of the presence of knots that exhibit differing ages; however, the oldest of these knots has been estimated to be $2300-3400$ years old \citep{Meaburn2008,Balick2023}.

{\it NGC 6445}: The projected elliptical shape and velocity extremes of the clumpy molecular emission ring within this PN yield an intermediate inclination of $i=50^\circ$ and a molecular torus dynamical age of $\sim$3700 years. This estimated dynamical age is within the range of previously determined dynamical ages for the molecular emitting region \citep[$\sim$4600 yr, based on assumed emitting region radius;][]{Schmidt2022} and the central $\sim$$40''\times30''$ ionized region \citep[$\sim$3400 yr;][]{vanHoof2000}, where we have corrected both age estimates for our adopted distance. We find no previous literature measurements of the expansion velocities and (hence) dynamical ages for NGC 6445's polar lobes. However, our inferred molecular torus inclination is in good agreement with the polar axis inclination of 50$^\circ$ listed in \citet{Schwarz2008}.

%This PN has the next oldest torus with a calculated age of $\sim4300$ years, determined with an ellipse with major axis $a=20.00''$, minor axis $b=13.90''$, and inclination $i=48^\circ$. \citealt{Schmidt2022} calculated an older age for the molecular torus at around 4700 years, while \citealt{vanHoof2000} had previously calculated an age of 3300 years. Although there are no derivations for the inclination angle of NGC 6445's molecular gas in past literature, the inclination angle for the optical shell has been well-determined with the most recent measurement being 50$^\circ$ \citep{Schwarz2008}. \citealt{Mata2016} made use of \textit{Spitzer} mid-IR images to derive an age of 8676 years for NGC 6445's the extended lobes using a distance of 1.39 kpc.

%Since NGC 2818's distance is not well-constrained, distances of 1.47 kpc \citep{Bucciarelli2023} and 4.2 kpc \citep{Derlopa2024} were used to set lower and upper limits. The central torus of NGC 2818 is estimated to have an age ranging from 19000$-$25000 years, using an ellipse with major axis $a=8.97''$, minor axis $b=4.87''$, and inclination $i=57^\circ$. Although there are no previous inclination angle and age calculations using molecular data in recent literature, the age derivation in this work is in agreement with optical lobe age estimates made in \citealt{Vazquez2012} ($i=60^\circ$) and \citealt{Derlopa2024} ranging from 5000$-$12000 years. 

{\it NGC 2899:} This PN is estimated to have the second oldest molecular torus in our ALMA Band 6 survey, with a dynamical age of $\sim$7600 years. 
%using an ellipse with major axis $a=16.42''$, minor axis $b=6.22''$, and 
We find a torus inclination of $i=70^\circ$, in good agreement with the polar axis inclination of $i=75^\circ$ listed in \citet{Schwarz2008}. Based on the polar lobe expansion velocities of $\sim$100-130 km s$^{-1}$ measured by \citet{Lopez1991} and the ($\sim$60$''$) extent of the lobes in their images, we roughly estimate a polar lobe dynamical age of 5400 yr, suggesting the lobes are somewhat younger than the molecular torus.
%There are no previous measurements of its inclination angle or derivations of dynamical age for NGC 2899 using radio data. However, \citealt{Schwarz2008} was able to determine an inclination angle $i=75^\circ$ using European Southern Observatory (ESO) H$\alpha$ images, and an age of 30000 years was estimated for NGC 2899's extended optical lobes with data provided in \citealt{Lopez1991}.

{\it NGC 2818:} We find an inclination of $i=60^\circ$ and a dynamical age for the molecular torus in the range 9000--13000 yr, adopting the (open-cluster-based) distance of 3.3 kpc. \citet{Vazquez2012} found a polar axis inclination of $i=60^\circ$ and a dynamical age for the polar lobes of $\sim$11000 yr, while \citealt{Derlopa2024} found $i=70^\circ$ and $\sim$7400 yr, respectively (where we have rescaled both  previous dynamical age estimates to our revised PN distance). Given either age estimate, it is apparent that the molecular torus dynamical age is far older than that of the lobes.

\subsection{Discussion} \label{sec:AgeDiscussion}

The results for dynamical age in Table \ref{tbl:allAge} are consistent with the progression of molecular torus morphologies described in \S\S % \ref{sec:molLineRatios})
\ref{sec:ResHb5}--\ref{sec:ResN2899}, wherein the dynamically younger PNe (Hubble 5, NGC 2440, NGC 6302, and NGC 6445) have the most coherent central molecular ring structures (see Figs.~\ref{fig:Hb5Mont}, \ref{fig:N2440Mont2}, \ref{fig:N6302Mont3}, and \ref{fig:N6445Mont2}, respectively) and the oldest PNe (NGC 2899 and NGC 2818) have more fragmented structures (see Figs.~\ref{fig:N2899Mont2} and \ref{fig:N2818Mont2}, respectively). Furthermore, this sequence also represents a monotonic decline in mm-wave continuum emission from ionized gas at the nebular cores, wherein Hubble 5, NGC 2440, and NGC 6302 display bright continuum emission from their tori, NGC 6445 displays relatively weak continuum emission, and NGC 2899 and NGC 2818 are undetected. 
Only the dynamically youngest PNe --- Hb 5, NGC 2440, NGC 6302 --- display emission from molecular gas within their optical (polar) lobes, in the form of bubble-like structures; NGC 2440 also displays twin knots of CO and trace molecule emission that coincide with optical condensations at the tips of one of its two main sets of polar lobes.

Furthermore, a pattern emerges across this sample of extreme, high-excitation bi-lobed PNe in which the molecular torus dynamical ages are consistently older than the dynamical ages of the optical lobes (Table~\ref{tbl:allAge}). \revision{The lone exception is the dynamically youngest sample PN, Hb 5, for which the torus and lobe ages are the same to within the uncertainties. The comparison of torus vs.\ lobe dynamical ages in Table~\ref{tbl:allAge} hence generally supports the schematic ``jet lag'' model describing the evolution of bipolar PNe \citep{Huggins2007}, wherein the ejection of a molecular torus precedes the fast, collimated outflows that 
%shaped the molecular torus and 
form the lobes of ionized gas observed in the optical/IR \citep[see also][and references therein]{Kastner2025b}.} Indeed, the dynamically younger sample PNe display evidence of multiple collimated outflows during their post-torus-ejection PN stages. Such episodic ejections are apparent both through determinations of the ages of point-symmetric lobe substructures \citep[e.g., NGC 6302;][and references therein]{Balick2023} as well as via inspection of their optical lobe morphologies \citep[e.g., Hb 5, NGC 2440, NGC 6445; see Fig.~\ref{fig:OptVsRadio} and discussions in][]{Lopez2012,Lago2016}. 
%\revision{It is important to point out that Hb 5 has a higher error percentage than all other PN in our sample, which is likely the reason that it does not appear to follow this model well.}

The strong implication is that molecular torus ejection represents the final major mass loss stage of the progenitor AGB stars of these pinched-waist bipolar PNe, and that this equatorially-directed mass loss then leads directly to a phase of collimated, episodic (and perhaps precessing) polar outflows from the vicinities of their hot, luminous central stars. We note that this general scenario is the same as that proposed to explain the structural evolution of ring-like PNe \citep[][and references therein]{Kastner2024,Kastner2025a} as well as the complex juxtaposition of structures observed in the young, molecule-rich, high-excitation PN NGC 7027 \cite[][and references therein]{Bublitz2023,Moraga2023}. Indeed, in NGC 7027, we are likely seeing the immediate precursor to somewhat (dynamically) older bipolar PNe Hb 5, NGC 2440, NGC 6302, and NGC 6445, which in turn represent precursors to the dynamically old --- yet still high-excitation and molecule-rich --- NGC 2899 and NGC 2818.
%This feature is further discussed in \S \ref{sec:conclusion} of this work. 
%\joel{agreed! And note that the torus ages are (almost?) always older than the lobes...this makes sense, if the torus was ejected first, and then effectively becomes the `lobe collimator'...also supports the late/great Pat Huggins's `jet lag' model, so we should cite his paper proposing that model...this discussion probably belongs in the Conclusions, see my comments there}

\section{H$_2$ Masses}\label{sec:massCalc}
%\joel{all values and errors in sci notation need `()', as in: $(5.8\pm0.4)\times10^{15}$}

\subsection{Mass Estimation from CO Lines}
As in most astrophysical environments, the masses of cold molecular gas in PNe are most easily measured via mm-wave observations of rotational lines of CO, the second most abundant molecule after H$_2$, as H$_2$ has no dipole moment and the low-lying rotational energy levels of CO are readily excited even in low-density ($n \gtrsim 10^2$ cm$^{-3}$) regions. We hence follow standard practice \cite[e.g.,][and references therein]{Kastner2024} and estimate H$_2$ masses from the ACA $^{12}$CO(2--1) and $^{13}$CO(2--1) observations of our target PNe by modeling these line intensities to obtain pixel-by-pixel CO column densities across each object, with the $^{12}$CO brightness temperatures serving to constrain the emitting region kinetic temperatures; integrating the resulting column densities over each PN's CO emission region, to obtain the total number of CO molecules; and, finally, applying assumptions for CO:H$_2$ number ratios to obtain the total number (hence mass) of H$_2$ molecules. 

To estimate the CO column densities, we used the RADEX statistical equilibrium radiative transfer code \citep{RADEXpaper}. For the purpose of these estimates, both $^{12}$CO(2--1) and $^{13}$CO(2--1) lines are assumed to be optically thin; for many or perhaps most of the sample objects, the former assumption may not hold (see below). 
%is likely valid for most if not all of the target PNe given both the RADEX model calculations and previous work modeling other CO-luminous PNe \cite[e.g.,][]{Bachiller1997,Kastner2024}. 
We use the peak main-beam antenna temperature $T_{mb}$ within the $^{12}$CO(2--1) data cube to provide a lower limit on the kinetic temperature $T_k$ via $T_{mb} \sim T_{k} - T_{bg}$ \citep{Dinh2008}, where $T_{bg}$ is the temperature of the cosmic background radiation and the main beam brightness temperature\footnote{Formula derived for radio interferometry in https://science.nrao.edu/facilities/vla/proposing/TBconv} T$_{mb}$ is obtained via
\begin{equation}
    T_{mb} = 1.222\times 10^{3} \frac{I}{\nu^2 \theta_{maj}\theta_{min}},
\end{equation}
with $I$ the line intensity in mJy beam$^{-1}$, $\nu$ the frequency of emission in GHz, and $\theta_{maj}$ and $\theta_{min}$ the major and minor axes of the beam in arcseconds. For most of the PNe, the resulting $T_{k}$ range from 18 to 40 K (Table \ref{tbl:allMass}), consistent with previous inferred molecular gas kinetic temperatures in PNe \citep[e.g.,][]{Bachiller1997}. The PN NGC 6302 is a notable outlier, with a far higher peak $T_{mb}$ within its brightest $^{12}$CO(2--1) emitting region, leading to a relatively high inferred $T_k$ of 84 K. %\joel{check the preceding for accuracy...I'm unclear as to whether you used $^{12}$CO(2--1) or $^{13}$CO(2--1) to set $T_k$...and whether you just used the peak $T_{mb}$ to set $T_k$ for all of a PN's $N_{CO}$ estimates -- which is probably the easiest and most accurate thing to do -- or allowed $T_k$ to be determined pixel by pixel (the latter approach might introduce systematics)}

With these $^{12}$CO-based estimates of T$_{k}$ as constraints, we used 
RADEX to obtain the column densities of $^{12}$CO and $^{13}$CO ($N_{\rm CO}$) from the observed $^{12}$CO(2--1) and $^{13}$CO(2--1) line brightness temperatures. \revision{We assume an H$_2$ number density of $n_{\rm H_2} = 10^4$ cm$^{-3}$, which is near the lower bound of the likely range of molecular gas densities in PNe \citep[e.g.,][and references therein]{Schmidt2016,Kastner2024}; the results for column densities are relatively insensitive to the assumed value of $n_{\rm H_2}$.} The resulting cubes of column density per pixel were computed for both $^{12}$CO and $^{13}$CO for regions above the 1$\sigma$ background (root mean square) noise level; the mean $^{12}$CO and $^{13}$CO column densities derived from these RADEX calculations are listed in Table~\ref{tbl:allMass}. 
%The same is done with the $^{13}$CO data cube using the same T$_{k}$ derived from $^{12}$CO, assuming that optically thick $^{12}$CO provides a more accurate measurement of T$_{k}$, and H$_2$ density of 10$^4$ cm$^{-3}$. 
%\joel{RADEX gives optical depth (tau) estimates along with the predicted brightness T's, for a given set of inputs/assumptions...try some RADEX calculations by hand, for a few typical cases and maybe an extreme case or two}
A caveat in interpreting these results is that the $^{12}$CO column density calculations are less reliable for lines of sight for which $^{12}$CO(2--1) is optically thick. While $^{12}$CO(2--1) emission appears to be optically thin for many if not most PNe \citep[e.g.,][]{Bachiller1997,Kastner2024}, $^{12}$CO(2--1) will invariably have greater opacity than $^{13}$CO(2--1), and the former may become optically thick in PN regions with especially large H$_2$ densities. Indeed the ratios of integrated $^{12}$CO(2--1) to $^{13}$CO(2--1) fluxes for our sample PNe fall in the range $\sim2-5$ (Table~\ref{tbl:allObs}). These ratios are much smaller than the expected $^{12}$CO/$^{13}$CO abundance ratio (see below), and are similar to the integrated H$^{12}$CO$^+$(3--2)/H$^{13}$CO$^+$(3--2) and H$^{12}$CN(3--2)/H$^{13}$CN(3--2) flux ratios for the two PNe for which we have measured emission lines from both C isotopologues of these molecules (NGC 6302 and NGC 2440; Table~\ref{tbl:allObs}), indicating that $^{12}$CO(2--1) is optically thick across some of the emitting regions in these PNe. We further consider this possibility below (\S~\ref{sec:massDiscussion}).

\revision{To estimate total molecular (H$_2$) gas masses from the resulting CO column densities, we assume $^{12}$CO:H$_2$ and $^{13}$CO:H$_2$ abundance ratios of 10$^{-4}$ and $10^{-5}$, respectively; both are well within the range of CO isotopologue abundances commonly assumed for PNe \citep[e.g.,][and references therein]{Healy1990,Schmidt2022,Kastner2024}. Adopting these ratios is, of course, also equivalent to assuming a $^{12}$CO:$^{13}$CO abundance ratio of 10. In \S~\ref{sec:massDiscussion}, we consider the implications of, and potential deviations from, this assumed $^{12}$CO:$^{13}$CO abundance ratio. }

The resulting H$_2$ masses obtained from the $^{12}$CO and $^{13}$CO column density calculations are listed in Table \ref{tbl:allMass}. 
Given the preceding caveat concerning the possibility that $^{12}$CO(2--1) emission from some or all of our sample PNe may be optically thick, 
%in comparing with previous estimates of the molecular masses contained in the sample PNe (\S~\ref{sec:massComparison}), 
we regard the estimates obtained from $^{12}$CO(2--1) 
%and $^{13}$CO(2--1) data 
as providing a lower limit %and upper limits, respectively, 
on each nebula’s molecular mass. These molecular mass estimates are compared with values from the literature in Table \ref{tbl:allMass} (last column), where we have rescaled these values according to our adopted distances to the sample PNe. The agreement between our ALMA-CO-based molecular masses and previously published molecular mass estimates is generally good, within a factor $\sim$2 in all cases. The discrepancies with previous estimates are likely reflective of the fact that many were based on single-dish CO measurements (with the exceptions of NGC 6302 and Hb 5, which were obtained via ALMA $^{12}$CO(3--2) and IR H$_2$ observations, respectively); %for which beam dilution may be significant;
discrepancies would also arise from differences in assumed or estimated gas kinetic temperatures.

%As an example, we find H$_2$ masses of $4.7\pm1.7\times10^{-3}$ M$_\odot$ and $6.1\pm2.2\times10^{-2}$ M$_\odot$ for NGC 6302's central torus from the $^{12}$CO and $^{13}$CO column density calculations, respectively. The same procedure was used to determine the mass calculations for all survey PNe, see results in Table \ref{tbl:allMass}.

\begin{table}
\begin{center}
\caption{\sc Molecular Mass Estimates}
\vspace{.2in}
%\begin{adjustwidth}{-2.4cm}{}
%\begin{adjustbox}{width=1.15\textwidth}
\label{tbl:allMass}
\scriptsize
\begin{tabular}{lccccccc}
\toprule
PN & Name & T$_{k}$ & \multicolumn{2}{c}{Column Density$^a$ (cm$^{-2}$)} & \multicolumn{3}{c}{Molecular Mass$^a$ (M$_\odot$)}\\
\cmidrule(lr){4-5}
\cmidrule(lr){6-8}
& & (K) & $^{12}$CO & $^{13}$CO & $^{12}$CO & $^{13}$CO & Lit.\\
\midrule
\midrule
359.3-00.9 & PN Hb 5 & 29 & $(5.8\pm0.4)\times10^{15}$ &  & $(2.3\pm0.4)\times10^{-3}$ & & $5.0\times10^{-3}$$^{\ b}$\\
234.8+02.4 & NGC 2440 & 40 & $(4.8\pm0.1)\times10^{15}$ & $(5.2\pm0.1)\times10^{15}$ & $(1.1\pm0.1)\times10^{-2}$ & $(7.4\pm1.0)\times10^{-2}$ & $\geq4.0\times10^{-3}$$^{\ c,d,e}$\\
349.5+01.0 & NGC 6302 & 84 & $(4.7\pm0.1)\times10^{15}$ & $(7.4\pm0.1)\times10^{15}$ & $(4.7\pm1.7)\times10^{-3}$ & $(6.1\pm2.2)\times10^{-2}$ & $1.0\times10^{-1}$$^{\ f}$\\
008.0+03.9 & NGC 6445 & 18 & $(6.3\pm0.1)\times10^{15}$ & & $(3.6\pm0.3)\times10^{-2}$ & & $(2.1-3.5)\times10^{-2}$$^{\ g,e}$\\
277.1-03.8 & NGC 2899 & 29 & $(3.7\pm0.1)\times10^{15}$ & $(1.1\pm0.1)\times10^{15}$ & $(3.6\pm1.6)\times10^{-3}$ & $(8.0\pm3.6)\times10^{-3}$ & $3.9\times10^{-3}$$^{\ e}$\\
261.9+08.5 & NGC 2818 & 18 & $(1.8\pm0.5)\times10^{15}$ & $(1.4\pm0.1)\times10^{15}$ & $(1.5\pm0.4)\times10^{-2}$ & $(8.6\pm2.3)\times10^{-2}$ & $1.5\times10^{-2}$$^{\ e}$\\
\end{tabular}
%\end{adjustbox}
%\end{adjustwidth}
\end{center}
{\sc Notes:} Error propagation of mass comes from computing additional upper and lower limit values using distance and column density errors. (a) Molecular masses and column densities derived from ACA observations, assuming an H$_2$ density of $1\times10^4$ cm$^{-3}$. Estimates for NGC 2440 include the N and S molecular lobes observed in ACA observations. Column density values represent the average column density over the molecular torus. References for literature values: (b) \citealt{BS2005}, from mid-IR H$_2$ emission lines; (c) \citealt{Dayal1996}; (d) \citealt{Wang2008}; (e) \citealt{Huggins1996}; (f) \citealt{Santander-Garcia2017}; (g) \citealt{Huggins1989}.
\end{table}

\subsection{Discussion}\label{sec:massDiscussion}

In the optically thin limit, the total molecular (H$_2$) masses as estimated from $^{12}$CO and $^{13}$CO should be consistent with one another, if our assumed $^{12}$CO/$^{13}$CO number ratio of 10 is accurate. However, in Table \ref{tbl:allMass}, it is clear that the mass estimates obtained from the $^{13}$CO data are systematically larger than those estimated from the $^{12}$CO data. %\paula{Selective photodissociation of $^{12}$CO in nebular environments does not seem to be affecting $^{12}$CO/$^{13}$CO abundance ratios, as ratios of H$^{12}$CN/H$^{13}$CN and H$^{12}$CO+/H$^{13}$CO+ (e.g. NGC 6302 and NGC 2440) exhibit similarly low ($<10$) values.} \joel{I mention this above, though w/ slightly different emphasis; we should ask Thierry about how he expects selective photodissociation to affect things, but maybe hold off mentioning it for the time being, as I noted in my email} 
This systematic discrepancy is hence evidence either of large $^{12}$CO opacities, $^{12}$CO/$^{13}$CO abundance ratios $<10$, or both. Selective photodissociation of $^{13}$CO may also affect the modeled $^{12}$CO/$^{13}$CO column density (hence abundance) ratios, although this should raise (as opposed to lower) the derived ratios. 

With regard to the possibility that $^{12}$CO is optically thick, we note that the mean $^{12}$CO and $^{13}$CO column densities obtained from the RADEX calculations are very similar for the PNe observed in both isotopologues; whereas, barring exceptional circumstances, one expects $^{12}$CO/$^{13}$CO $> 1$ and, hence, the derived $^{12}$CO column densities should be larger than those of $^{13}$CO, for pure optically thin emission. Indeed, for the mean column densities in Table~\ref{tbl:allMass}, the RADEX calculations indicate the $^{12}$CO(2--1) optical depths typically lie in the range $\tau \sim 0.2$ to $\sim$1.3 for the sample PNe, given the adopted values of $T_k$ and an assumed H$_2$ number density of $10^4$ cm$^{-3}$. Thus, it is likely that $^{12}$CO(2--1) is at least marginally optically thick, with values $\tau > 1.0$ likely along specific lines of sight --- a prime example being the region of NGC 6302 marked `a' in Fig.~\ref{fig:N6302Mont3}. 

%\joel{here's where I tried to modify our conclusions about the origin/nature of the low $^{12}$CO/$^{13}$CO ratios}

However, the results for mean $^{12}$CO and $^{13}$CO column densities are indicative of low intrinisic $^{12}$C/$^{13}$C ratios in the PN molecular gas. \revision{This inference is supported by the low integrated H$^{12}$CN/H$^{13}$CN and H$^{12}$CO$^+$/H$^{13}$CO$^+$ flux ratios we have measured for NGC 6302 and NGC 2440 ($\sim$3--14; see Table~\ref{tbl:allObs}), as well as the results of previous single-dish studies of C-bearing isotopologues in molecule-rich PNe, including NGC 2440 \citep[e.g.,][and references therein]{Ziurys2020}}. The likelihood of low $^{12}$C/$^{13}$C ratios across our sample has interesting implications for the potential common origin and evolution of these PNe; this C isotopic ratio is a potentially sensitive indicator of hot-bottom burning in the more massive progenitors of AGB stars \citep[e.g.,][]{Karakas2016} or binary interactions during the primary star's post-main sequence evolution \citep[e.g.,][and references therein]{Khouri2022}. According to the models presented in \citet{Karakas2016}, $^{12}$C/$^{13}$C is predicted to be $\lesssim$10 for initial masses in the range $\sim$4--8 $M_\odot$, with a precipitous drop in $^{12}$C/$^{13}$C from $>$40 to $<$10 as progenitor mass exceeds 4 $M_\odot$ (or even slightly higher, for super-solar-metallicity progenitors). \revision{For reference, the $^{12}$C/$^{13}$C isotopic ratio in the local ISM typically ranges from about 40 to 80, while regions of increased stellar density near the Galactic Center exhibit $^{12}$C/$^{13}$C ratios in the range $\sim15-30$, with the $^{13}$C enhancement in these regions attributed to the accumulated effects of CN processing in relative massive stars \citep[and references therein]{Wilson1994,Halfen2017,HB2019}. }

However, anomalously low $^{12}$CO/$^{13}$CO ratios are also observed in some bipolar pre-PNe that are thought to be descended from relatively low-mass progenitor stars, with the low $^{12}$CO/$^{13}$CO ratios in such cases indicative of truncation of AGB evolution via onset of the common envelope stage of an interacting binary system \citep[e.g.,][and references therein]{Khouri2022}. \revision{Indeed, while the estimated progenitor masses of sample PNe Hb 5 and NGC 6302 exceed $\sim$4 $M_\odot$ \citep[e.g.,][and references therein]{Steffen2013,Kastner2022}, the progenitor masses of NGC 2440, NGC 2818, and NGC 2899 have been estimated as $\lesssim$3 $M_\odot$ \secrevise{\citep{Kaler1989,Miller2019,Fragkou2025}.}}
% Selective photodissociation of 12C does not seem to be affecting 12CO/13CO, as ratios of HCN/H13CN and HCO+/H13CO+ exhibit similarly low values.
%The discrepancies are likely due to an assumed $^{12}$CO/$^{13}$CO abundance ratio that is inconsistent with real values that can be attributed to our sample's progenitors. Another possibility is that $^{12}$CO emission is much more optically thick than our initial assumption, thus decreasing the observed $^{12}$CO/$^{13}$CO abundance ratio due to $^{12}$CO self-shielding. A clear example of the latter is NGC 6302 (see Fig.~\ref{fig:N6302Mont3}), where the molecular torus exhibits a feature of low emission (denoted as feature `a') only in $^{12}$CO emission, revealing that feature `a' is fairly optically thick and that in this region of the molecular torus $^{12}$CO exhibits self-shielding. These effects are not so obvious in the other target PNe, however we have plans to make use of molecular line ratio diagnostic diagrams to disentangle the influence of optical depth versus intrinsic $^{12}$CO/$^{13}$CO abundances in a subsequent paper.
%Measurements of $^{12}$CO/$^{13}$CO integrated flux are used to constrain the progenitor mass of NGC 2440, NGC 6302, NGC 2899, and NGC 2818, assuming the integrated flux ratio is consistent with measured relative isotropic abundances, with models of AGB stars ranging between $1-8$ M$_\odot$ and Z$=$0.007$-$0.03 metallicity, that include hot-bottom burning (HBB) effects \citep{Karakas2016}. 
Hence --- barring the possibility of large $^{13}$CO(2--1) as well as $^{12}$CO(2--1) optical depths --- both the ALMA data (i.e., measurements of integrated $^{12}$CO(2--1)/$^{13}$CO(2--1) line intensity ratios in the range $\sim$2--5) and RADEX modeling ($^{12}$CO/$^{13}$CO column density ratios near unity) suggest that the PNe in our sample are either descended from relatively massive AGB progenitors (initial masses $\sim$4--8 $M_\odot$) or underwent AGB-truncating binary interactions, perhaps via common envelope evolution. 
%We conclude that the class of high-excitation bipolar PNe from which our sample is drawn represents the near-final evolutionary stages of stars at the upper end of the range of initial masses that can avoid exploding as Type II supernovae, via copious, asymmetric mass loss on and beyond their AGB evolutionary stages. 

%\paula{this was javier's comment: Note that for O-rich pPNe 12CO/13CO ratios of about 10 are also typical. I do not thinck that this ratio necessarily implies that our targets are massive (of masses larger than 4 Msun). In addition, HBB also predicts the destruction of 18O and we have detected C18O in some sources. There are even cases of PNe where 12C/13C as low a a factor 3 have been estimated. I think this result must be revised.}
%More specifically, their low $^{12}$C/$^{13}$C abundances, ranging between $2-6$, suggest HBB has occurred in their progenitor stars which signifies that they have descended from AGB stars with at least 3 M$_\odot$, but no greater than 8 M$_\odot$ \citep{Karakas2016}.

\section{Conclusions}\label{sec:conclusion}

We have presented results from extensive mm-wave interferometric molecular line mapping observations of a sample of a half-dozen molecule-rich, high-excitation, bipolar PNe. The data were obtained with the ALMA ACA and 12-meter arrays operating at 1.3 mm (Band 6). The sample we observed with ALMA comprises the PNe Hubble 5, NGC 2440, NGC 6302, NGC 6445, NGC 2818, and NGC 2899. Our primary objective is to gain a deeper understanding of the progenitor systems, shaping histories, and irradiation geometries of these nebulae. 
In this paper, we have presented the survey data and initial survey results, focusing on the molecular species and transitions detected and the molecular gas kinematics and masses of the sample PNe. 

High-quality, subarcsecond-resolution maps of $^{12}$CO(2--1) emission were obtained with the 12-m array for all six of the sample PNe, conclusively demonstrating that the bulk of their molecular gas masses are confined to the central, dusty, pinched waists of these nebulae. We interpret the resulting position-velocity $^{12}$CO(2--1) data cubes as revealing, in each case, the presence of an expanding molecular torus that generally traces the bipolar PN's equatorial plane. These equatorial torus structures range from relatively coherent, for Hb 5, NGC 2440, NGC 6302, and NGC 6445, to highly fragmented, for NGC 2899 and NGC 2818. This apparent evolution of molecular torus structure is consistent with the dynamical age sequence represented by the sample PNe; we infer torus dynamical ages ranging from $\sim$500 yr for Hb 5, to a few thousand yr for NGC 2440, NGC 6302, and NGC 6445, to $\sim$10$^4$ yr for NGC 2899 and NGC 2818, respectively (Table~\ref{tbl:allAge}). 

Our results hence demonstrate how, as a molecule-rich bipolar PN evolves, its molecular torus expands and fragments. In the youngest such nebulae, e.g., Hb 5 \citep[and NGC 7027;][]{Bublitz2023}, the molecular torus forms a nearly continuous, compact structure surrounding the nebular core that lies more or less perpendicular to the optical lobes. Evidently, however, the torus begins to fragment, on a timescale of a few thousand years. This fragmentation process is reflected in the rings of detached, knot-like structures observed in the cores of NGC 2440 and NGC 6445, and in the distortions and azimuthal asymmetries of the NGC 6302 molecular torus. Eventually, the torus becomes highly disrupted, with the molecular gas fragmenting into isolated structures or filaments, as observed in NGC 2899 and NGC 2818.
%reflecting the expansion of the torus.
%, which may have once been connected to the main molecular tori.
%As observed in NGC 6302, as a PN stellar system evolves, a multi-ringed structure of expanding molecular gas can be produced. A similar, albeit more fragmented, structure is observed in NGC 2818. Additionally, the mm-wave continuum emission becomes weaker with age until it is virtually nonexistent, as observed in NGC 2818. 
%Our results hence demonstrate that as bipolar PNe evolve, the molecular torus expands and fragments, causing the optical thickness of the molecular gas to decrease.  
%This is reflected in the detached knot-like structures observed in the molecular tori of NGC 2440 and NGC 6445, which may have once been connected to the main molecular tori. 

The ALMA observations of the six sample PNe also yielded detections and maps of CO isotopologues and the trace molecules HCN, HNC, HCO$^+$, CN, CS, and SO, as well as 1.3 mm continuum emission, in some or all of these nebulae (see Table \ref{tbl:allObs}). The HCN(3--2), CN(2--1), and HCO$^+$(3--2) transitions are detected in all sample PNe observed (the only exception being NGC 6445, for which no HCN or HCO$^+$ data were obtained), while HNC(3--2) is detected in all PNe except Hb 5. Emission from CS(5--4) is detected in NGC 6302, NGC 6445, and NGC 2818. Lines of SO covered in our spectrometer setups are detected only in \revision{NGC 2440 and} NGC 6302. Emission from CO$^+$(2--1) was not detected from any of the four nebulae for which the ALMA data included coverage of this line, leaving NGC 7027 as the only PN detected to date in CO$^+$ \citep[][and references therein]{Bublitz2023}. Bright, compact 1.3 mm continuum emission is detected in Hb 5, NGC 2440, and NGC 6302, while weak continuum emission is detected in NGC 6445.
In a subsequent paper, we investigate the molecular chemistries of the PNe as revealed by molecular line ratio images and line ratio diagnostic diagrams, focusing on the potential effects of UV and X-ray irradiation of PN molecular gas by the hot, luminous central stars powering these nebulae and the impacts of fast stellar winds.

%To estimate the molecular masses, we used observations of $^{12}$CO and $^{13}$CO, which serve as proxies for the molecular content. 
We analyzed the $^{12}$CO(2--1) and $^{13}$CO(2--1) emission-line data cubes to obtain estimates for $^{12}$CO and $^{13}$CO column densities and, hence, the total masses of molecular gas in these nebulae, under an assumed $^{12}$CO:H$_2$ \secrevise{ratio} of $10^{-4}$ and an assumed $^{13}$CO:H$_2$ \secrevise{ratio} of $10^{-5}$. The resulting molecular mass estimates for our target PNe lie in the range $\sim 0.002-0.09$ M$_\odot$ (\S \ref{sec:massCalc}). Interestingly, there is no clear correlation of the estimated PN molecular masses with the inferred torus dynamical ages. This lack of dependence of PN molecular mass on age suggests that the molecular mass of a PN is determined by its progenitor system properties, rather than by its UV irradiation (hence molecular dissociation) history. The low measured $^{12}$CO/$^{13}$CO integrated flux ratios and resulting low inferred $^{12}$CO/$^{13}$CO column density ratios, in addition to suggesting large $^{12}$CO opacities, likely reflect intrinsically low $^{12}$C/$^{13}$C isotopic ratios. Such low C isotopic ratios, in turn, would suggest that the progenitor masses of our sample PNe generally lie in the range $\sim$4--8 $M_\odot$ and/or that the central stars of these PNe reside in interacting binary systems (\S~\ref{sec:massDiscussion}). 
%\joel{that last sentence tweaked to address Javier's comment}

%\citep{Karakas2016}.
%\joel{need to mention/briefly discuss whether/how low $^{12}$CO/$^{13}$CO supports the case for high-mass progenitors}

%\joel{defer the next (commented) text to Paper II, I think...replace with a summary of the new molecules/isotopologues detected in each PN?}
%By analyzing a series of line ratio images, we explore the molecular abundances and optical depth properties of the sample, which provide crucial insights into the physical conditions and chemical compositions of the surrounding molecular gas. Among the nebulae in our sample, NGC 6302, NGC 2440, and NGC 2818 stand out as having S-rich environments, with NGC 6302 also exhibiting an O-rich environment. This dual enrichment in O and S may provide important clues about the stellar processes at play during the nebula's evolution.

This ALMA molecular line mapping study of a sample of high-excitation, molecule-rich, bipolar PNe  
%--- the descendants of progenitor AGB stars of initial mass $\geq4$ $M_\odot$ ---
hence well illustrates how the presence and evolution of molecular gas in a PN is inextricably linked to its progenitor star system as well as the resultant nebula's formation and structural evolution.
%Our target PNe descend from progenitors of higher mass than the ``classic'' Ring-like PNe \citep[e.g.,][]{Kastner2025,Kastner2024}. 
%with molecular torus ages varying from $1000-19000$ years 
Our analysis bolsters the assertion \citep[][]{Huggins2007} that ejection of a dense equatorial torus precedes the onset of the collimated polar outflows (jets) that, along with large molecular gas masses, represent the defining characteristics of the class of extreme bi-lobed PNe.
The sequential nature of the formation of torus and lobes (\S~\ref{sec:AgeDiscussion}), along with the multi-polar, episodic polar ejections observed in optical emission-line imaging, are indicative of mass loss processes far more complex than can be explained via single-star AGB and post-AGB models \citep[see, e.g., ][and references therein]{Hofner2018}, adding to the evidence for the presence of interacting binary companions at the cores of the sample PNe. 
%single final mass ejection in the progenitors' post-AGB phases. 
%Further proof that validates this statement, are 

Our results further demonstrate that the high CSPN $T_{\mathrm{eff}}$ and $L_{\mathrm{bol}}$ of molecule-rich bipolar PNe are maintained over $\sim$10--20 kyr timescales, post-AGB; such timescales far exceed expected post-AGB core cooling times \citep[e.g.,][]{Miller2016}. Perhaps mass accretion at the central stars and/or companion stars in these systems is both sustaining their intense UV fields and driving polar flows well into post-AGB evolution. Such a long-timescale yet episodic accretion/outflow process, along with the delay we and others have deduced between the formation of the molecular tori and optical lobes of extreme bi-lobed PNe, appears consistent with a model of bipolar PN formation via Intermediate-Luminosity Optical Transient (ILOT) events \citep{Soker2012}. This model offers an explanation for the formation of bi-lobed PNe that display evidence for multiple, paired mass ejections. Such a model, which has been considered to explain the formation and structural evolution of NGC 6302 \citep[][] {Soker2012,Kastner2022}, is worth exploring in more detail for the other PNe in our ALMA survey sample.

%\joel{just noticed that the first reference in the list is truncated...might have to fix it by hand}

%\joel{briefly summarize results, i.e. range of masses across the sample...is there any clear trend of decreasing molecular mass w/ age? doesn't look like it, to me...that is definitely worth pointing out, as it implies that a PN's molecular mass is determined by its progenitor (AGB plus companion) system properties and not its UV irradiation (i.e. dissociation) history...perhaps this conflicts w/ the statement below?}

%In addition, we investigated the potential evolutionary sequence represented by our sample of bipolar PNe by imaging the molecular tori of the sample objects and ascertaining their molecular gas kinematics. 
\section*{Acknowledgments}
The authors thank Letizia Stanghellini and the anonymous referee for their many valuable comments, which greatly improved this paper. Major support for the work of PMB and JHK on mm-wave interferometric mapping of molecular line emission from planetary nebulae has been provided by NSF grant AST-2206033 to RIT. JA and MSG acknowledge the financial support of I+D+i projects PID2019-105203GB-C21 and PID2023-146056NB-C21, funded by the Spanish MCIN/AEI/10.13039/501100011033 and EU/ERDF. This paper makes use of ALMA data from programs 2021.1.00456.S, 2021.2.00004.S, and 2022.1.00401.S (PI: J. Kastner). ALMA is a partnership of ESO (representing its member states), NSF (USA) and NINS (Japan), together with NRC (Canada), NSTC and ASIAA (Taiwan), and KASI (Republic of Korea), in cooperation with the Republic of Chile. The Joint ALMA Observatory is operated by ESO, AUI/NRAO and NAOJ. The National Radio Astronomy Observatory is a facility of the National Science Foundation operated under cooperative agreement by Associated Universities, Inc.

\bibliography{references}

\appendix 
\section{ALMA Band 6 Observations: Molecules Detected}\label{sec:app1}

%\joel{add brief text like:}

Tables~\ref{tbl:detections}--\ref{tbl:N2818detections} present compilations of results for molecular lines detected in each PN.  %\joel{move Hb 5 line profile fig here too, and make sure tables and figures appear in order of molecular torus dynamical age}

\begin{table}[h]
\caption{\sc Detected Molecular Transitions}
%\vspace{.2in}
\label{tbl:detections}
\begin{center}
\scriptsize
\begin{tabular}{lllrccccccc}
\toprule
PN G & Name & ALMA\ & Molecule & QN\ & $\nu$ & $\log{A_{ul}}$ & $E_u/k$  & $I$ ($\sigma_I^a$) & $V_{LSR}$ & $\Delta V_{1/2}$ \\
& & Config. & & & (GHz) & (s$^{-1}$) & (K) & (Jy km s$^{-1}$) & (km s$^{-1}$) & (km s$^{-1}$)\\
\midrule
\midrule
359.3-00.9 & PN Hb 5 & ACA & $^{12}$CO & $J=2-1$  & 230.538 & $-6.1605$ & 16.5961 & 360.709 (1.053)$^c$ & $-10.777$ & $53.022$\\
& & & \tbf{CN} & $\bm{N=2-1^b}$ & \tbf{226.697} & $\bm{-4.2783}$ & \tbf{16.3089} & \tbf{118.239 (1.044)} & $\bm{-5.028}$ & $\bm{49.081}$\\
& & &    & $\bm{N=2-1^b}$ & \tbf{226.874} & $\bm{-3.9419}$ & \tbf{16.3349} & \tbf{279.184 (0.931)} & $\bm{-2.136}$ & $\bm{44.237}$\\
& & & HCN & $J=3-2$ & 265.886 & $-3.0780$ & 25.5209 & 101.701 (0.302) & $-9.529$ & $46.657$\\
& & & HCO$^+$ & $J=3-2$ & 267.557 & $-2.8376$ & 25.6817 & 140.549 (0.696) & $-9.583$ & $49.596$ \\
& & C43-1 & $^{12}$CO & $J=2-1$  & 230.538 & $-6.1605$ & 16.5961 & 517.688 (1.053) & $-22.211$ & $55.562$\\
& & & \tbf{CN} & $\bm{N=2-1^b}$ & \tbf{226.697} & $\bm{-4.2783}$ & \tbf{16.3089} & \tbf{90.101(1.044)} & $\bm{-0.922}$ & $\bm{45.918}$\\
& & &    & $\bm{N=2-1^b}$ & \tbf{226.874} & $\bm{-3.9419}$ & \tbf{16.3349} & \tbf{222.419(0.931)} & $\bm{-3.702}$ & $\bm{46.210}$ \\
& & & \tbf{H$\bm{^{13}}$CO$^{\bm{+}}$} & $\bm{N=3-2}$ & \tbf{260.255} & $\bm{-2.8737}$ & \tbf{24.9807} & \tbf{4.529 (0.050)} & $\bm{-0.256}$ & $\bm{15.124}$\\
\end{tabular}
\end{center}
\vspace{.1in}
{\sc Notes:} 
New molecule detections are denoted by bold font. a) Integrated line flux within $8''$ x $8''$ synthetic beam centered on (17:44:43.970, $-$29:58:39.709) position. b) Hyperfine complex; representative transition frequency, $A_{ul}$, and upper level energy listed retrieved from CDMS. c) The ACA observation of $^{12}$CO reflects flux loss due to continuum subtraction errors that occurred during the pipeline cleaning process. 
\end{table}

%\begin{comment}

\begin{table}
\begin{center}
\caption{\sc NGC 2440: Detected Molecular Transitions}
\vspace{.2in}
\label{tbl:N2440detections}
\footnotesize
\begin{tabular}{lrcccccccc}
\toprule
ALMA\ & Molecule & QN\ & $\nu$ & $\log{A_{ul}}$ & $E_u/k$  & $I^a$ ($\sigma_I$) & $V_{LSR}$ & $\Delta V_{1/2}$ \\
Config. & & & (GHz) & (s$^{-1}$) & (K) & (Jy km s$^{-1}$) & (km s$^{-1}$) & (km s$^{-1}$)\\
\midrule
\midrule
ACA & $^{12}$CO & $J=2-1$  & 230.538 & $-6.1605$ & 16.5961 & 93.656 (0.872) & $40.235$ & $28.485$ \\
& $^{13}$CO & $N=2-1$  & 220.398 & $-6.2164$ & 15.8662 & 44.422 (0.391) & $38.138$ & $30.225$ \\
&  \tbf{CN} & $\bm{N=2-1^b}$ & \tbf{226.697} & $\bm{-4.2783}$ & \tbf{16.3089} & \tbf{22.484 (1.150)} & $\bm{55.195}$ & $\bm{12.272}$\\
&    & $\bm{N=2-1^b}$ & \tbf{226.874} & $\bm{-3.9419}$ & \tbf{16.3349} & \tbf{57.031 (0.868)} & $\bm{42.410}$ & $\bm{29.040}$ \\
& HCN & $J=3-2$ & 265.886 & $-3.0780$ & 25.5209 & 22.935 (0.629) & $44.828$ & $23.264$\\
& H$^{13}$CN & $J=3-2$  & 259.011 & $-3.1121$ & 24.8616 & 7.993 (0.653) & $41.750$ & $19.200$ \\
& \tbf{HNC} & $\bm{J=3-2}$ & \tbf{271.981} & $\bm{-3.0298}$ & \tbf{26.1069} & \tbf{8.056 (0.606)} & $\bm{41.202}$ & $\bm{18.842}$ \\
& HCO$^+$ & $J=3-2$ & 267.557 & $-2.8376$ & 25.6817 & 10.357 (0.887) & $42.634$ & $20.793$\\
C43-3 & $^{13}$CO & $N=2-1$  & 220.398 & $-6.2164$ & 15.8662 & 9.501 (0.206) & $41.881$ & $40.851$\\
& HCN & $J=3-2$ & 265.886 & $-3.0780$ & 25.5209 & 25.143 (0.290) & $43.556$ & $38.402$\\
& H$^{13}$CN & $N=3-2$  & 259.011 & $-3.1121$ & 24.8616 & 9.736 (0.294) & $42.604$ & $20.571$\\
& \tbf{HNC} & $\bm{J=3-2}$ & \tbf{271.981} & $\bm{-3.0298}$ & \tbf{26.1069}  & \tbf{0.042 (0.00788)} & $\bm{38.918}$ & $\bm{24.220}$\\
& HCO$^+$ & $J=3-2$ & 267.557 & $-2.8376$ & 25.6817 & 9.247 (0.409) & $42.690$ & $28.520$\\
& \tbf{H$\bm{^{13}}$CO$^{\bm{+}}$} & \tbf{\tit{N}}$\bm{=3-2}$ & \tbf{260.255} & $\bm{-2.8737}$ & \tbf{24.9807} & \tbf{2.298 (0.196)} & $\bm{39.125}$ & $\bm{16.242}$\\
& \tbf{SO} & $\bm{N_J=6_5-5_4}$$\bm{^b}$ & \tbf{219.949} & $\bm{-3.8744}$ & \tbf{16.9790} & \tbf{1.775 (0.163)} & $\bm{52.990}$ & $\bm{10.650}$\\
C43-4 & $^{12}$CO & $J=2-1$  & 230.538 & $-6.1605$ & 16.5961 & 104.282 (0.525) & $42.059$ & $40.957$ \\
& \tbf{CN} & $\bm{N=2-1^b}$ & \tbf{226.697} & $\bm{-4.2783}$ & \tbf{16.3089} & \tbf{20.843 (0.483)} & $\bm{38.790}$ & $\bm{37.137}$\\
&    & $\bm{N=2-1^b}$ & \tbf{226.874} & $\bm{-3.9419}$ & \tbf{16.3349} & \tbf{50.520 (0.526)} & $\bm{41.516}$ & $\bm{40.974}$\\

\end{tabular}
\end{center}

{\sc Notes:} 
New molecule detections are denoted by bold font. a) Integrated line flux within a $9''$ x $14''$
synthetic beam centered on (7:39:41.414, $-$18:05:24.739) position for the central torus and a $5''$ x $5''$ synthetic beam centered on the NE and SW lobes (ACA data). b) Hyperfine complex; representative transition frequency, $A_{ul}$, and upper level energy listed retrieved from CDMS.
\end{table}

\begin{table}
\begin{center}
\caption{\sc NGC 6302: Detected Molecular Transitions}
\vspace{.2in}
\label{tbl:N6302detections}
\footnotesize
\begin{tabular}{lrccccccc}
\toprule
ALMA\ & Molecule & QN\ & $\nu$ & $\log{A_{ul}}$ & $E_u/k$  & $I^a$ ($\sigma_I$) & $V_{LSR}$ & $\Delta V_{1/2}$\\
Config. & & & (GHz) & (s$^{-1}$) & (K) & (Jy km s$^{-1}$) & (km s$^{-1}$) & (km s$^{-1}$)\\
\midrule
\midrule
ACA & $^{12}$CO & $J=2-1$  & 230.538 & $-6.1605$ & 16.5961 & 802.263 (0.805) & $-40.622$ & $27.302$\\
& \tbf{CS} & $\bm{N=5-4}$  & \tbf{244.936} & $\bm{-3.5257}$ & \tbf{35.2660} & \tbf{14.338 (0.866)} & $\bm{-38.120}$ & $\bm{2.988}$ \\
& \tbf{C}$\bm{^{34}}$\tbf{S} & $\bm{N=5-4}$  & \tbf{241.016} & $\bm{-3.5467}$ & \tbf{27.7763} & \tbf{3.054 (0.660)} & $\bm{-39.076}$ & $\bm{3.644}$ \\
& CN & $N=2-1^b$ & 226.697 & $-4.2783$ & 16.3089 & 36.676 (0.618) & $-28.600$ & $34.2265$\\
&    & $N=2-1^b$ & 226.874 & $-3.9419$ & 16.3349 & 79.996 (0.738) & $-38.211$ & $24.217$\\
& \tbf{H}$\bm{^{13}}$\tbf{CN} & $\bm{J=3-2}$ & \tbf{259.011} & $\bm{-3.1121}$ & \tbf{24.8616} & \tbf{17.766 (0.413)} & $\bm{-30.4065}$ & $\bm{10.9515}$ \\
& \tbf{HNC} & $\bm{J=3-2}$ & \tbf{271.981} & $\bm{-3.0298}$ & \tbf{26.1069} & \tbf{23.545 (0.696)} & $\bm{-30.967}$ & $\bm{13.455}$\\
& \tbf{H}$\bm{^{13}}$\tbf{CO}$\bm{^+}$ & $\bm{N=3-2}$ & \tbf{260.255} & $\bm{-2.8737}$ & \tbf{24.9807} & \tbf{9.844 (0.434)} & $\bm{-38.586}$ & $\bm{3.1645}$ \\
C43-1 & $^{13}$CO & $N=2-1$  & 220.398 & $-6.2164$ & 15.8662 & 286.858 (0.363) & $-40.898$ & $30.886$\\
& \tbf{C}$\bm{^{18}}$\tbf{O} & $\bm{J=2-1}$ & \tbf{219.560} & $\bm{-6.2210}$ & \tbf{15.8058} & \tbf{5.813 (0.104)} & $\bm{-32.304}$ & $\bm{13.002}$ \\
& SO & $N_J=6_5-5_4$$^b$ & 219.949 & $-3.8744$ & 16.9790 & 6.343 (0.0955) & $-30.668$ & $11.314$\\
& \tbf{H30}$\bm{\alpha}$ & \tbf{...} & \tbf{231.901} & \tbf{...} & \tbf{...} & \tbf{25.827 (0.235)} & \tbf{...} & \tbf{...} \\
C43-3 & HCN & $J=3-2$  & 265.886 & $-3.0780$ & 25.5209 & 97.156 (0.477) & $-34.122$ & $18.995$ \\
& \tbf{H}$\bm{^{13}}$\tbf{CN} & $\bm{J=3-2}$ & \tbf{259.011} & $\bm{-3.1121}$ & \tbf{24.8616} & \tbf{23.432 (0.547)} & $\bm{-29.976}$ & $\bm{11.517}$ \\
& \tbf{HNC} & $\bm{J=3-2}$ & \tbf{271.981} & $\bm{-3.0298}$ & \tbf{26.1069} & \tbf{19.634 (0.500)} & $\bm{-30.262}$ & $\bm{22.069}$\\
& HCO$^+$ & $J=3-2$ & 267.557 & $-2.8376$ & 25.6817 & 140.159 (0.963) & $-31.180$ & $24.349$\\
& \tbf{H}$\bm{^{13}}$\tbf{CO}$\bm{^+}$ & $\bm{N=3-2}$ & \tbf{260.255} & $\bm{-2.8737}$ & \tbf{24.9807} & \tbf{11.111 (0.661)} & $\bm{-37.251}$ & $\bm{4.852}$\\
& SO & $N_J=5_6-4_5$$^b$ & 251.857 & $-3.6737$ & 26.8110 & 0.377 (0.268) & $-37.160$ & $2.906$ \\
&    & $N_J=6_5-5_4$$^b$ & 219.949 & $-3.8744$ & 16.9790 & 4.556 (0.302) & $-38.130$ & $11.002$ \\
C43-4 & $^{12}$CO & $J=2-1$  & 230.538 & $-6.1605$ & 16.5961 & 818.523 (0.442) & $-38.885$ & $30.799$ \\
& $^{13}$CO & $N=2-1$ & 220.398 & $-6.2164$ & 15.8662 & 262.820 (0.412) & $-39.927$ & $27.898$ \\
& \tbf{C}$\bm{^{18}}$\tbf{O} & $\bm{J=2-1}$ & \tbf{219.560} & $\bm{-6.2210}$ & \tbf{15.8058} & \tbf{2.545 (0.185)} & $\bm{-32.326}$ & $\bm{13.002}$ \\
& CN & $N=2-1^b$ & 226.697 & $-4.2783$ & 16.3089 & 15.674 (0.413) & $-32.632$ & $20.991$ \\
&    & $N=2-1^b$ & 226.874 & $-3.9419$ & 16.3349 & 59.984 (0.399) & $-38.372$ & $29.064$ \\
\end{tabular}
\end{center}

{\sc Notes:} 
New molecule detections are denoted by bold font. a) Integrated line flux within $8''$ x $12''$ synthetic beam centered on (17:10:21.263, $-$37:02:43.793) position. b) Hyperfine complex; representative transition frequency, $A_{ul}$, and upper level energy listed retrieved from CDMS.
\end{table}

\begin{table}
\begin{center}
\caption{\sc NGC 6445: Detected Molecular Transitions}
\vspace{.2in}
\label{tbl:N6445detections}
\begin{adjustwidth}{-1.cm}{}    
\footnotesize
\begin{tabular}{lrccccccc}
\toprule
ALMA\ & Molecule & QN\ & $\nu$ & $\log{A_{ul}}$ & $E_u/k$  & $I^a$ ($\sigma_I$) & $V_{LSR}$ & $\Delta V_{1/2}$\\
Config. & & & (GHz) & (s$^{-1}$) & (K) & (Jy km s$^{-1}$) & (km s$^{-1}$) & (km s$^{-1}$)\\
\midrule
\midrule
ACA & $^{12}$CO & $J=2-1$ & 230.538 & $-6.1605$ & 16.5961 & 830.698 (1.053) & $18.065$ & $38.731$\\
& \tbf{CN} & $\bm{N=2-1^b}$ & \tbf{226.697} & $\bm{-4.2783}$ & \tbf{16.3089} & \tbf{103.930 (1.044)} & $\bm{18.217}$ & $\bm{39.393}$\\
&          & $\bm{N=2-1^b}$ & \tbf{226.874} & $\bm{-3.9419}$ & \tbf{16.3349} & \bf{277.213 (0.931)} & $\bm{20.578}$ & $\bm{37.456}$\\
& HNC & $J=3-2$ & 271.981 & $-3.0298$ & 26.1069 & 13.285 (0.819) & 23.347 & 18.567 \\
C43-1 & $^{12}$CO & $J=2-1$  & 230.538 & $-6.1605$ & 16.5961 & 978.148 (1.108) & $21.750$ & $44.260$\\
& CS  & $N=5-4$  & 244.936 & $-3.5257$ & 35.2660 & 5.953 (0.487) & $14.968$ & $29.286$\\
& \tbf{CN} & $\bm{N=2-1^b}$ & \tbf{226.697} & $\bm{-4.2783}$ & \tbf{16.3089} & \tbf{92.246 (1.047)} & $\bm{12.744}$ & $\bm{45.208}$\\
&          & $\bm{N=2-1^b}$ & \tbf{226.874} & $\bm{-3.9419}$ & \tbf{16.3349} & \tbf{222.366 (2.187)} & $\bm{32.694}$ & $\bm{46.824}$\\
C43-4 & $^{12}$CO & $J=2-1$  & 230.538 & $-6.1605$ & 16.5961 & 396.250 (3.156) & $16.048$ & $39.680$\\
& \tbf{CN} & $\bm{N=2-1^b}$ & \tbf{226.697} & $\bm{-4.2783}$ & \tbf{16.3089} & \tbf{43.218 (2.916)} & $\bm{21.677}$ & $\bm{28.735}$\\
&          & $\bm{N=2-1^b}$ & \tbf{226.874} & $\bm{-3.9419}$ & \tbf{16.3349} & \tbf{145.479 (2.978)} & $\bm{22.716}$ & $\bm{34.546}$\\
\end{tabular}
\end{adjustwidth}
\end{center}

{\sc Notes:} 
New molecule detections are denoted by bold font. a) Integrated line flux within $21''$ x $25''$ synthetic beam centered on (17:46:16.993, $-$19:59:41.126) position. b) Hyperfine complex; representative transition frequency, $A_{ul}$, and upper level energy listed retrieved from CDMS.
\end{table}

\begin{table}
\begin{center}
\caption{\sc NGC 2899: Detected Molecular Transitions}
\vspace{.2in}
\label{tbl:N2899detections}
\begin{adjustwidth}{-1.cm}{}    
\footnotesize
\begin{tabular}{lrccccccc}
\toprule
ALMA\ & Molecule & QN\ & $\nu$ & $\log{A_{ul}}$ & $E_u/k$  & $I^a$ ($\sigma_I$) & $V_{LSR}$ & $\Delta V_{1/2}$\\
Config. & & & (GHz) & (s$^{-1}$) & (K) & (Jy km s$^{-1}$) & (km s$^{-1}$) & (km s$^{-1}$)\\
\midrule
\midrule
ACA & $^{12}$CO & $J=2-1$  & 230.538 & $-6.1605$ & 16.5961 & 84.664 (0.322) & $-15.294$ & $14.288$\\
& $\bm{^{13}}$\tbf{CO} & $\bm{N=2-1}$  & \tbf{220.398} & $\bm{-6.2164}$ & \tbf{15.8662} & \tbf{23.410 (0.187)} & $\bm{-13.888}$ & $\bm{12.953}$\\
& \tbf{CN} & $\bm{N=2-1^b}$ & \tbf{226.697} & $\bm{-4.2783}$ & \tbf{16.3089} & \tbf{8.063 (0.409)} & $\bm{-4.138}$ & $\bm{5.813}$\\
&    & $\bm{N=2-1^b}$ & \tbf{226.874} & $\bm{-3.9419}$ & \tbf{16.3349} & \tbf{25.200 (0.416)} & $\bm{-7.374}$ & $\bm{7.104}$ \\
& \tbf{HCN} & $\bm{J=3-2}$ & \tbf{265.886} & $\bm{-3.0780}$ & \tbf{25.5209} & \tbf{6.836 (0.302)} & $\bm{-8.018}$ & $\bm{4.268}$ \\
& \tbf{HNC} & $\bm{J=3-2}$ & \tbf{271.981} & $\bm{-3.0298}$ & \tbf{26.1069} & \tbf{2.533 (0.354)} &  $\bm{-7.220}$ & $\bm{3.499}$\\
& \tbf{HCO}$\bm{^+}$ & $\bm{J=3-2}$ & \tbf{267.557} & $\bm{-2.8376}$ & \tbf{25.6817} & \tbf{1.115 (0.450)} & $\bm{-7.216}$ & $\bm{3.010}$\\
C43-4 & $^{12}$CO & $J=2-1$  & 230.538 & -6.1605 & 16.5961 & 21.848 (0.212) & $-17.996$ & $18.732$ \\
& $\bm{^{13}}$\tbf{CO} & $\bm{N=2-1}$  & \tbf{220.398} & $\bm{-6.2164}$ & \tbf{15.8662} & \tbf{5.831 (0.0753)} & $\bm{-17.346}$ & $\bm{15.941}$\\
& \tbf{CN} & $\bm{N=2-1^b}$ & \tbf{226.697} & $\bm{-4.2783}$ & \tbf{16.3089} & \tbf{1.326 (0.227)} & $\bm{-4.246}$ & $\bm{5.812}$ \\
&    & $\bm{N=2-1^b}$ & \tbf{226.874} & $\bm{-3.9419}$ & \tbf{16.3349} &  \tbf{8.361 (0.299)} & $\bm{-7.006}$ & $\bm{15.504}$ \\
C43-3 & \tbf{HCN} & $\bm{J=3-2}$ & \tbf{265.886} & $\bm{-3.0780}$ & \tbf{25.5209} & \tbf{3.540 (0.149)} & $\bm{-14.047}$ & $\bm{11.149}$ \\
& \tbf{HCO}$\bm{^+}$ & $\bm{J=3-2}$ & \tbf{267.557} & $\bm{-2.8376}$ & \tbf{25.6817} & \tbf{0.532 (0.213)} & $\bm{-14.028}$ & $\bm{8.754}$\\
\end{tabular}
\end{adjustwidth}
\end{center}

{\sc Notes:} 
New molecule detections are denoted by bold font. a) Integrated line flux within $15''$ x $25''$ synthetic beam centered on (9:25:30.467, $-$55:53:13.217) position. b) Hyperfine complex; representative transition frequency, $A_{ul}$, and upper level energy listed retrieved from CDMS.
\end{table}

\begin{table}
\begin{center}
\caption{\sc NGC 2818: Detected Molecular Transitions}
\vspace{.2in}
\label{tbl:N2818detections}
\begin{adjustwidth}{-1.cm}{}    
\footnotesize
\begin{tabular}{lrccccccc}
\toprule
ALMA\ & Molecule & QN\ & $\nu$ & $\log{A_{ul}}$ & $E_u/k$  & $I^a$ ($\sigma_I$) & $V_{LSR}$ & $\Delta V_{1/2}$\\
Config. & & & (GHz) & (s$^{-1}$) & (K) & (Jy km s$^{-1}$) & (km s$^{-1}$) & (km s$^{-1}$)\\
\midrule
\midrule
ACA & $^{12}$CO & $J=2-1$  & 230.538 & $-6.1605$ & 16.5961 & 17.167 (0.779) & $3.431$ & $14.604$\\
& $\bm{^{13}}$\tbf{CO} & $\bm{N=2-1}$  & \tbf{220.398} & $\bm{-6.2164}$ & \tbf{15.8662} & \tbf{3.466 (0.351)} & $\bm{9.942}$ & $\bm{9.632}$\\
& \tbf{CN} & $\bm{N=2-1^b}$ & \tbf{226.874} & $\bm{-3.9419}$ & \tbf{16.3349} & \tbf{10.459 (0.508)} & $\bm{1.646}$ & $\bm{15.501}$\\
& \tbf{HCN} & $\bm{J=3-2}$ & \tbf{265.886} & $\bm{-3.0780}$ & \tbf{25.5209} & \tbf{3.271 (0.238)} & $\bm{10.356}$ & $\bm{10.186}$\\
& \tbf{HNC} & $\bm{J=3-2}$ & \tbf{271.981} & $\bm{-3.0298}$ & \tbf{26.1069} & \tbf{0.949 (0.208)} & $\bm{3.060}$ & $\bm{2.700}$\\
& \tbf{HCO}$\bm{^+}$ & $\bm{J=3-2}$ & \tbf{267.557} & $\bm{-2.8376}$ & \tbf{25.6817} & \tbf{0.561 (0.122)} & $\bm{2.671}$ & $\bm{3.284}$\\
&    & $\bm{N=2-1^b}$ & \tbf{226.874} & $\bm{-3.9419}$ & \tbf{16.3349} & \tbf{1.609 (0.141)} & $\bm{1.509}$ & $\bm{19.374}$ \\
C43-3 & \tbf{HCN} & $\bm{J=3-2}$ & \tbf{265.886} & $\bm{-3.0780}$ & \tbf{25.5209} & \tbf{1.962 (0.139)} & $\bm{4.468}$ & $\bm{16.104}$\\
& \tbf{HNC} & $\bm{J=3-2}$ & \tbf{271.981} & $\bm{-3.0298}$ & \tbf{26.1069} & \tbf{2.446 (0.056)} & $\bm{2.242}$ & $\bm{14.017}$\\
& \tbf{HCO}$\bm{^+}$ & $\bm{J=3-2}$ & \tbf{267.557} & $\bm{-2.8376}$ & \tbf{25.6817} & \tbf{0.778 (0.210)} & $\bm{-2.226}$ & $\bm{6.018}$ \\
C43-4 & $^{12}$CO & $J=2-1$  & 230.538 & $-6.1605$ & 16.5961 & 5.291 (0.253) & $7.004$ & $22.859$\\
& $\bm{^{13}}$\tbf{CO} & $\bm{N=2-1}$  & \tbf{220.398} & $\bm{-6.2164}$ & \tbf{15.8662} & \tbf{1.187 (0.0886)} & $\bm{4.066}$ & $\bm{17.270}$ \\
& \tbf{CS} & $\bm{N=5-4}$  & \tbf{244.936} & $\bm{-3.5257}$ & \tbf{35.2660} & \tbf{0.552 (0.138)} & $\bm{5.554}$ & $\bm{13.746}$ \\
& \tbf{CN} & $\bm{N=2-1^b}$ & \tbf{226.697} & $\bm{-4.2783}$ & \tbf{16.3089} & \tbf{0.587 (0.113)} & $\bm{9.007}$ & $\bm{17.114}$ \\
\end{tabular}
\end{adjustwidth}
\end{center}

{\sc Notes:} 
New molecule detections are denoted by bold font. a) Integrated line flux within $12''$ x $18''$ synthetic beam centered on (9:14:00.276, $-$36:25:04.002) position. b) Hyperfine complex; representative transition frequency, $A_{ul}$, and upper level energy listed retrieved from CDMS.
\end{table}

%\end{comment}

%\include{Table}
%\include{appendix_ratImages}
\newpage
\section{ALMA Band 6 Observations: Molecular Line Spectra}\label{sec:app4}

\begin{figure}[h!]
    \centering
    \includegraphics[width=0.9\textwidth]{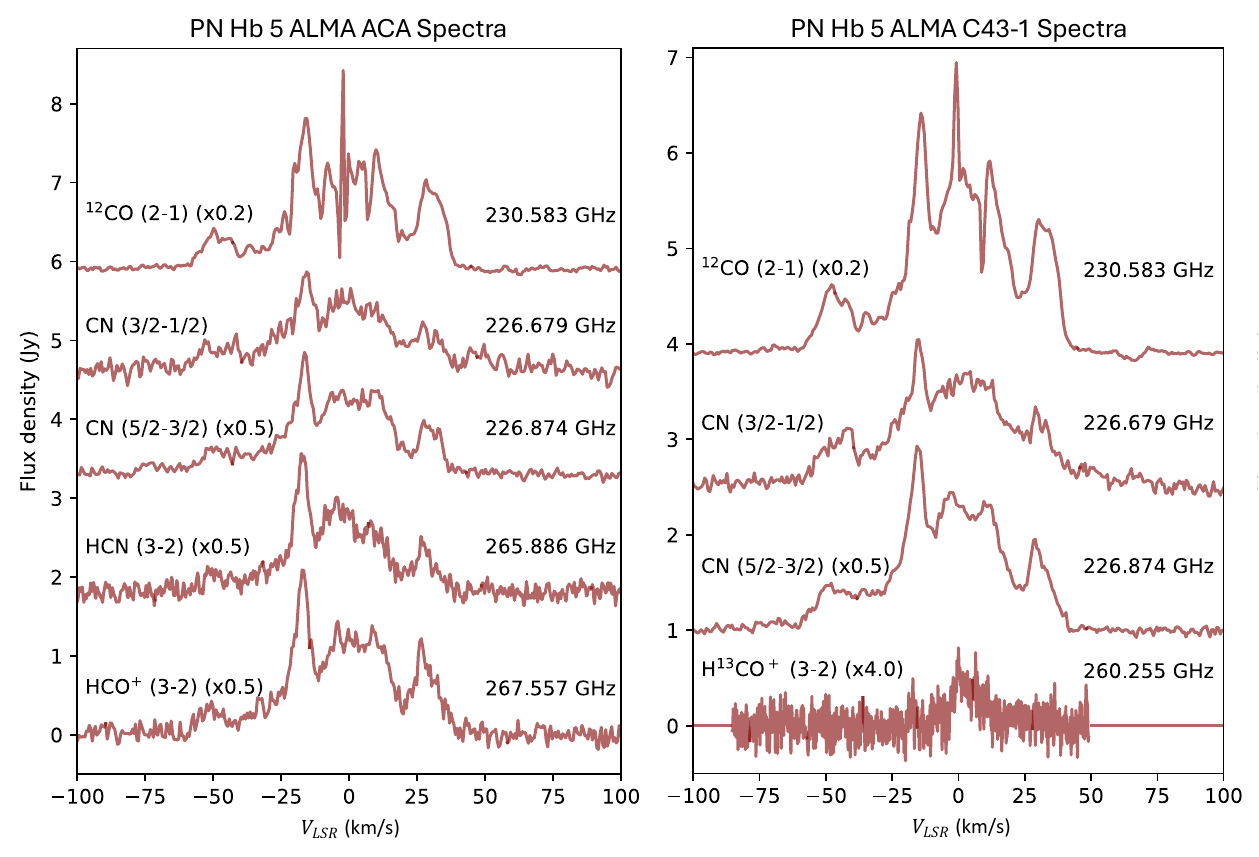}
    \caption{\textit{Left panel}: spectral profiles of PN Hb 5 ALMA Band 6 ACA data. The \textit{Right panel}: spectral profiles of PN Hb 5 ALMA Band 6 12-m data. In some cases, molecular spectra have been magnified or de-magnified to fit in the same plot. These magnifications are stated next to the molecule name and transition.}
    \label{fig:Hb5Spect}
\end{figure}

\begin{figure}[h!]
    \centering
    \includegraphics[width=1.0\textwidth]{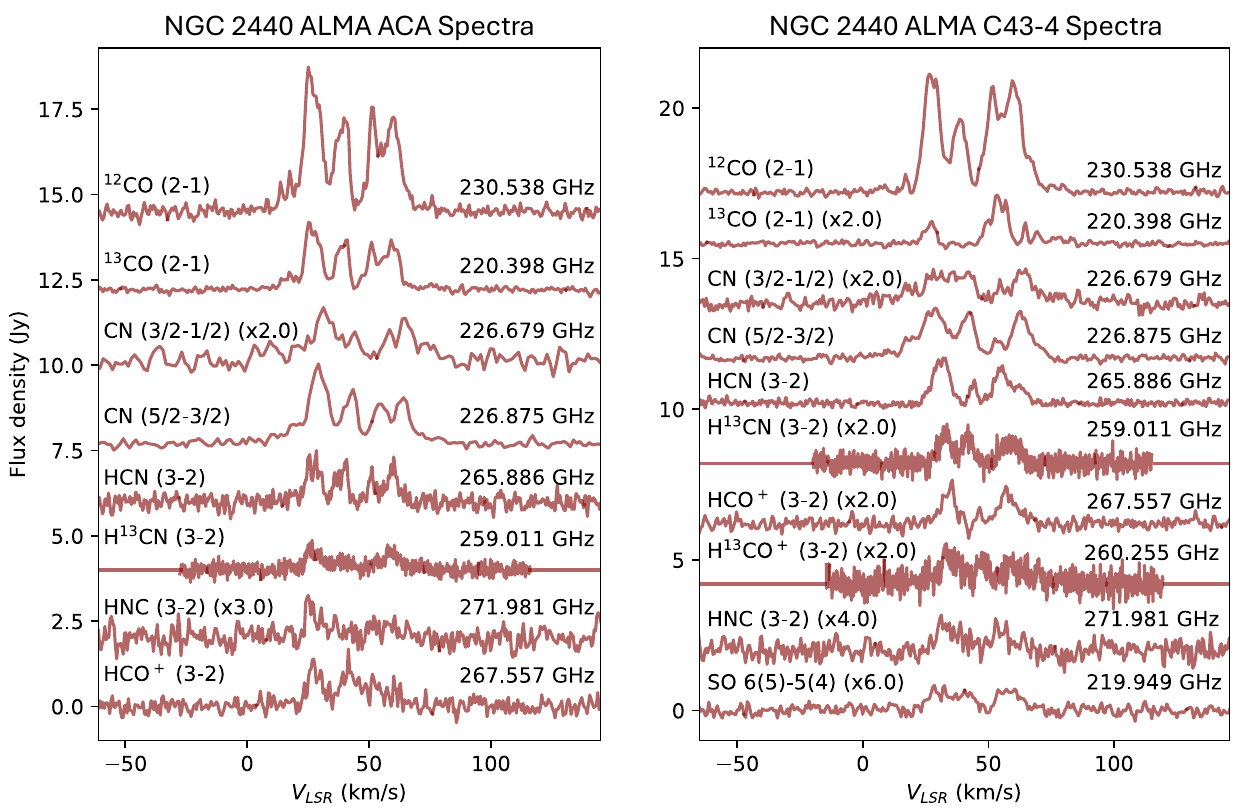}
    \caption{\textit{Left panel}: spectral profiles of NGC 2440 ALMA Band 6 ACA data. The \textit{Right panel}: spectral profiles of NGC 2440 ALMA Band 6 12-m data. Magnifications are stated next to the molecule name and transition if applicable.}
    \label{fig:N2440Spect}
\end{figure}

\begin{figure}[h!]
    \centering
    \includegraphics[width=1.0\textwidth]{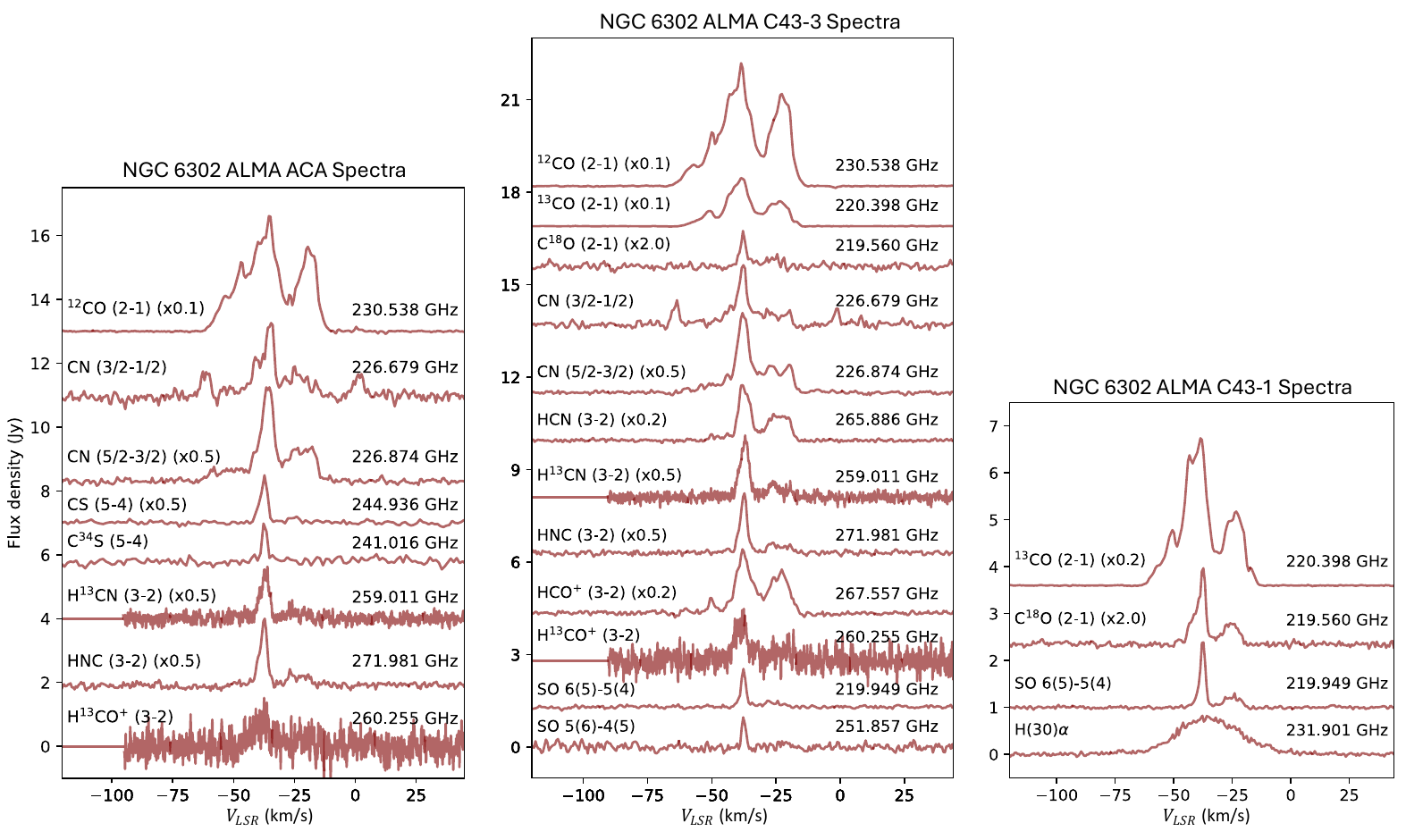}
    \caption{\textit{Left panel}: spectral profiles of NGC 6302 ALMA Band 6 ACA data. \textit{Middle panel}: spectral profiles of NGC 6302 ALMA Band 6 12-m (C43-4) data. \textit{Right panel}: spectral profiles of NGC 6302 ALMA Band 6 12-m (C43-1) data. Magnifications are stated next to the molecule name and transition if applicable.}
    \label{fig:N6302Spect}
\end{figure}

\begin{figure}[h!]
    \centering
    \includegraphics[width=1.0\textwidth]{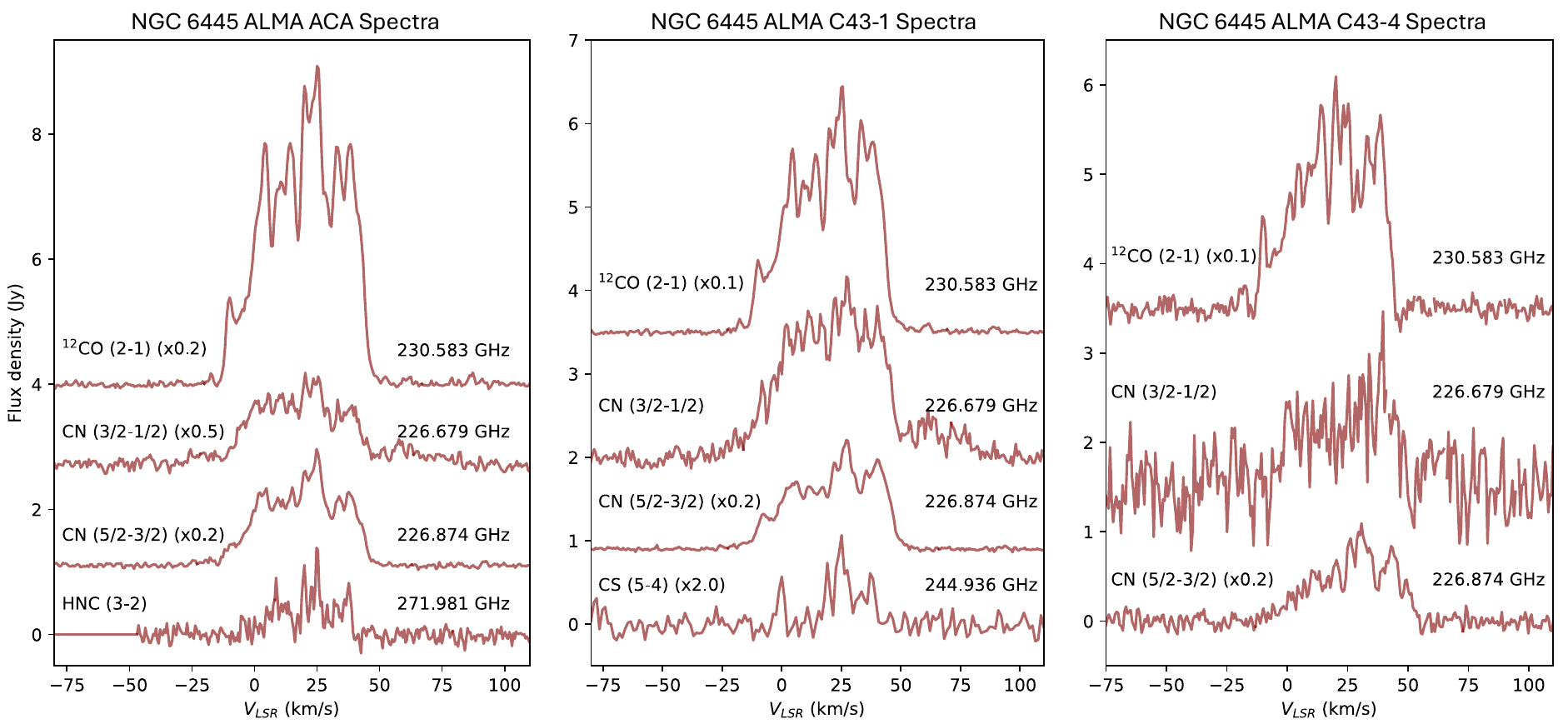}
    \caption{\textit{Left panel}: spectral profiles of NGC 6445 ALMA Band 6 ACA data. \textit{Middle panel}: spectral profiles of NGC 6445 ALMA Band 6 12-m (C43-1) data. \textit{Right panel}: spectral profiles of NGC 6445 ALMA Band 6 12-m (C43-4) data. Magnifications are stated next to the molecule name and transition if applicable.}
    \label{fig:N6445Spect}
\end{figure}

\begin{figure}[h!]
    \centering
    \includegraphics[width=0.9\textwidth]{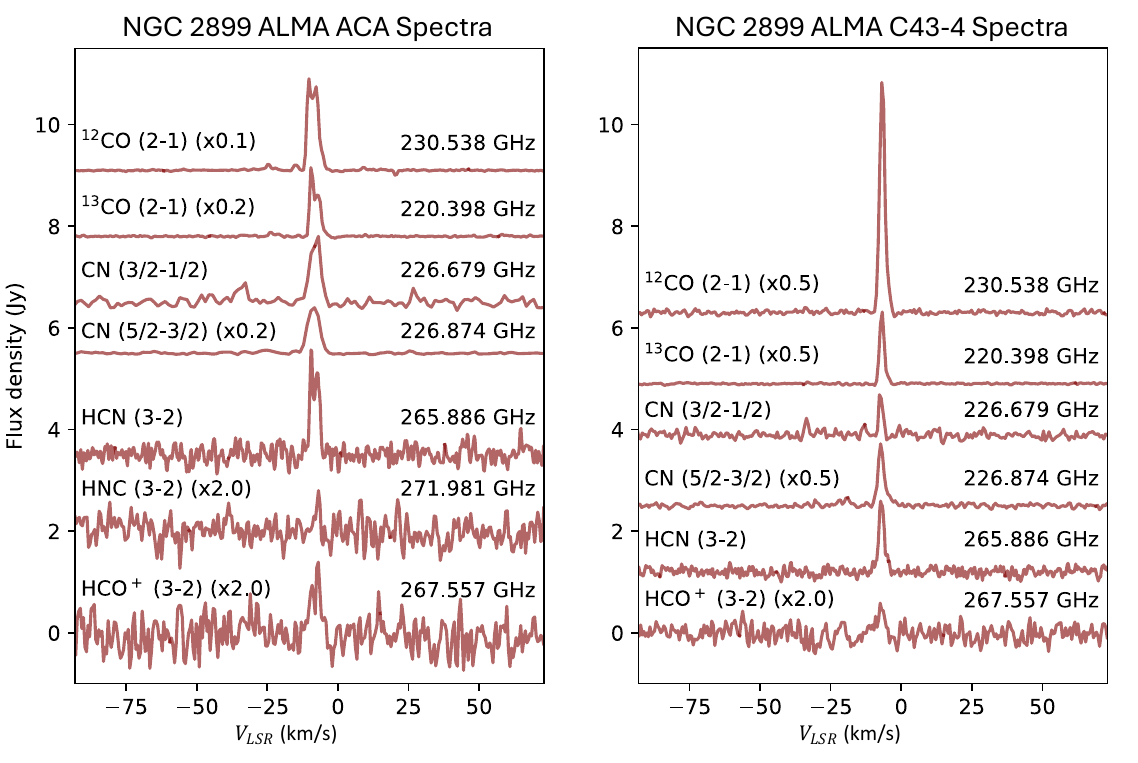}
    \caption{\textit{Left panel}: spectral profiles of NGC 2899 ALMA Band 6 ACA data. \textit{Right panel}: spectral profiles of NGC 2899 ALMA Band 6 12-m (C43-4) data. Magnifications are stated next to the molecule name and transition if applicable.}
    \label{fig:N2899Spect}
\end{figure}

\begin{figure}[h!]
    \centering
    \includegraphics[width=0.9\textwidth]{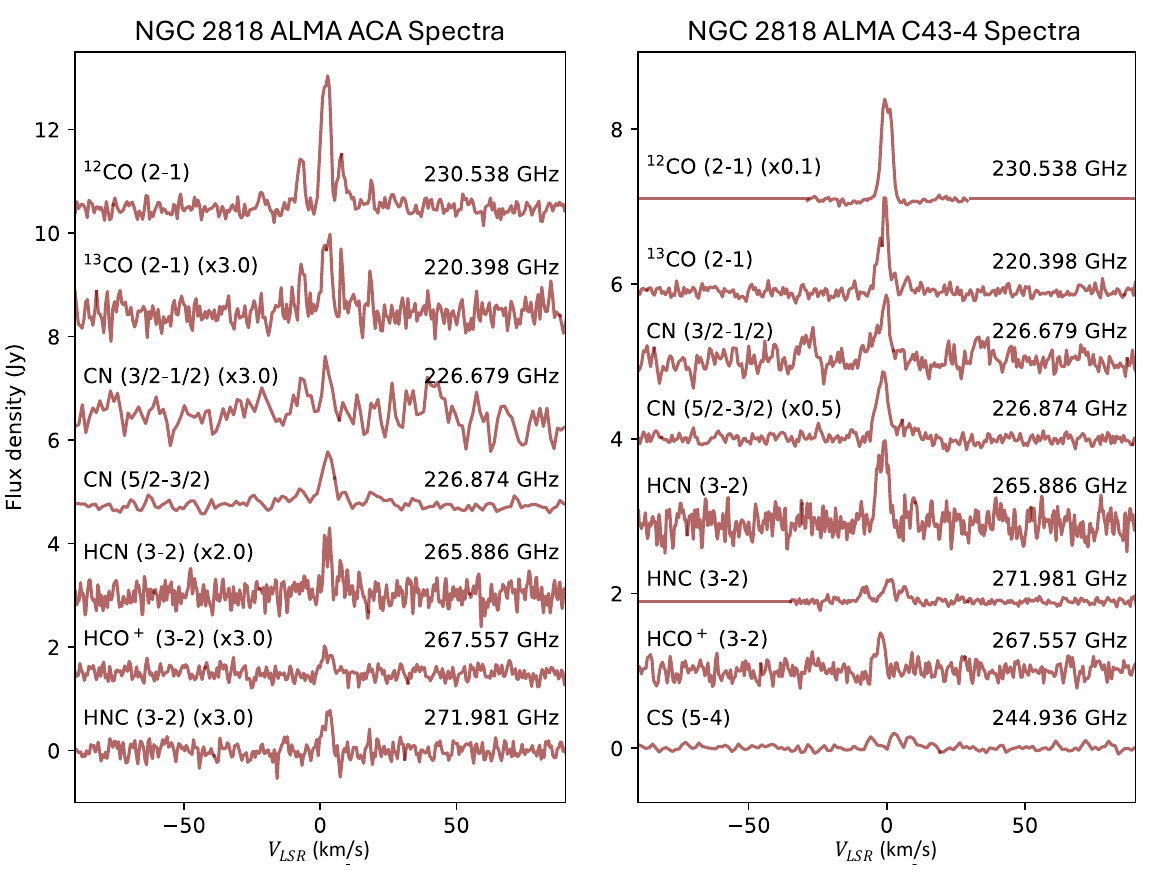}
    \caption{\textit{Left panel} spectral profiles of NGC 2818 ALMA Band 6 ACA data. \textit{Right panel} spectral profiles of NGC 2818 ALMA Band 6 12-m (C43-4) data. Magnifications are stated next to the molecule name and transition if applicable.}
    \label{fig:N2818Spect}
\end{figure}

Molecular emission line profiles extracted from the data cubes for the lines detected in each PN are presented in Figures~\ref{fig:Hb5Spect}--\ref{fig:N2818Spect}. In some cases, molecular spectra have been magnified or de-magnified to fit in the same plot. Magnifications are stated next to the molecule name and transition if applicable.

%\include{FigureSet}
%\begin{figure}[h!]
%    \centering
%    \includegraphics[width=0.9\textwidth]{}
%    \caption{Each panel corresponds to the molecular line spectra of each PN in our sample observed with a specific ALMA Band 6 configuration. This is a place holder for the figure set that will be created with help of the editor.}
%    \label{fig:allSpect}
%\end{figure}

%\begin{comment}
%\end{comment}

\end{document}